\documentclass{article}
\usepackage[dvips]{graphicx}

\usepackage[centertags]{amsmath}
\usepackage{amsfonts}
\usepackage{amssymb}
\usepackage{amsthm}
\usepackage{newlfont}
\usepackage[english]{babel}
\usepackage{pstcol}
\usepackage{pst-node}
\usepackage{pst-plot}
\numberwithin{equation}{section}

\newcommand{\R}{\mathbb{R}}

\newcommand{\Z}{\mathbb{Z}}

\newcommand{\N}{\mathbb{N}}

\newcommand{\C}{\mathbb{C}}

\newcommand{\tr}{{\mathrm tr}}

\newcommand{\I}{\mbox{Im }}
\newcommand{\E}{\mathcal{E}}
\newcommand{\Ra}{\mbox{Re }}

\newcommand{\G}{\mathcal{G}}
\theoremstyle{plain}
\newtheorem{thm}{Theorem}
\newtheorem{prop}{Proposition}
\newtheorem{lem}{Lemma}[section]
\newtheorem{cor}{Corollary}
\newtheorem{defn}{Definition}[section]
\newenvironment{dem}{{\bf Proof} \rm }{$\diamondsuit$}
\theoremstyle{definition}

\makeindex

\begin{document}
\title{On the eigenvalues for slowly varying perturbations of a periodic Schr{\"o}dinger operator}
\author{Magali Marx}
\date{}
\maketitle


\begin{abstract}
In this paper, I consider one-dimensional periodic Schr{\"o}dinger
operators perturbed by a slowly decaying potential. In the
adiabatic limit, I give an asymptotic expansion of the eigenvalues
in the gaps of the periodic operator. When one slides the
perturbation along the periodic potential, these eigenvalues
oscillate. I compute the exponentially small amplitude of the
oscillations.
\end{abstract}
\section{Introduction}
\label{intro} In this paper, we study the spectrum of
one-dimensional perturbed periodic Schr{\"o}dinger operators.
Precisely, we consider the Schr{\"o}dinger operator defined on
$L^{2}(\R)$ by:
\begin{equation}
\label{eqpa}
H_{\varphi,\varepsilon}=-\frac{d^{2}}{dx^{2}}+[V(x)+W(\varepsilon
x+\varphi)],
\end{equation}
where $\varepsilon>0$ is a small positive parameter, $\varphi$ is
a real parameter, and $V$ is a real valued 1-periodic function. We
also assume that $V$ is $L^{2}_{\textrm{loc}}$ and that $W$ is a
fast-decaying function.\\
The operator $H_{\varphi,\varepsilon}$ can be regarded as an
adiabatic perturbation of the periodic operator $H_{0}$:
\begin{equation}
\label{opper} H_{0}=-\triangle+V.
\end{equation}
The spectrum of the periodic operator $H_{0}$ is absolutely
continuous and consists of intervals of the real axis called the
spectral bands, separated by the gaps.\\
If the perturbation $W$ is relatively compact with respect to
$H_{0}$, there are in the gaps of $H_{0}$ some eigenvalues
\cite{Zhe, RB}. We intend to locate these eigenvalues, called
impurity levels.\\
The equation
\begin{equation}
\label{eqp} H_{\varphi,\varepsilon}\psi=E\psi
\end{equation}
depends on two parameters $\varepsilon$ et $\varphi$. We study the
operator $H_{\varphi,\varepsilon}$ in the adiabatic limit, i.e as
$\varepsilon\rightarrow 0$. The periodicity of $V$ implies that
the eigenvalues of $H_{\varphi,\varepsilon}$ are
$\varepsilon$-periodic in $\varphi$. We shall shift $\varphi$ in
the complex plane and we shall assume that $W$ is analytic in a
strip of the complex plane.\\
If $V=0$, there are many results. The case when $W$ is a well has
been studied; in the interval $]\inf\limits_{\R} W,0[$, there is a
quantified sequence of eigenvalues \cite{Fe}. We shall give an
analogous description of the eigenvalues of
$H_{\varphi,\varepsilon}$ in an interval $J$ out of the spectrum
of $H_{0}$. Precisely, when $W$ and $J$ satisfy some additional
conditions described in sections \ref{assW1}, \ref{assW2} et
\ref{assJ}, we show that the eigenvalues of
$H_{\varphi,\varepsilon}$ oscillate around some quantized
energies. The quantization is given by a Bohr-Sommerfeld
quantization rule; the amplitude of oscillation is exponentially
small and is determined by a tunneling coefficient.\\
\subsection{Physical motivation}
The operator $H_{\varphi,\varepsilon}$ is an important model of
solid state physics. The function $\psi$ is the wave function of
an electron in a crystal with impurities. $V$ represents the
potential of the perfect crystal; as such it is periodic. The
potential $W$ is the perturbation created by impurities. In the
semiconductors, this perturbation is slow-varying \cite{Zi}. It is
natural to consider the semi-classical limit.
\subsection{Perturbation of periodic operators}
In $\R^{d}$, the spectral theory of the perturbations of a
periodic operator
\begin{equation}
\label{hp} H_{P}=H_{0}+P
\end{equation}
has motivated numerous studies with different view points.\\
The characterization of the existence of eigenvalues is not easy:
particularly, in any dimension, \cite{KuVa2} deals with the
existence of embedded eigenvalues in the bands. On the real axis,
the situation is simpler. When the perturbation is integrable, the
eigenvalues are necessarily in the adherence of the gaps (\cite{RoBe, HiSh1}).\\
To count the eigenvalues in the gaps, many results have been
obtained thanks to trace formulas. In the large coupling constant
limit, i.e when $P=\lambda U$, with $\lambda\rightarrow+\infty$,
\cite{ADH, Bi1, So2} have studied
$\lim\limits_{\lambda\mapsto+\infty}\textrm{tr}(P_{[E,E']}^{(\lambda)})$,
where $P_{[E,E']}^{(\lambda)}=1_{[E,E']}H_{\lambda}$ (spectral
projector of $H_{\lambda}$ on an interval $[E,E']$ of a gap of
$H_{0}$). In the semi-classical case, \cite{Di1} has given, under
assumptions close to mine, an asymptotic expansion of $\textrm{tr}
[f(H_{\varphi,\varepsilon})]$, for $f\in C_{0}^{\infty}(\R)$ and
$\textrm{Supp }f$ in a gap of $H_{0}$. These formulas are valid in
any dimension but are less accurate. For example, in the expansion
obtained in \cite{Di1}, the accuracy depends on the successive
derivatives of the function $f$; the formula does not
give an exponentially precise localization of the eigenvalues.\\
In the one-dimensional case, the scattering theory, well-known in
the case $V=0$, has been developed in \cite{Fi1, New} for the
periodic case. Precisely, we construct some particular solutions
of equation \eqref{hp}, which tend to zero as $x$ tends to
infinity. We call these functions recessive functions. The
eigenvalues of equation \eqref{eqp} are given by a relation of
linear dependence between these solutions.
\subsection{Main steps of the study}
\label{ppalet} We give here the main ideas of the paper. An
important difficulty is the dependence of the equation on the
parameters $\varepsilon$ and $\varphi$; particularly, one has to
decouple the ``fast'' variable $x$ and the ``slow'' variable
$\varepsilon x$. The new idea developed in \cite{FK1, FK2} is the
following : we construct some particular solutions of \eqref{eqp},
satisfying an additional relation called the consistency
condition:
\begin{equation}
\label{coh}
f(x+1,\varphi,E,\varepsilon)=f(x,\varphi+\varepsilon,E,\varepsilon).
\end{equation}
This condition relates their
behavior in $x$ and their behavior in $\varphi$.\\
To find a recessive solution of \eqref{eqp}, it suffices to
construct a solution of \eqref{eqp} which satisfies \eqref{coh}
and which tends to $0$ as $|\Ra\varphi|$ tends to $+\infty$.
First, we build on the horizontal half-strip $\{\varphi\in \C\ ;\
\varphi\in]-\infty,-A]+i[-Y,Y]\}$ a solution $h_{-}^{g}$ of
equation \eqref{eqp} which is consistent and which tends to $0$ as
$\Ra\varphi$ tends to $-\infty$. Similarly, we construct
$h_{+}^{d}$ for $\{\varphi\in \C\ ;\
\varphi\in[A,+\infty[+i[-Y,Y]\}$ (Theorem \ref{jostthm}). These
functions are recessive for the variable $x$. The characterization
of the eigenvalues is given by the relation of linear dependence
between $h_{-}^{g}$ and $h_{+}^{d}$:
$$w(h_{-}^{g},h_{+}^{d})=0.$$
In the above-mentioned equation, $w$ represents the Wronskian
whose definition is recalled in \eqref{wronsk}.\\
It remains to compute $w(h_{-}^{g},h_{+}^{d})$. To do that, we use
the complex WKB method developed by A. Fedotov and F. Klopp. This
method consists in describing some complex domains, called
canonical domains, on which we construct some functions satisfying
\eqref{coh} and having a particular asymptotic behavior:
\begin{equation}
\label{stdas}
f_{\pm}(x,\varphi,E,\varepsilon)=e^{\pm\frac{i}{\varepsilon}\int^{\varphi}\kappa}(\psi_{\pm}(x,\varphi,E)+o(1)),\quad
\varepsilon\rightarrow 0.
\end{equation}
In equation \eqref{stdas}, the function $\kappa$ is a analytic
multi-valued function, defined in \eqref{momcompa}; the functions
$\psi_{\pm}$ are some particular solutions of equation
$$H_{0}\psi=(E-W(\varphi))\psi,$$ analytic in $\varphi$ on these canonical domains and called Bloch solutions. We will prove the existence of such functions
in section \ref{cansolbloch}.\\
A. Fedotov and F. Klopp prove the existence of functions with
standard asymptotic only on compact domains of the complex plane.
We shall extend some results on infinite strips of the complex
plane. The consistency condition implies that the function
$h_{-}^{g}$ satisfies the standard asymptotic \eqref{stdas} to the
left of $-A$ and that $h_{+}^{d}$ satisfies an analogous property
to the right of $A$. Thus, the computation of
$w(h_{-}^{g},h_{+}^{d})$ is similar to the calculations of A.
Fedotov and F. Klopp. We must find a sufficiently large domain of
the complex plane, in which we know the Wronskian of $h_{-}^{g}$ and $h_{+}^{d}$.\\
The methods used in their works underline some topological
obstacles, which change the standard asymptotic \eqref{stdas};
these obstacles depend on $W$ and $E$. We give precise assumptions
in sections \ref{assW} and \ref{assJ}.
\section{The main results}
\label{resppaux} In this section, we describe the general context
and the main results of the paper.\\
First, we present the assumptions on the potentials $V$ and $W$,
and on the interval $J$. There are mainly three kinds of
assumptions. Firstly, the study requires some assumptions on the
decay of $W$ to develop the scattering theory. Then, in view of
the hypotheses of the complex WKB method of \cite{FK1}, we assume
that $W$ is analytic in some domain of the complex plane. Finally,
we shall depict the geometric framework and particularly the
subset
$(E-W)^{-1}(\R)$.\\
We obtain an equation for the eigenvalues in terms of geometric
objects depending on $H_{0}$, $W$ and $E$: the phases and action
integrals, defined in sections \ref{splecross}.
\subsection{The
potential $V$}
\label{assV} We assume that $V$ has the following properties:\\
\\
{\bf ($\mathbf{H_{V,p}}$) $\mathbf{V}$ is
$\mathbf{L^{2}_{\textrm{loc}}}$, $\mathbf{1}$-periodic.}\\
\\
We consider \eqref{eqp} as a perturbation of the periodic
equation:
\begin{equation}
-\frac{d^{2}}{dx^{2}}\psi(x)+V(x)\psi(x)=(E-W(\varphi))\psi(x).\label{esp}
\end{equation}
We shall use some well known facts about periodic Schr{\"o}dinger
operators. They are described in detail in section \ref{opepera}.\\
We just recall elementary results on $H_{0}$. The operator $H_{0}$
defined in \eqref{opper} is a self-adjoint operator on
$H^{2}(\R)$. The spectrum of $H_{0}$ consists of intervals of the
real axis:
\begin{equation}
\label{band}
\sigma(H_{0})=\bigcup\limits_{n\in\N}[E_{2n+1},E_{2n+2}],
\end{equation}
such that:
$$ E_{1}<E_{2}\leq E_{3}< E_{4}...E_{2n}\leq E_{2n+1}<
E_{2n+2}...,\quad E_{n}\rightarrow + \infty,n\rightarrow
+\infty.$$
 These intervals $[E_{2n+1},E_{2n+2}]$ are called the
{\it spectral bands}. We set $E_{0}=-\infty$. The intervals
$(E_{2n},E_{2n+1})$ are called the {\it spectral gaps}. If
$E_{2n}\neq E_{2n+1}$, we say that the gap is open.\\
Furthermore, we assume that $V$ satisfies:
\\
{\bf ($\mathbf{H_{V,g}}$) Every gap of $\mathbf{H_{0}}$ is not empty.}\\
\\
This assumption is ``generic'', we refer to \cite{ReSi4} section
XIII.16. An important object of the theory of one-dimensional
periodic operators is the Bloch quasi-momentum $k$ (see section
\ref{qm}). This function is a multi-valued analytic function; its
branch points are the ends of the spectrum, they are of square
root type. We shall give a few details about this function in
section
\ref{opepera}. Finally, we suppose:\\
\\
{\bf ($\mathbf{H_{V}}$) $\mathbf{V}$ satisfy ($\mathbf{H_{V,p}}$)
and
($\mathbf{H_{V,g}}$).}\\
\subsection{The perturbation $W$}
\label{assW}
\subsubsection{Smoothness assumptions}
\label{assW1} We assume that $W$ is such that:\\
\\
{\bf ($\mathbf{H_{W,r}}$) There exists $\mathbf{Y>0}$ such that
$\mathbf{W}$ is analytic in the strip
$\mathbf{S_{Y}=\{|\I(\xi)|\leq Y\}}$ and there exists
$\mathbf{s>1}$ et $C>0$ such that for $\mathbf{z\in S_{Y}}$, we
have:}
\begin{equation}
\mathbf{|W(z)|\leq\frac{C}{1+|z|^{s}}}.\end{equation} These
assumptions are essential to develop the complex WKB method. The
analyticity of the perturbation is crucial in the theory of
\cite{FK1}. The decay of $W$ replaces the
compactness resulting from periodicity in \cite{FK1}.\\
We begin with presenting the complex momentum. This main object of
the complex WKB method shows the importance of $W^{-1}(\R)$.
\subsubsection{The complex momentum and its branch points}
\label{momcompb} We put:
$$\C_{+}=\{\varphi\in\C\ ;\ \I\varphi\geq
0\}\textrm{ and } \C_{-}=\{\varphi\in\C\ ;\ \I\varphi\leq 0\}.$$
For equation \eqref{eqp}, we consider the analytic function
$\kappa$ defined by
\begin{equation}
\label{momcompa} \kappa(\varphi)=k(E-W(\varphi)).
\end{equation}
We recall that the function $k$ is presented in section
\ref{assV}. The function $\kappa$ is called the complex momentum.
It plays a
crucial role in adiabatically perturbed problems, see \cite{Bu, FK1}.\\
$\N$ is the set of non-negative integers. We define:
\begin{equation}
\label{nupsilon} \Upsilon(E)=\{\varphi\in S_{Y}\ ;\ \exists\
n\in\N^{*}\ /\ E-W(\varphi)=E_{n}\}
\end{equation}
The set of branch points of $\kappa$ is clearly a subset of
$\Upsilon(E)$. The following result gives a characterization of
the branch points of $\kappa$ among the points of $\Upsilon(E)$:
\begin{lem}
Let $\varphi$ be a point of $\Upsilon(E)$. If $\inf\{q\ ;\
W^{(q)}(\varphi)\neq 0\}\in 2\N+1$, then $\varphi$ is a branch
point of $\kappa$.
\end{lem}
This result follows from the fact that the ends of the spectrum
are of square root type.
\subsubsection{Geometric assumptions}
\label{assW2} \label{descw} The spectrum $\sigma(H_{0})$ consists
of real intervals. Fix $E\in\R$. If $E-W(\varphi)$ is in the
spectrum $\sigma(H_{0})$, then $W(\varphi)$ is real. The spectral
study of (\ref{eqp}) is then tightly connected with the geometry
of $W^{-1}(\R)$.\\
We state now the geometric assumptions for $W$. These assumptions
are mainly a description of $W^{-1}(\R)$ in a strip containing the
real axis. We call strictly vertical a line whose slope does not
vanish; for precise definitions, we refer to section
\ref{vertdef}.
\\
{\bf ($\mathbf{H_{W,g}}$)\begin{enumerate}
\item $\mathbf{W_{|\R}}$ is real and has a finite number of extrema, which are non-degenerate. \item There exists $\mathbf{Y>0}$
and a finite sequence of strictly vertical lines containing an
extremum of $\mathbf{W}$, such that:
\begin{equation}
\mathbf{W^{-1}(\R)\cap S_{Y}=\bigcup\limits_{i\in \{1\ldots
p\}}(\Sigma_{i})\cup\R}.
\end{equation}\end{enumerate}}
\subsection{Some remarks}
\begin{itemize}
\item Since $W$ is real analytic, we know that $W(\overline{\varphi})=\overline{W(\varphi)}$; this implies that $W^{-1}(\R)$ is symmetric with respect to the real axis.
\item
We define $\Sigma_{i}^{+}=\Sigma_{i}\cap\C_{+}$ and
$\Sigma_{i}^{-}=\Sigma_{i}\cap\C_{-}$. \item Figure \ref{exW1}
shows an example of the pre-image of the real axis by such a
potential.
\end{itemize}
As we have explained in section \ref{ppalet}, we cover the strip
$S_{Y}$ with local canonical domains. On these domains, we
construct consistent functions with standard behavior (ie
satisfying
\eqref{coh} and \eqref{stdas}).\\
To compute the connection between the bases associated with
different domains, we get round the branch points (for analog
studies, we refer the reader to \cite{FR, FK2}). We will now state
some more accurate assumptions about the configuration of the
branch points; in particular, these assumptions specify
$(E-W)^{-1}(\sigma(H_{0}))$ when $E$ is real. The spectral results
of A. Fedotov and F. Klopp on perturbed periodic equation have
shown the
importance of the relative positions of $J$ and $\sigma(H_{0})$.\\
\psset{unit=1em,linewidth=.05}
\begin{center}
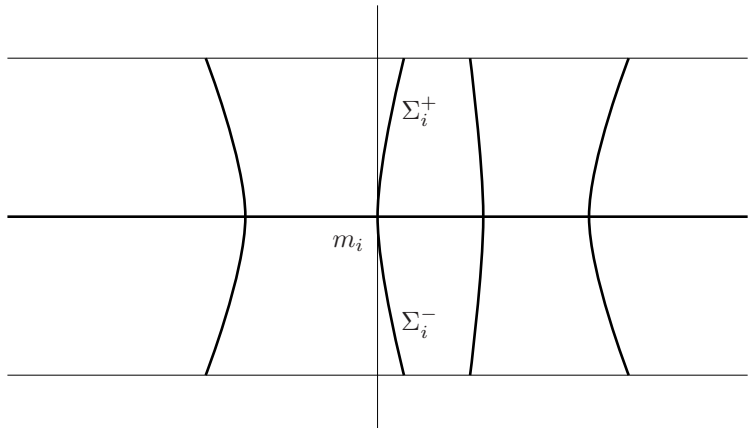
\begin{figure}
\begin{pspicture}(-20,-10)(20,10) \psline[linewidth=0.1](-14,0)(14,0)
\psline[linewidth=0.01](0,8)(0,-8)\psline[linewidth=0.01](-14,6)(14,6)
\psline[linewidth=0.01](-14,-6)(14,-6)
\pscurve[linewidth=0.1](-6.5,-6)(-5,0)(-6.5,6)
\pscurve[linewidth=0.1](1,-6)(0,0)(1,6)
\pscurve[linewidth=0.1](3.5,-6)(4,0)(3.5,6)
\pscurve[linewidth=0.1](9.5,-6)(8,0)(9.5,6)
\uput[180](0,-1){$m_{i}$}
\uput[180](2.8,4){$\Sigma_{i}^{+}$}\uput[180](2.8,-4){$\Sigma_{i}^{-}$}
\end{pspicture}
\caption{A subset of $W^{-1}(\R)$}\label{exW1}
\end{figure}
\end{center}
\subsection{Assumptions on the interval $J$} \label{assJ}
Now, we describe the interval $J$ on which we study equation
\eqref{eqp}.
\subsubsection{Hypotheses}
We assume that the interval $J$ is a compact interval satisfying:\\
\\
{\bf $\mathbf{(H_{J})}$ \begin{enumerate}
  \item For any $\mathbf{E\in J}$, there exists only one band $\mathbf{B}$ of
  $\mathbf{\sigma(H_{0})}$ such that the pre-image $\mathbf{C:=(E-W)^{-1}(B)}$ is not empty.
  \item For any $\mathbf{E\in J}$, $\mathbf{C:=(E-W)^{-1}(B)}$ is connected and compact and $\mathbf{(E-W)^{-1}(\stackrel{\circ}{B})}$ contains exactly one real extremum of $\mathbf{W}$.
\end{enumerate}}
\subsubsection{Consequences}
\begin{itemize}
\item $(H_{J})$ implies that $J$ is included in a gap.\item The band $B$ in $(H_{J})$ (1) depends a priori on $E$.
But, since $J$ is connected, the band $B$ is fixed for any $E\in
J$.\item Similarly, the extremum of $W$ in assumption $(H_{J})$
(2) depends on $E$, but by connectedness, it is the same for any
$E\in J$.\end{itemize}
\subsubsection{Notations}
Put $B=[E_{2n-1},E_{2n}]$, for $n\in\N^{*}$. Moreover, we can
always change $W$ or
$\varphi$ so that the extremum of $W$ in $(2)$ is $0$.\\
Then $(H_{J})$ has the following consequences:
\begin{enumerate}\item For any $E\in J$, $(E-W)^{-1}(\sigma(H_{0}))\cap
S_{Y}=(E-W)^{-1}(B)\cap S_{Y}$\item Let
$E_{r}\in\{E_{2n-1},E_{2n}\}$ be the end of $B$ satisfying
$E_{r}\in(E-W)(\R)$ for any $E\in J$. We define $E_{i}$ such that
$\{E_{i},E_{r}\}=\{E_{2n-1},E_{2n}\}$.\item There are exactly four
branch points $(\varphi_{r}^{-},\varphi_{r}^{+})\in\R^{2}$ and
$(\varphi_{i}, \overline{\varphi_{i}})$ in $S_{Y}$ related to
$E_{r}$ and $E_{i}$. They satisfy:
$$E-W(\varphi_{r}^{+})=E_{r},E-W(\varphi_{r}^{-})=E_{r},\
\varphi_{r}^{-}<0<\varphi_{r}^{+},$$
$$E-W(\varphi_{i})=E-W(\overline{\varphi_{i}})=E_{i},\
\I\varphi_{i}>0.$$\item There exists a strictly vertical line
$\sigma$ containing $0$ and  connecting $\overline{\varphi_{i}}$
to $\varphi_{i}$, such that $(E-W)^{-1}(B)\cap
S_{Y}=[\varphi_{r}^{-},\varphi_{r}^{+}]\cup\sigma$. We define
$\sigma_{+}=\sigma\cap\C_{+}$ and $\sigma_{-}=\sigma\cap\C_{-}$.
We let $\Sigma=(E-W)^{-1}(\R)\backslash\R$, $\sigma\subset\Sigma$.
\end{enumerate} These objects are described in figure \ref{pbf}.
\psset{unit=1em,linewidth=.05}

 \psset{unit=0.5em,linewidth=.05}
\begin{center}
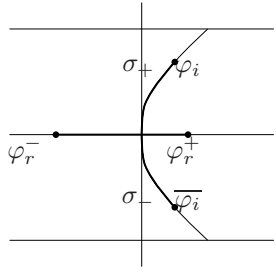
\begin{figure}
\begin{pspicture}(-50,-10)(10,10)
\psline[linewidth=0.05](-10,0)(10,0)\psline[linewidth=0.05](0,-10)(0,10)
\psline(-10,8)(10,8) \psline(-10,-8)(10,-8)
\psdots[dotstyle=*](-6.5,0)\uput[180](-6.5,-1.2){$\varphi_{r}^{-}$}
\psdots[dotstyle=*](3.5,0)\uput[180](5.5,-1.2){$\varphi_{r}^{+}$}
\psline(-6.5,0)(3.5,0)
\pscurve[linewidth=0.05](0,0)(0.25,2.3)(2.5,5.5)(5,8)\psdots[dotstyle=*](2.5,5.5)\uput[180](5.5,5){$\varphi_{i}$}\uput[180](2,5){$\sigma_{+}$}
\pscurve(0,0)(0.25,2.3)(2.5,5.5)
\pscurve[linewidth=0.05](0,0)(0.25,-2.3)(2.5,-5.5)(5,-8)\psdots[dotstyle=*](2.5,-5.5)\uput[180](5.5,-5){$\overline{\varphi_{i}}$}\uput[180](2,-5){$\sigma_{-}$}
\pscurve(0,0)(0.25,-2.3)(2.5,-5.5)
\pscurve[linewidth=0.15](0,0)(0.25,2.3)(2.5,5.5)
\pscurve[linewidth=0.15](0,0)(0.25,-2.3)(2.5,-5.5)
\psline[linewidth=0.15](-6.5,0)(3.5,0)
\end{pspicture}
\caption{$(E-W)^{-1}(\sigma(H_{0}))$}\label{pbf}
\end{figure}
\end{center}
\subsubsection{Remarks and examples}
We first give a few comments on assumption $(H_{J})$.
\begin{itemize}
\item
We call $C$ the cross. \item This assumption means intuitively
that, in $S_{Y}$, we see the band $B$ only near the extremum $0$.
\end{itemize}
To illustrate these technical assumptions, we give a few examples
of potentials $W$ and intervals $J$. We have depicted some
examples in figure \ref{exW2}.\begin{itemize}\item The simplest
case is when $W$ has only a non-degenerate minimum
$W_{-}$ (see figure \ref{exW2} A).\\
in concrete terms, we can think of the example:
$$ W(x)=\frac{-\alpha}{1+x^{2}},\quad \alpha>0,$$
Then, if we fix $B=[E_{2n-1},E_{2n}]$ and $Y<1$, we can choose
$J=[a,b]$ such that:
$$\max\{E_{2n-2},E_{2n-1}-\alpha,E_{2n}-\frac{\alpha}{1-Y^{2}}\}< a<b<\min\{E_{2n-1},E_{2n}-\alpha,E_{2n+1}-\frac{\alpha}{1-Y^{2}}\}$$
\item We can assume that $W$ has a maximum $W_{+}$ and
a minimum $W_{-}$, if $J$ is chosen to see the band only near the maximum (see figure \ref{exW2} B).\\
$$W(x)=\frac{2}{1+x^{2}}-\frac{1}{1+(x-5)^{2}} $$
$$J\subset]E_{2n-1}+W_{+},E_{2n-2}+W_{+}[\cup]E_{2n},E_{2n+1}+W_{-}[,\quad |J|\leq|E_{2n-2}-E_{2n-1}|$$
Consider this example a little further. The choice of $Y$ is more
complicated in this case. The study of equation $W(u)=w$ for
$w>W_{+}$ shows that there exists only one solution in the strip
$\{\I u\in]0,1[\}$ that we call $Z(w)$ ; we choose
$Y\in]\sup\limits_{E\in J}Z(E-E_{2l-1}),\inf\limits_{E\in
J}Z(E-E_{2l-2})[$.
\item In fact, we could adapt our method to weaker assumptions. For example, we can assume that we do not see the branch points $\varphi_{i}$
and $\overline{\varphi_{i}}$ (incomplete cross), which means that
the vertical line $\sigma$ does not contain any branch points of
$\kappa$. We refer to section \ref{unccross} for some details.
\item For the sake of simplicity, we have assumed that all the
extrema of $W$ are non degenerate. Actually, it suffices to assume
that only the extremum of $W$ in $0$ is non degenerate. \item
Similarly, we could weaken assumption $(H_{V,g})$. We only have to
assume that the gaps adjoining the band $B$ of $(H_{J})$ are not
empty.
\end{itemize}
\psset{unit=1em,linewidth=.05}

\psset{unit=0.5em,linewidth=.05}
\begin{center}
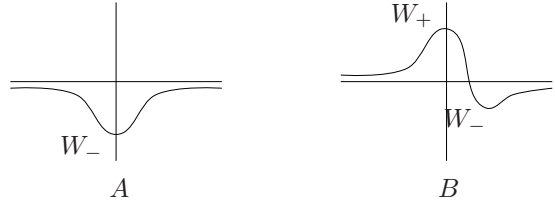
\begin{figure}
\begin{pspicture}(-30,-10)(35,10)
\psline(-8,0)(8,0)\psline(17,0)(33,0)
\psline(0,-6)(0,6)\psline(25,-6)(25,6)
\pscurve(-8,-0.5)(-3,-1)(0,-4)(3,-1)(8,-0.5)
\uput[180](0,-5){$W_{-}$} \uput[180](2,-8){$A$}
\uput[180](25,5){$W_{+}$}\uput[180](29,-3){$W_{-}$}
\pscurve(17,0.5)(22,1)(25,4)(28,-2)(30,-1)(33,-0.5)
\uput[180](27,-8){$B$}
\end{pspicture}
\caption{Some examples of potential $W$}\label{exW2}
\end{figure}
\end{center}
\subsection{Phases and action}
\label{splecross} In this section, we define the tunneling
coefficient $t$ and the phases $\Phi$ et $\Phi_{d}$; these
analytic objects play an essential role in the location of the
eigenvalues. These coefficients are represented as integrals of
the complex momentum $\kappa$ in the $\varphi$ plane.\\
In the strip $S_{Y}$, we consider $\kappa$ a branch of the complex
momentum, continuous on $C$.
\subsubsection{Definition and properties}
\label{chem} We introduce the action $S$ and the phases $\Phi$ and
$\Phi_{d}$ related to the branch $\kappa$.
\begin{defn}
We define the phase:
\begin{equation}
\label{phi}
\Phi(E)=\int_{\varphi_{r}^{-}}^{\varphi_{r}^{+}}(\kappa(u)-\kappa(\varphi_{r}^{-}))du,
\end{equation}
the action:
\begin{equation}
\label{action}
S(E)=i\int_{\sigma}(\kappa(u)-\kappa(\varphi_{i}))du,
\end{equation}
the second phase: \begin{equation} \label{phid}
\Phi_{d}(E)=\int_{\varphi_{r}^{-}}^{0}(\kappa(u)-\kappa(\varphi_{r}^{-}))du+\int_{\varphi_{r}^{+}}^{0}(\kappa(u)-\kappa(\varphi_{r}^{+}))du+\int_{\sigma_{+}}(\kappa(u)-\kappa(\varphi_{i}))du-\int_{\sigma_{-}}(\kappa(u)-\kappa(\overline{\varphi_{i}}))du.
\end{equation}
\end{defn}
In section \ref{anares}, we prove the following result on the
behavior of the coefficients $\Phi$, $S$ and $\Phi_{d}$.
\begin{lem}
\label{phaseactint} There exists a branch $\tilde{\kappa}_{i}$
such that the phases and action integrals have the following
properties:
\begin{enumerate}
\item $\Phi$, $S$, $\Phi_{d}$ are analytic in $E$ in a complex neighborhood of the interval $J$.
\item $\Phi$, $S$, $\Phi_{d}$ take real values on $J$. $\Phi$ and $S$ are positive on $J$.
\item $\forall E\in J,\quad \Phi'(E)(E_{i}-E_{r})>0,\quad S(E)\leq
2\pi\ \I(\varphi_{i}(E)).$
\end{enumerate}
\end{lem}
We define {\it the tunneling coefficient} :
\begin{equation}
t(E,\varepsilon)=\exp(-S(E)/\varepsilon).
\end{equation}
$t$ is exponentially small.
\subsubsection{Remark} The phase and action are simply a generalization
of the coefficients of the form $\int\sqrt{E-W(\varphi)}d\varphi$,
well-known in the case $V=0$ (we refer to \cite{Fe, FR, Ra}).\\
We point out that the coefficient $\Phi$ depend only on the value
of $W$ on the real axis, whereas $S$ and $\Phi_{d}$ depend on the
values of $W$ in the complex plane. The phase $\Phi$ is
independent of the analyticity of $W$ unlike $S$ and $\Phi_{d}$.\\
Now, we state the equation for eigenvalues for \eqref{eqp}.
\subsubsection{The main result}
\begin{thm}
\label{eigenloc}
Equation for eigenvalues.\\
Let $V$, $W$ and $J$ satisfy assumptions $(H_{V})$, $(H_{W,r})$, $(H_{W,g})$ and $(H_{J})$. Fix $Y_{0}\in]0,Y[$.\\
There exists a complex neighborhood $\mathcal{V}$ of $J$, a real
number $\varepsilon_{0}>0$ and two functions $\widetilde{\Phi}$
and $\widetilde{\Phi_{d}}$ with complex values, defined on
$\mathcal{V}\times]0,\varepsilon_{0}[$ such that:
\itemize{\item The functions $\widetilde{\Phi}(\cdot,\varepsilon)$
and $\widetilde{\Phi_{d}}(\cdot,\varepsilon)$ are analytic on
$\mathcal{V}$. Moreover, $\widetilde{\Phi}$ and
$\widetilde{\Phi_{d}}$ satisfy:
$$\widetilde{\Phi}(E,\varepsilon)=\Phi(E)+h_{0}(E,\varepsilon)\quad\textrm{
and
}\quad\widetilde{\Phi_{d}}(E,\varepsilon)=\Phi_{d}(E)+h_{1}(E,\varepsilon),$$
where $\rho$ is a real coefficient,
$h_{0}(E,\varepsilon)=o(\varepsilon)$ and
$h_{1}(E,\varepsilon)=o(\varepsilon)$ uniformly in
$E\in\mathcal{V}$. \item If we define the energy levels
$\{E^{(l)}(\varepsilon)\}$ in $J$ by:
\begin{equation}
\frac{\widetilde{\Phi}(E^{(l)}(\varepsilon),\varepsilon)}{\varepsilon}=l\pi+\frac{\pi}{2},\quad\quad\forall
l\in\{L_{-}(\varepsilon),\ldots,L_{+}(\varepsilon)\},
\end{equation}
then, for any $\varepsilon\in]0,\varepsilon_{0}[$,\itemize{\item
the spectrum of $H_{\varphi, \varepsilon}$ in $J$ consists in a
finite number of eigenvalues, that is to say
\begin{equation} \sigma(H_{\varphi, \varepsilon})\cap J
=\bigcup\limits_{l\in\{L_{-}(\varepsilon),\ldots,L_{+}(\varepsilon)\}}\{E_{l}(\varphi,\varepsilon)\},
\end{equation}\item these eigenvalues satisfy
{\footnotesize\begin{equation}\label{compcross}
E_{l}(\varphi,\varepsilon)=E^{(l)}(\varepsilon)+\varepsilon(-1)^{l+1}\frac{t(E^{(l)}(\varepsilon),\varepsilon)}{\Phi'(E^{(l)}(\varepsilon))}\left[\cos\left(\frac{\widetilde{\Phi_{d}}(E^{(l)}(\varepsilon),\varepsilon)+2\pi\varphi+\rho\varepsilon}{\varepsilon}\right)+t(E^{(l)}(\varepsilon),\varepsilon)r(E^{(l)}(\varepsilon),\varphi,\varepsilon)\right],
\end{equation}}
where there exists $c>0$ such that
$$\sup\limits_{E\in\mathcal{V},\varphi\in\R}r(E,\varphi,\varepsilon)<\frac{1}{c}e^{-\frac{c}{\varepsilon}}.$$}}
\end{thm}
We prove this result in section \ref{anares}.
\subsubsection{Remark}
\label{unccross} If we only assume that $\sigma$ does not contain
any branch points, asymptotic \eqref{compcross} is replaced by the
estimate:
$$|E_{l}(\varphi,\varepsilon)-E^{(l)}(\varepsilon)|< C e^{-\frac{2\pi Y}{\varepsilon}}$$
where $2 Y$ is the width of the strip $S_{Y}$.
\subsubsection{Application : asymptotic expansion of the trace}  By using the previous result,
we can compute the first terms in the asymptotic expansion of the
trace formula , and partially recover a result of \cite{Di1}.
\begin{cor}
\label{cordim} Let $f\in C_{0}^{\infty}(\R)$ be a real function
such that $\textrm{Supp }f\in J$. Then the function
$f(H_{\varphi,\varepsilon})$ is $\varepsilon$-periodic in
$\varphi$ and its Fourier expansion satisfies:
\begin{equation}
\textrm{tr
}[f(H_{\varphi,\varepsilon})]=\frac{1}{\varepsilon}\int_{0}^{\varepsilon}[f(H_{u,\varepsilon})]du+O(e^{-S/\varepsilon})
\end{equation}
\begin{equation}
\int_{0}^{\varepsilon}[f(H_{u,\varepsilon})]du=\frac{1}{2\pi}\int_{\R_{u}}\int_{[-\pi,\pi]}f(W(u)+E(\kappa))d\kappa
du+o(\varepsilon)
\end{equation}
where $S=\inf\limits_{e\in\textrm{Supp }f}S(e)>0$
\end{cor}
We give more details and the proof of this corollary in section
\ref{trform2}.

\section{Main steps of the study}
\label{schem} Here, we explain the main ideas of the paper.
\subsection{One-dimensional perturbed periodic operators}
\subsubsection{} \label{rapp}
We consider equation \eqref{eqp} as a perturbation of the periodic
equation
\begin{equation}
\label{espa}
 H_{0}\psi=E\psi
\end{equation}
where the operator $H_{0}$ is defined in \eqref{opper}. To do
that, we shall describe the spectral theory of periodic operators
in section \ref{opepera}.
\begin{itemize}
\item
\label{rappa} For the moment, we simply introduce the Bloch
solutions of equation \eqref{esp}. We call a {\it Bloch solution}
of \eqref{esp} a function $\Psi$ satisfying \eqref{esp} and:
\begin{equation}
\label{blochsola} \forall
x\in\R,\quad\Psi(x+1,E)=\lambda(E)\Psi(x,E),
\end{equation}
with $\lambda\neq 0$ independent of $x$. The coefficient
$\lambda(E)$ is called {\it Floquet multiplier}. We represent
$\lambda(E)$ in the form $\lambda(E)=e^{i k(E)}$; $k$ is the
quasi-momentum presented in section \ref{assV} and described in
section \ref{qm}. If $E\notin\sigma(H_{0})$, there exist two
linearly independent Bloch solutions of \eqref{espa}(see section
\ref{bloch}). We call them $\widetilde{\Psi}_{+}$ et
$\widetilde{\Psi}_{-}$; the associated Floquet multipliers are
inverse of each other and the functions $\widetilde{\Psi}_{\pm}$
are represented in the form:
$$ \widetilde{\Psi}_{\pm}(x,E)=e^{\pm ik(E)x}p_{\pm}(x,E)\quad\textrm{avec}\quad p_{\pm}(x+1,E)=p_{\pm}(x,E).$$
For $\I k(E)>0$, the function $\widetilde{\Psi}_{+}(x,E)$ tends to
$0$ as $x$ tends to $+\infty$ and the function
$\widetilde{\Psi}_{-}(x,E)$ tends to $0$ as $x$ tends to
$-\infty$. Actually, equation \eqref{blochsola} defines the
functions $\widetilde{\Psi}_{+}$ and $\widetilde{\Psi}_{-}$ except
for a multiplicative coefficient. Precisely, equation
\eqref{blochsola} defines two one-dimensional vector spaces that
we call {\it Bloch sub-spaces}.\\
To study the eigenvalues of perturbations of periodic operators,
\cite{Fi1} and \cite{New} introduce, for $\I k(E)>0$, two
functions $(x,\varphi,E,\varepsilon)\mapsto
F_{+}(x,\varphi,E,\varepsilon)$ and
$(x,\varphi,E,\varepsilon)\mapsto F_{-}(x,\varphi,E,\varepsilon)$
solutions of \eqref{eqp} satisfying:
\begin{equation}
\label{jostcond} \lim\limits_{x\rightarrow
+\infty}[F_{+}(x,\varphi,E,\varepsilon)-\widetilde{\Psi}_{+}(x,E)]=0,\quad
\lim\limits_{x\rightarrow
-\infty}[F_{-}(x,\varphi,E,\varepsilon)-\widetilde{\Psi}_{-}(x,E)]=0
\end{equation}
Condition \eqref{jostcond} guarantees the uniqueness of $F_{+}$
(resp. of $F_{-}$) since the function $\widetilde{\Psi}_{+}$
(resp. $\widetilde{\Psi}_{-}$) tends to $0$ as $x$ tends to
$+\infty$ (resp. $-\infty$). These functions are called Jost
functions; they are generally constructed as solutions of a
Lippman-Schwinger integral equation. This construction is an
adaptation of the usual theory of scattering (chapter XI of
\cite{ReSi3}) for a perturbation of laplacian; it consists in
looking for particular solutions of
\eqref{eqp} from the solutions of the periodic equation.\\
We call {\it Jost sub-spaces} the sub-spaces $\mathcal{J}_{+}$ and
$\mathcal{J}_{-}$ generated by $F_{+}$ and $F_{-}$.\\
$\mathcal{J}_{+}$ (resp $\mathcal{J}_{-}$) is the set of solutions
of \eqref{eqp} being a member of $L^{2}([0,\infty))$ (resp.
$L^{2}((-\infty,0])$).
\item Let $f$ and $g$ be two derivable functions, the {\it
Wronskian} of $f$ and $g$ called $w(f,g)$ is defined by:
\begin{equation}
\label{wronsk} w(f,g)=f'g-fg'
\end{equation}
We recall that if $f$ and $g$ are the solutions of a second-order
differential equation, their Wronskian is independent of $x$. The
spectral interest of the Jost sub-spaces is the following:
\begin{prop}\label{carvp} We assume that $\I k(E)>0$. Let $h^{-}_{g}\in\mathcal{J}_{-}$
and $h^{+}_{d}\in\mathcal{J}_{+}$ be two nontrivial Jost solutions
of \eqref{eqp}. $E$ is an eigenvalue of $H_{\varphi,\varepsilon}$
if and only if: \begin{equation}
 w(h^{+}_{d},h^{-}_{g})=0
\end{equation}
\end{prop}
To compute the eigenvalues, it suffices to construct the Jost
sub-spaces.
\end{itemize}
\subsection{Construction of consistent Jost solutions} We denote by $(H_{J}^{0})$ the following assumption:
\\ \\
{\bf $\mathbf{(H_{J}^{0})}$ There exists $n\in\N$ such that
$\mathbf{J}$ is a compact interval of $\mathbf{]E_{2n},E_{2n+1}[}$.}\\
\\
Clearly, $(H_{J}^{0})$ is weaker than $(H_{J})$.\\
We introduce
a new notation.\\
For a function $f:\ \mathcal{U}\subset\C^{n}\rightarrow\C^{p}$, we
define the function $f^{*}:\
\overline{\mathcal{U}}\rightarrow\C^{p}$:
\begin{equation}
\label{eqconj1} f^{*}(Z)=\overline{f(\overline{Z})}.
\end{equation}
As we have explained in section \ref{intro}, an useful idea to
study \eqref{eqp} is the construction of consistent solutions,
i.e. satisfying \eqref{coh}. First, we choose in $\mathcal{J}_{-}$
and $\mathcal{J}_{+}$ some consistent bases. We shall prove the
following result:
\begin{thm}\label{jostthm}
 We assume that $(H_{V})$, $(H_{W,r})$ and
$(H_{J}^{0})$ are satisfied. Fix $X>1$. Then, there exist a
complex neighborhood $\mathcal{V}=\overline{\mathcal{V}}$ of $J$,
a real $\varepsilon_{0}>0$, two points $m_{g}$ and $m_{d}$ in
$\C$, two real numbers $A_{g}$ and $A_{d}$ and two functions
$(x,\varphi,E,\varepsilon)\mapsto
h_{-}^{g}(x,\varphi,E,\varepsilon)$,
$(x,\varphi,E,\varepsilon)\mapsto
h_{+}^{d}(x,\varphi,E,\varepsilon)$ such that:
\begin{itemize}
\item The functions $(x,\varphi,E,\varepsilon)\mapsto
h_{-}^{g}(x,\varphi,E,\varepsilon)$ and
$(x,\varphi,E,\varepsilon)\mapsto
h_{+}^{d}(x,\varphi,E,\varepsilon)$ are defined and consistent on
$\R\times S_{Y}\times \mathcal{V}\times]0,\varepsilon_{0}[$.
\item For any $x\in[-X,X]$ and $\varepsilon\in]0,\varepsilon_{0}[$, $(\varphi,E)\mapsto
h_{-}^{g}(x,\varphi,E,\varepsilon)$ and $(\varphi,E)\mapsto
h_{+}^{d}(x,\varphi,E,\varepsilon)$ are analytic on $S_{Y}\times
\mathcal{V}$.
\item The function $x\mapsto h_{-}^{g}(x,\varphi,E,\varepsilon)$ (resp. $x\mapsto h_{+}^{d}(x,\varphi,E,\varepsilon)$)
is a basis of $\mathcal{J}_{-}$ (resp. $\mathcal{J}_{+}$).
\item The functions $h_{-}^{g}$ and $h_{+}^{d}$ have the following asymptotic behavior:
\begin{equation}
\label{asj1}
h_{-}^{g}(x,\varphi,E,\varepsilon)=e^{\frac{-i}{\varepsilon}\int_{m_{g}}^{\varphi}\kappa(u)du}\psi_{-}(x,\varphi,E)(1+R_{g}(x,\varphi,E,\varepsilon)),
\end{equation}
and
\begin{equation}
\label{asjda1}
h_{+}^{d}(x,\varphi,E,\varepsilon)=e^{\frac{i}{\varepsilon}\int_{m_{d}}^{\varphi}\kappa(u)du}\psi_{+}(x,\varphi,E)(1+R_{d}(x,\varphi,E,\varepsilon)),
\end{equation}
where
\begin{itemize}
\item $R_{g}$ and $R_{d}$ satisfy:
$$\sup\limits_{x\in]-X,X[,\ \Ra\varphi<A_{g},\\ E\in\mathcal{V}}\max\{|R_{g}(x,\varphi,E,\varepsilon)|,|\partial_{x}R_{g}(x,\varphi,E,\varepsilon)|\}\leq
r(\varepsilon).$$
$$\sup\limits_{x\in]-X,X[,\ \Ra\varphi>A_{d},\\ E\in\mathcal{V}}\max\{|R_{d}(x,\varphi,E,\varepsilon)|,|\partial_{x}R_{d
}(x,\varphi,E,\varepsilon)|\}\leq r(\varepsilon),$$ with
$$\lim\limits_{\varepsilon\rightarrow 0}r(\varepsilon)=0.$$
\item The functions $\psi_{+}$ and $\psi_{-}$ are the Bloch canonical solutions
of the periodic equation \eqref{espa} defined in section
\ref{cansolbloch}.
\end{itemize}
\item There exist two real numbers $\sigma_{g}\in\{-1,1\}$, $\sigma_{d}\in\{-1,1\}$, an integer $p$ and two functions $E\mapsto\alpha_{g}(E)$ and $E\mapsto\alpha_{d}(E)$ such
that:
\begin{enumerate}
\item For any $\varepsilon\in]0,\varepsilon_{0}[$, $x\in\R$,
$E\in\mathcal{V}$,et $\varphi\in S_{Y}$ ,we have:
\begin{equation}
\label{stargj}
\alpha_{g}^{*}(E)(h_{-}^{g})^{*}(x,\varphi,E,\varepsilon)=i\sigma_{g}e^{-\frac{i}{\varepsilon}2p\pi
x} \alpha_{g}(E)h_{-}^{g}(x,\varphi,E,\varepsilon)
\end{equation}
\begin{equation}
\label{stardj}
\alpha_{d}^{*}(E)(h_{+}^{d})^{*}(x,\varphi,E,\varepsilon)=i\sigma_{d}e^{\frac{i}{\varepsilon}2p\pi
x} \alpha_{d}(E)h_{+}^{d}(x,\varphi,E,\varepsilon)
\end{equation}
\item The functions $\alpha_{g}$ and $\alpha_{d}$ are analytic
and given by \eqref{renormconstg} and \eqref{renormconstd}. They
do not vanish on $\mathcal{V}$.
\end{enumerate}
\end{itemize}
\end{thm}
We immediately deduce from Theorem \ref{jostthm} and Proposition
\ref{carvp} that the eigenvalues of $H_{\varphi,\varepsilon}$ are
characterized by:
\begin{equation}
w(h_{-}^{g}(\cdot,\varphi,E,\varepsilon),h_{+}^{d}(\cdot,\varphi,E,\varepsilon))=0
\end{equation}
Theorem \ref{jostthm} is the consequence of two main ideas:
\begin{itemize}
\item First, we adapt the construction of Jost functions developed
by \cite{Fi1,Ne}. Indeed, this construction proves that asymptotic
\eqref{as1} is only satisfied on domains which depend on
$\varepsilon$ (see section \ref{scattheory}).
\item We must understand how this asymptotic evolves on a domain
which does not depend on $\varepsilon$. To do that, we extend the
continuation results of Fedotov and Klopp in a non compact frame
(see section \ref{infwkb}).
\end{itemize}
\subsection{Conclusion}
To finish the computations, it suffices to apply the methods of
\cite{FK1}. We have to go through the cross (see figure
\ref{pbf}). We will show that there exists in the neighborhood of
the cross a consistent basis $f_{\pm}^{i}$ with standard
asymptotic. We shall express the functions $h_{-}^{g}$ and
$h_{+}^{d}$ on this basis (section \ref{calcmattransf}).

\section{Periodic Schr{\"o}dinger operators on the real line}
\label{opepera} We now discuss the periodic operator \eqref{opper}
where $V$ is a 1-periodic, real-valued,
$\mathbf{L^{2}_{\textrm{loc}}}$-function. We collect known results
needed in the present paper (see \cite{Ma, McK, Ti}).
\subsection{Bloch solutions}
\label{bloch} Let $\widetilde{\Psi}$ be a solution of the equation
\begin{equation}
\label{espc} H_{0}\widetilde{\Psi}=\E\widetilde{\Psi}
\end{equation}
satisfying the relation
\begin{equation}
\widetilde{\Psi}(x+1)=\lambda\widetilde{\Psi}(x),\quad\forall
\in\R
\end{equation}
for some complex number $\lambda\neq 0$ independent of $x$. Such a
solution is called a {\it Bloch solution}, and the number
$\lambda$ is called the {\it Floquet multiplier}. Let us discuss
the analytic properties of Bloch solutions.\\
In \eqref{band}, we have denoted by $[E_{1}, E_{2}],\ldots
,[E_{2n+1}, E_{2n+2}],\ldots$ the spectral bands of the periodic
Schr{\"o}dinger equation. Consider $\Gamma_{\pm}$ two copies of
the complex plane $\E\in \C$ cut along the spectral bands. Paste
them together to get a Riemann surface with square root
branch points. We denote this Riemann surface by $\Gamma$.\\
One can construct a Bloch solution $\widetilde{\Psi}(x,\E)$ of
equation \eqref{espc} meromorphic on $\Gamma$. The poles of this
solution are located in the spectral gaps. Precisely, each
spectral gap contains precisely one simple pole. This pole is
situated either on $\Gamma_{+}$ or on $\Gamma_{-}$. The position
of the pole is
independent of $x$. For the details, we refer to \cite{Fi1}.\\
Except at the edges of the spectrum (i.e. the branch points of
$\Gamma$), the restrictions $\widetilde{\Psi}_{\pm}$ of
$\widetilde{\Psi}$ on $\Gamma_{\pm}$ are linearly independent
solutions of \ref{espc}. Along the gaps, these functions are real
and satisfy :
\begin{equation}
\label{gaprel}
\overline{\widetilde{\Psi}_{\pm}(x,\E-i0)}=\widetilde{\Psi}_{\pm}(x,\E+i0),\quad\forall
\E\in]E_{2n},E_{2n+1}[,\ n\in\N.
\end{equation}
Along the bands, we have :
\begin{equation}
\label{bandrel}
\overline{\widetilde{\Psi}_{\pm}(x,\E-i0)}=\widetilde{\Psi}_{\mp}(x,\E+i0),\quad\forall
\E\in]E_{2n+1},E_{2n+2}[,\ n\in\N.
\end{equation}
\subsection{Bloch quasi-momentum}
\subsubsection{}
Consider the Bloch solution $\widetilde{\Psi}(x,\E)$ introduced in
the previous subsection. The corresponding Floquet multiplier
$\lambda(\E)$ is analytic on $\Gamma$. Represent it in the form:
\begin{equation}
\lambda(\E)=\exp(ik(\E)).
\end{equation}
The function $k(\E)$ is called Bloch quasi-momentum. It has the
same branch points as $\widetilde{\Psi}(x,\E)$, but the
corresponding Riemann surface is more complicated. \\ To describe
the main properties of $k$, consider the complex plane cut along
the real line from $E_{1}$ to $+\infty$. Denote the cut plane by
$\C_{0}$. One can fix there a single valued branch of the
quasi-momentum by the condition
\begin{equation}
ik_{0}(\E)<0,\quad \E<E_{1}.
\end{equation}
All the other branches of the quasi-momentum have the form $\pm
k_{0}(\E)+2\pi m, m\in\Z$. The $\pm$ and the number $m$ are
indexing these branches. The image of $\C_{0}$ by $k_{0}$ is
located in the upper half of the complex plane.
\begin{equation}
\I k_{0}(\E)>0,\quad \E\in\C_{0}.
\end{equation}
In figure \ref{qm}, we drew several curves in $\C_{0}$ and their
images under transformation $E\mapsto k_{0}(E)$. The
quasi-momentum $k_{0}(E)$ is real along the spectral zones, and,
along the spectral gaps, its real part is constant; in particular,
we have
\begin{equation}\label{qm1}
k_{0}(E_{1})=0\quad  k_{0}(E_{2l}\pm i0)=k_{0}(E_{2l+1}\pm
i0)=\pm\pi l,\ \ \ l\in\N.
\end{equation}
All the branch points of $k$ are of square root type. Let $E_{m}$
be one of the branch points of $k$. Then, each function:
\begin{equation} \label{qm2}
 f_{m}^{\pm}(\E)=(k_{0}(\E\pm i0)-k_{0}(E_{m}\pm
i0))/\sqrt{\E-E_{m}},\ \ \ E\in\R
\end{equation}
can be analytically continued in a small vicinity of the branch
point $E_{m}$.\\ Finally, we note that
\begin{equation}
k_{0}(\E)=\sqrt{\E}+O(1/\sqrt{\E}),\ \ \ |\E|\rightarrow\infty
\end{equation}
where $E\in\C_{0}$ and $0<\arg E<2\pi$.

 The values of the quasi-momentum $k_{0}$ on the
two sides of the cut $[E_{1},+\infty)$ are related to each other
by the formula:
\begin{equation}
\label{ko}
 \forall \E\in]E_{1},+\infty[,\quad k_{0}(\E+i0)=-\overline{k_{0}(\E-i0)},\ \ \ E_{1}\leq
 \E.
\end{equation}
Consider the spectral gap $(E_{2l},E_{2l+1}),l \in\N$. Let
$\C_{l}$ be the complex plane cut from $-\infty$ to $E_{2l}$ and
from $E_{2l+1}$ to $+\infty$. Denote by $k_{l}$ the branch of the
quasi-momentum analytic on $\C_{l}$ and coinciding with $k_{0}$
for $\I E >0$. Then, one has:
\begin{equation}
\label{kl}
 \forall\E\in]-\infty,E_{2l}[\cup]E_{2l+1},+\infty[,\quad\quad k_{l}(\E+i0)+\overline{k_{l}(\E-i0)}=2\pi
 l.
\end{equation}
\psset{unit=1em,linewidth=.05}

\psset{unit=1em,linewidth=.1}
\begin{center}
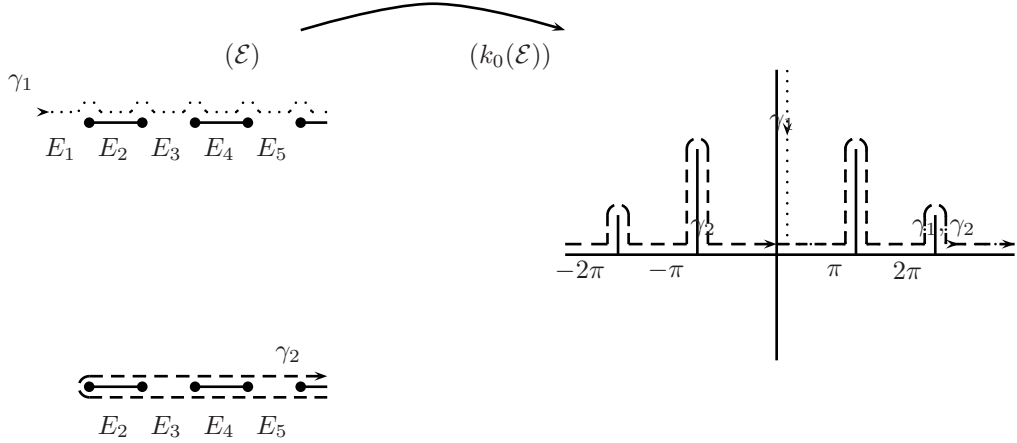
\begin{figure}
\begin{pspicture}(-10,-10)(30,10)

\pscurve{->}(2,8.5)(7,9.5)(12,8.5) \uput[180](1,7.5){$(\E)$}
\uput[180](12,7.5){$(k_{0}(\E))$}

\psline(-6,5)(-4,5) \psline(-2,5)(0,5) \psline(2,5)(3,5)
\psdots[dotstyle=*](-6,5)(-4,5)(-2,5)(0,5)(2,5)
\uput[180](-6,4){$E_{1}$}

\uput[180](-4,4){$E_{2}$}
\uput[180](-2,4){$E_{3}$}\uput[180](0,4){$E_{4}$}\uput[180](2,4){$E_{5}$}
\psline[linestyle=dotted]{>-}(-8,5.4)(-7,5.4)(-6.4,5.4)
\psarc[linestyle=dotted](-6,5.4){0.4}{0}{180}
\psarc[linestyle=dotted](-4,5.4){0.4}{0}{180}
\psarc[linestyle=dotted](-2,5.4){0.4}{0}{180}
\psarc[linestyle=dotted](0,5.4){0.4}{0}{180}
\psarc[linestyle=dotted](2,5.4){0.4}{0}{180}
\psline[linestyle=dotted](-5.6,5.4)(-4.4,5.4)
\psline[linestyle=dotted](-3.6,5.4)(-2.4,5.4)
\psline[linestyle=dotted](-1.6,5.4)(-.4,5.4)
\psline[linestyle=dotted](0.4,5.4)(1.6,5.4)
 \psline[linestyle=dotted](2.4,5.4)(3,5.4)

\uput[180](-7.6,6.5){$\gamma_{1}$}

\psline(-6,-5)(-4,-5) \psline(-2,-5)(0,-5) \psline(2,-5)(3,-5)
\psdots[dotstyle=*](-6,-5)(-4,-5)(-2,-5)(0,-5)(2,-5)
\uput[180](-4,-6.5){$E_{2}$}
\uput[180](-2,-6.5){$E_{3}$}\uput[180](0,-6.5){$E_{4}$}\uput[180](2,-6.5){$E_{5}$}
\psline[linestyle=dashed](3,-5.4)(-6,-5.4)

\psarc[linestyle=dashed](-6,-5){0.4}{90}{270}

\psline[linestyle=dashed]{->}(-6,-4.6)(3,-4.6)
\uput[180](2.5,-3.8){$\gamma_{2}$}

\psline(20,-4)(20,7) \psline(12,0)(29,0) \psline(17,0)(17,4)
\psline(14,0)(14,1.5)\psline(23,0)(23,4)\psline(26,0)(26,1.5)
\uput[180](14,-0.6){$-2\pi$}\uput[180](26,-0.6){$2\pi$}
\uput[180](17,-0.6){$-\pi$}\uput[180](23,-0.6){$\pi$}
\psline[linestyle=dashed](12,0.4)(13.6,0.4)
\psline[linestyle=dashed](13.6,0.4)(13.6,1.5)
\psarc[linestyle=dashed](14,1.5){0.4}{0}{180}
\psline[linestyle=dashed](14.4,1.5)(14.4,0.4)

\psline[linestyle=dashed](14.4,0.4)(16.6,0.4)
\psline[linestyle=dashed](16.6,0.4)(16.6,4)
\psarc[linestyle=dashed](17,4){0.4}{0}{180}
\psline[linestyle=dashed](17.4,0.4)(17.4,4)

 \psline[linestyle=dashed]{->}(17.4,0.4)(20,0.4)
 \psline[linestyle=dashed](20,0.4)(22.6,0.4)
\psline[linestyle=dashed](22.6,0.4)(22.6,4)
\psarc[linestyle=dashed](23,4){0.4}{0}{180}
 \psline[linestyle=dashed](23.4,4)(23.4,0.4)
\psline[linestyle=dashed](23.4,0.4)(25.6,0.4)
\psline[linestyle=dashed](25.6,0.4)(25.6,1.5)
\psarc[linestyle=dashed](26,1.5){0.4}{0}{180}
 \psline[linestyle=dashed](26.4,0.4)(26.4,1.5)
 \psline[linestyle=dashed]{>-}(26.4,0.4)(29,0.4)

\psline[linestyle=dotted](20.4,7)(20.4,5)
\psline[linestyle=dotted]{>-}(20.4,5)(20.4,0.4)
\psline[linestyle=dotted](20.4,0.4)(22.6,0.4)
 \psline[linestyle=dotted](22.6,0.4)(22.6,4)
\psarc[linestyle=dotted](23,4){0.4}{0}{180}
 \psline[linestyle=dotted](23.4,4)(23.4,0.4)
\psline[linestyle=dotted](23.4,0.4)(25.6,0.4)
\psline[linestyle=dotted](25.6,0.4)(25.6,1.5)
\psarc[linestyle=dotted](26,1.5){0.4}{0}{180}
 \psline[linestyle=dotted](26.4,0.4)(26.4,1.5)
 \psline[linestyle=dotted]{->}(26.4,0.4)(29,0.4)

\uput[180](18.2,1){$\gamma_{2}$} \uput[180](21.2,5){$\gamma_{1}$}
\uput[180](28,1){$\gamma_{1},\gamma_{2}$}

\end{pspicture}
\caption{The quasi-momentum $k$} \label{qm}
\end{figure}
\end{center}
\subsection{Periodic components of the Bloch solution}
Let $D$ a simply connected domain that does not contain any branch
point of $k$. On $D$, we fix an analytic branch of $k$. Consider
two copies of $D$, denoted by $D_{\pm}$, corresponding to two
sheets of $\G$. Now we redefine $\widetilde{\Psi}_{\pm}$ to be the
restrictions of $\widetilde{\Psi}$ to $D_{\pm}$. They can be
represented in the form:
\begin{equation}
\widetilde{\Psi}_{\pm}(x,\E)=e^{\pm i k(\E)x}p_{\pm}(x,\E),\ \ \
\E\in D
\end{equation}
where $p_{l}^{\pm}(x,\E)$ are 1-periodic in $x$,
\begin{equation}
p_{\pm}(x+1,\E)=p_{\pm}(x,\E),\ \ \ \forall x\in\R
\end{equation}
\subsection{Analytic solutions of \eqref{espc}}
To describe the asymptotic formulas of the complex WKB method for
equation \eqref{eqp}, one needs specially normalized Bloch
solutions of the equation \eqref{espc}.\\
Let $\mathcal{D}$ be a simply connected domain in the complex
plane containing no branch point of the quasi-momentum $k$. We fix
on $\mathcal{D}$ a continuous determination of $k$. We fix
$\mathcal{E}_{0}\in\mathcal{D}$. We recall the following result
(\cite{FK1, FK4}).
\begin{lem}\label{anasol}
We define the functions $g_{\pm}$ :
\begin{equation}
\label{fu} g_{\pm}\ :\ \mathcal{D}\rightarrow\C \ ;\
\mathcal{E}\mapsto
-\frac{\int_{0}^{1}p_{\mp}(x,\mathcal{E})\partial_{\mathcal{E}}p_{\pm}(x,\mathcal{E})dx}{\int_{0}^{1}p_{+}(x,\mathcal{E})p_{-}(x,\mathcal{E})dx},
\end{equation}
and the functions $\psi_{\pm}^{0}$ :
\begin{equation}
\label{blochsol} \psi_{\pm}^{0}\ :\
\R\times\mathcal{D}\rightarrow\C\ ;\
(x,\mathcal{E})\mapsto\sqrt{k'(\mathcal{E})}e^{\int_{\mathcal{E}_{0}}^{\mathcal{E}}g_{\pm}(e)de}\widetilde{\Psi}_{\pm}(x,\mathcal{E}).
\end{equation}
The functions $\E\mapsto\psi_{\pm}^{0}(x,\E)$ are analytic on
$\mathcal{D}$, for any $x\in\R$. The functions $\psi_{\pm}^{0}$
are called {\it analytic Bloch solutions normalized at the point
$\mathcal{E}_{0}$} of \eqref{espc}.
\end{lem}
Sometimes, we shall denote $\psi_{\pm}^{0}(x,\E,\mathcal{E}_{0})$
to specify the normalization. We refer to section 1.4.4 of
\cite{FK4} for the details of the proof. The proof follows from
the study of the poles of $\widetilde{\Psi}_{\pm}$ and the zeros
of $k'$. The poles of $g_{\pm}$ are simple and exactly situated at
the singularities of $\sqrt{k'}\widetilde{\Psi}_{\pm}$. The
computation of the residues of $g_{\pm}$ at these points completes
the proof.
\subsection{Useful formulas}
We end this section with some useful formulas. We recall that the
functions $g_{\pm}$ are given in \eqref{fu}. Fix $n\in\N$.
Equations \eqref{gaprel} and \eqref{bandrel} lead to the following
relations:
\begin{equation}\label{symgap}
g_{\pm}^{*}(\E)=g_{\pm}(x,\E),\quad\forall \E\in]E_{2n},E_{2n+1}[.
\end{equation}
\begin{equation}\label{symband}
g_{\pm}^{*}(\E)=g_{\mp}(x,\E),\quad\forall
\E\in]E_{2n+1},E_{2n+2}[.
\end{equation}

\section{Main tools of the complex WKB method}
\label{wkbconst} In this section, we recall the main tools of the
complex WKB method on compact domains. The idea of the method is
to construct some consistent functions of \eqref{eqp} with
asymptotic behavior \eqref{stdas}. This construction is not
possible on any domain of the complex plane but on some domains called canonical.\\
We apply the results of \cite{FK1, FK2, FK3} to the assumptions
$(H_{W,g})$ and $(H_{J})$. We build a neighborhood of the cross,
in which we construct a consistent basis with standard behavior
\eqref{stdas}. In this section, we fix $Y$ such that the
assumptions $(H_{W,g})$ and $(H_{J})$ are satisfied in the strip
$S_{Y}$.
\subsection{Canonical domains}
The canonical domain is the main geometric notion of the complex
WKB method.
\subsubsection{The complex momentum}
\label{vertdef}
 The canonical domains can be described in terms of
the complex momentum $\kappa(\varphi)$. Remind that this function
is defined by formula \eqref{momcompa}. We have described $\kappa$
in section \ref{momcompb}. The properties of $\kappa$ depend on
the spectral
parameter $E$ and of the analytic properties of $W$.\\
We first formulate some definitions (\cite{FK1}).
\subsubsection{Vertical, strictly vertical curves}
\begin{defn}
We say that a curve $\gamma$ is $\textit{vertical}$ if it
intersects the lines $\I z=\textrm{Const}$ at non-zero angles
$\theta$.\\
We say that a curve $\gamma$ is $\textit{strictly vertical}$ if
there is a positive number $\delta$ such that, at any point of
$\gamma$, the intersection angle $\theta$ satisfies the
inequality:
\begin{equation}
\delta<\theta<\pi-\delta.
\end{equation}
\end{defn}
\subsubsection{Canonical, strictly canonical curves}
\label{lc} Let $\gamma$ be a vertical curve which does not contain
any branch point. On $\gamma$, fix a continuous branch of the
momentum of $\kappa$.
\begin{defn}
We call $\gamma$ $\textit{canonical}$ if, along $\gamma$,
\begin{itemize}
\item $\I\varphi\mapsto\I\int^{\varphi}\kappa(u)du$ is strictly increasing. \item
$\I\varphi\mapsto\I\int^{\varphi}(\kappa(u)-\pi)du$ is strictly
decreasing.
\end{itemize}
\end{defn}
Assume that $\gamma$ is strictly vertical. If there is a positive
number $\delta$ such that, along $\gamma$:
\begin{equation}
\label{integra}
\I\int_{\varphi}^{\varphi'}\kappa(u)du\geq\delta\I(\varphi'-\varphi)\quad\forall(\varphi,\varphi')\in\gamma^{2},
\end{equation}
and
\begin{equation}\label{integrb}
\I
\int_{\varphi}^{\varphi'}(\pi-\kappa(u))du\geq\delta\I(\varphi'-\varphi)\quad\forall(\varphi,\varphi')\in\gamma^{2},
\end{equation}
we call $\gamma$ $\delta-\textit{strictly canonical}$.\\ We
identify the complex numbers with vectors in $\R^{2}$. To
construct canonical lines, we have to study the vector fields
$\kappa$ and $\kappa-\pi$, or rather their integral curves. For
$\varphi\in D$, $S(\varphi)$ denotes the sector of apex $\varphi$
such that, for any vector $z\in S(\varphi)$, we have:
\begin{equation}
\I(i\overline{\kappa(\varphi)}(z-\varphi))>0\textrm{ et }
\I(i(\overline{\kappa(\varphi)}-\pi)(z-\varphi))<0.
\end{equation}
Let $\gamma\in D$ a curve which does not contain any branch point.
For all $\varphi\in\gamma$, we denote $t(\varphi)$ the vector
tangent to $\gamma$ in $\varphi$ and oriented upward. The curve
$\gamma\in D$ is canonical for the determination $\kappa$ if and
only if for any $\varphi\in\gamma$, the vector $t(\varphi)$
belongs to $S(\varphi)$ (see figure \ref{can}). The cone
$S(\varphi)$ depends on the determination of $\kappa$. For
example, if $\kappa$ satisfies $\Ra\kappa\in]0,\pi[$, this cone is
not empty.
\subsubsection{}
In what follows, $\xi_{1}$ and $\xi_{2}$ are two points in $\C$
such that
$$ \I \xi_{1}<\I\xi_{2}.$$
We shall denote by $\gamma$ a smooth curve going from $\xi_{1}$ to
$\xi_{2}$; this curve will always be oriented from $\xi_{1}$ to
$\xi_{2}$.
\psset{unit=1em,linewidth=.05}

\psset{unit=1em,linewidth=.05}
\begin{center}
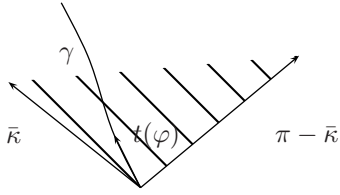
\begin{figure}
\begin{pspicture}(-20,-1)(10,10)
\psline{->}(0,0)(-5,4) \psline{->}(0,0)(6,5)
\pscurve(0,0)(-1,2)(-2,5)(-3,7) \psline{->}(0,0)(-1,2)
\uput[180](-4,2){$\bar{\kappa}$}
\uput[180](8,2){$\pi-\bar{\kappa}$}
\pscustom[linestyle=none,fillstyle=vlines,hatchsep=1.2,hatchangle=45]{\pscurve(-5,4)(-2.5,2)(0,0)
\pscurve(0,0)(3,2.5)(6,5)} \uput[180](2,2){$t(\varphi)$}
\uput[180](-2,5){$\gamma$}
\end{pspicture}
\caption{The cone $S(\varphi$)}\label{can}
\end{figure}
\end{center}
\subsubsection{Definition of the canonical domain}
Let $K$ be a simply connected domain in
$\left\{\I\varphi\in[\I\xi_{1},\I\xi_{2}]\right\}$ containing no
branch points of the complex momentum. On $K$, fix a continuous
branch $\kappa$.
\begin{defn}
We call $K$ a $\textit{canonical domain for }\kappa,\
\xi_{1}\textit{ and }\xi_{2}$ if it is the union of curves that
are connecting $\xi_{1}$ and $\xi_{2}$ and that are canonical with respect to $\kappa$.\\
If there is $\delta>0$ such that $K$ is a union of
$\delta$-strictly canonical curves, we call $K$
$\delta-\textit{strictly canonical}$.
\end{defn}
\subsubsection{}
Assume that $K$ is a canonical domain. Denote by $\partial K$ its
boundary. Fix a positive number $\delta$. We call the domain
$$\mathcal{C}=\{z\in K\ ;\  \textrm{dist}(z,\partial K)>\delta\}$$
an admissible sub-domain of $K$.\\
Note that the branch points of the complex momentum are outside of
$\mathcal{C}$, at a distance greater than $\delta$.
\subsection{Canonical Bloch solutions}
\label{cansolbloch}To describe the asymptotic formulas of the
complex WKB method for equation \eqref{eqp}, we shall use the
analytic Bloch solutions of \eqref{espc}, defined in Lemma
\ref{anasol} for the parameter $\E=E-W(\varphi)$. Precisely, we
consider the unperturbed periodic equation:
\begin{equation}
\label{espb} H_{0}\psi=(E-W(\varphi))\psi.
\end{equation}
\subsection{}
Let $D$ be a simply connected domain in $S_{Y}$, containing no
branch points of $\kappa$. The mapping $\varphi\mapsto
E-W(\varphi)$ maps $D$ onto a domain $\mathcal{D}\subset\C$. The
domain $\mathcal{D}$ does not contain any branch point of $k$.\\
Fix $\varphi_{0}\in D$, such that $k'(E-W(\varphi_{0}))\neq 0$. In
Lemma \ref{anasol}, we have built the analytic Bloch solutions
$\{\psi_{\pm}^{0}\}$ of equation (\ref{espc}), normalized in
$E-W(\varphi_{0})$. For $\varphi\in D$, we define:
\begin{equation}
\psi_{\pm}(x,\varphi,E)=\psi_{\pm}^{0}(x,E-W(\varphi)),\quad\forall
u\in\R,\quad\forall \varphi\in D.
\end{equation}
In \cite{FK1}, it is proved that the functions
$\varphi\mapsto\psi_{\pm}(x,\varphi,E)$ can be analytically
continued to $D$. $\psi_{\pm}$ are called the {\it canonical Bloch
solutions} of equation \eqref{espb}. Sometimes, we shall precise $\psi_{\pm}(x,\varphi,E,\varphi_{0})$ to specify the normalization.\\
We define
\begin{equation}
\label{omega}
\omega_{\pm}(\varphi,E)=-W'(\varphi)g_{\pm}(E-W(\varphi)).
\end{equation}
We also define:
\begin{equation}
\label{racq}
 q(\varphi)=\sqrt{k'(E-W(\varphi)}
\end{equation}
\subsection{The consistency relation}
\subsubsection{Consistent functions and consistent bases}
We recall that we say that $f$ is a {\it consistent} function if
it satisfies \eqref{coh}.\\
We say also that a basis $\{f_{\pm}\}$ of solutions of \eqref{eqp}
is {\it a consistent basis} if:
\begin{itemize}
\item The functions $f_{+}$ and $f_{-}$ are consistent.
\item Their Wronskian is independent of $\varphi$.
\end{itemize}
\subsubsection{Analyticity and consistency}
First, we define the width of a set.
\begin{defn}
Fix $Y_{0}>0$ and $M\subset S_{Y_{0}}$ a set of points. We define
$l(M,Y_{0})$:
\begin{equation}
\label{larga}
l(M,Y_{0})=\inf\limits_{y\in[-Y_{0},Y_{0}]}\sup\left\{|\Ra\varphi-\Ra\varphi'|
; (\varphi,\varphi')\in M^{2}\textrm{ such that }
\I\varphi=\I\varphi'=y \right\}
\end{equation}
$l(M,Y_{0})$ is called the {\it width} of $M$ in $S_{Y_{0}}$
\end{defn}
One has:
\begin{lem}
\label{anacoh} Fix $E$. We consider $X>0$, $\tilde{Y}\in]0,Y[$,
$\varepsilon_{0}>0$ and $K$ a complex domain such that
$l(K,\tilde{Y})>\varepsilon_{0}$. We assume that for any
$\varepsilon\in]0,\varepsilon_{0}[$,
$f(\cdot,\varphi,E,\varepsilon)$ is a consistent solution of
\eqref{eqp} for $\varphi\in K$ and that for any $x\in[-X,X]$, the
function $\varphi\mapsto f(x,\varphi,E,\varepsilon)$ is analytic
on $K$. Then, for any $\varepsilon\in]0,\varepsilon_{0}[$ and any
$x\in[-X,X]$, the function $\varphi\mapsto
f(x,\varphi,E,\varepsilon)$ is analytic on $S_{\tilde{Y}}$.
\end{lem}
This result is proved in \cite{FK3, FK4}.
\subsection{The theorem of the complex WKB method on a compact domain}
\label{wkbth} In this section, we recall the main result of the
complex WKB method.
\subsubsection{Standard asymptotic behavior}
\label{cptmtasstd} We briefly introduce the notion of standard
asymptotic behavior (see \cite{FK4}). Speaking about a solution
having standard asymptotic behavior, we mean first of all that
this solution has the asymptotics \eqref{stdas} and other
properties that we present now.\\
Fix $E_{0}\in\C$. Let $D\subset\C$ a simply connected domain
containing no branch points. Let $\kappa$ be a branch of the
complex momentum continuous in $D$ and $\psi_{\pm}$ the canonical
Bloch solutions normalized in $\varphi_{0}\in D$.\\
We say that a consistent solution $f$ has standard behavior $f\sim
e^{\frac{i}{\varepsilon}\int^{\varphi}\kappa(u)du}\psi_{+}(x,\varphi,E)$,
respectively $f\sim
e^{-\frac{i}{\varepsilon}\int^{\varphi}\kappa(u)du}\psi_{-}(x,\varphi,E)$
in $D$ if
\begin{itemize}\item there exists a complex neighborhood $\mathcal{V}_{0}$ of $E_{0}$ and
$X>0$ such that $f$ is a consistent solution of equation
\eqref{eqp} for any $(x,\varphi,E)\in[-X,X]\times D\times
\mathcal{V}_{0}$;
\item for any $x\in[-X,X]$, the function $((\varphi,E)\mapsto
f(x,\varphi,E,\varepsilon))$ is analytic on $D\times
\mathcal{V}_{0}$;
\item for any $A$, a sub-admissible domain of $D$, there is a
neighborhood $\mathcal{V}_{A}$ of $E_{0}$ such that
\begin{equation}
\label{cptmtasstda} \forall(x,\varphi,E)\in[-X,X]\times D\times
\mathcal{V}_{A},\quad
f(x,\varphi,E,\varepsilon)=e^{\frac{i}{\varepsilon}\int^{\varphi}\kappa(u)du}(\psi_{+}(x,\varphi,E)+o(1)),\quad\varepsilon\rightarrow
0
\end{equation}
respectively
\begin{equation}
\label{cptmtasstdb} \forall(x,\varphi,E)\in[-X,X]\times D\times
\mathcal{V}_{A},\quad
f(x,\varphi,E,\varepsilon)=e^{-\frac{i}{\varepsilon}\int^{\varphi}\kappa(u)du}(\psi_{-}(x,\varphi,E)+o(1)),\quad\varepsilon\rightarrow
0
\end{equation}
\item the asymptotics \eqref{cptmtasstda} and \eqref{cptmtasstdb}
are uniform on $[-X,X]\times D\times \mathcal{V}_{A}$;
\item the asymptotics \eqref{cptmtasstda} and \eqref{cptmtasstdb}
can be differentiated once in $x$.
\end{itemize}
\subsubsection{}
Let us formulate the Theorem WKB on a compact domain.
\begin{thm}\cite{FK1,FK4}\\
\label{finwkbthm} We assume that $V$ satisfies $(H_{V})$ and that
$W$ satisfies $(H_{W,r})$. Fix $X>1$ and $E_{0}\in\C$. Let
$K\subset S_{Y}$ be a bounded canonical domain with respect to
$\kappa$. There exists $\varepsilon_{0}>0$ and a consistent basis
$\{f_{+}(x,\varphi,E,\varepsilon),f_{-}(x,\varphi,E,\varepsilon)\}$
of solutions of \eqref{eqp}, having the standard behavior
\eqref{cptmtasstda} et \eqref{cptmtasstdb} in $K$.\\
For any fixed $x\in\R$, the functions $\varphi\mapsto
f_{\pm}(x,\varphi,E,\varepsilon)$ are analytic in $K$.
\end{thm}
\subsection{The main geometric tools of the complex WKB method}
\label{constgeo} In this section, we introduce the main geometric
tools of the complex WKB method. To do that, we recall some ideas
of \cite{Fe, FK1, FK3, Wa}.
\subsubsection{Stokes lines}
The definition of the Stokes lines is fairly standard,
\cite{Fe,FK1}. The integral
$\varphi\mapsto\int^{\varphi}\kappa(u)du$ has the same branch
points as the complex momentum. Let $\varphi_{0}$ be one of them.
Consider the curves beginning at $\varphi_{0}$, and described by
the equation
\begin{equation}
\I\int_{\varphi_{0}}^{\varphi}(\kappa(\xi)-\kappa(\varphi_{0}))d\xi=0
\end{equation}
These curves are the {\it Stokes lines} beginning at
$\varphi_{0}$. According to equation \eqref{ko} and equation
\eqref{kl}, the Stokes line definition is independent of the
choice of the branch of $\kappa$.\\
Assume that $W'(\varphi_{0})\neq 0$. Equation \eqref{qm2} implies
that there are exactly three Stokes lines beginning at
$\varphi_{0}$. The angle between any two of them at this point is
equal to $\frac{2\pi}{3}$.
\subsection{Lines of Stokes type}
\label{ligtypsto} We recall that $D\subset S_{Y}$ is a simply
connected domain containing no branch points. Let $\gamma\subset
D$ be a smooth curve. We say that $\gamma$ is a line of Stokes
type with respect to $\kappa$ if, along $\gamma$, we have
$$\textrm{ either }\I\left(\int^{\varphi}\kappa(u)
du\right)=\textrm{Const}\quad\textrm{ or
}\quad\I\left(\int^{\varphi}(\kappa(u)-\pi)
du\right)=\textrm{Const}$$
\subsection{Pre-canonical lines}
Let $\gamma\subset D$ be a vertical curve. We call $\gamma$
$\textit{pre-canonical}$ if it consists of union of bounded
segments of canonical curves and/or lines of Stokes type.
\subsection{Some branches of the complex momentum}
In this section, we describe different branches of $\kappa$ near
the branch points described in \ref{assJ}. The geometrical
configuration is similar to the one studied in \cite{FK2}.

\subsubsection{Different cases}
\label{poss} We assume that $(H_{W,r})$, $(H_{W,g})$, and
$(H_{J})$ are satisfied. To study the geometrical tools of the WKB
complex method, one needs to specify the properties of $\I\kappa$
and $\Ra\kappa$. We know that $\kappa(\varphi_{r}^{\pm})\equiv
0[\pi]$ (see section \ref{opepera}). We consider two cases: either
$\kappa(\varphi_{r}^{\pm})\equiv 0[2\pi]$ or
$\kappa(\varphi_{r}^{\pm})\equiv \pi[2\pi]$.\\
We define $S_{-}$ the open domain delimited by the real line at
the bottom and by $\Sigma_{+}$ to the right:
\begin{equation}
\label{smoins} S_{-}=\{\varphi-r\ ;\
\varphi\in\Sigma_{+}^{*},r\in\R_{+}^{*}\}\cap S_{Y}
\end{equation}
Similarly, we define $S_{+}$ the open domain delimited by the real
line at the bottom and by $\Sigma_{+}$ to the left:
\begin{equation}
\label{splus} S_{+}=\{\varphi+r\ ;\
\varphi\in\Sigma_{+}^{*},r\in\R_{+}^{*}\}\cap S_{Y}
\end{equation}
The domains $S_{+}$ and $S_{-}$ are shown in figure
\ref{smoinsa}.\\ We prove the following result.
\begin{lem}
\label{detpos}\label{dtepos}There exists a branch $\kappa_{i}$ of
the complex momentum such that
\begin{enumerate}
\item $\I\kappa_{i}(\varphi)>0$ for $\varphi\in S_{-}$, $\kappa_{i}(\varphi_{r}^{-}+i0)=0$ and $\kappa_{i}(\varphi_{i}-0)=\pi$,\\
ou
\item $\I\kappa_{i}(\varphi)<0$ for $\varphi\in S_{-}$, $\kappa_{i}(\varphi_{r}^{-}+i0)=\pi$ and $\kappa_{i}(\varphi_{i}-0)=0$.
\end{enumerate}
\end{lem}
\begin{dem}
\begin{itemize}\item First, we specify the sign of $\I\kappa_{i}$. The set
$(E-W)(\R-[\varphi_{r}^{-},\varphi_{r}^{+}]))$ belongs to a gap
$G$. We define $\Lambda_{-}=(E-W)(S_{-})$. We prove that
$\Lambda_{-}$ is a connected domain which intersects with $\R$
only in the gap $G$. According to assumption $(H_{W,g})$
(\ref{assW2}), there exists a sequence of vertical curves
$\widetilde{\Sigma}_{k}$ such that:
$$\Lambda_{-}\cap\R=(E-W)((-\infty,\varphi_{r}^{-}]\cup[\varphi_{r}^{+},+\infty))\cup(E-W)(\widetilde{\Sigma}_{k}^{+}).$$
$(E-W)(\widetilde{\Sigma}_{k}^{+})$ is a connected domain of $\R$;
it contains at least a point of $G$ and does not intersect with
$\partial\sigma(H_{0})$. Consequently,
$(E-W)(\widetilde{\Sigma}_{k}^{+})$ belongs to $G$ and:
$$\Lambda_{-}\cap\R=G.$$
We fix on $\Lambda_{-}$ a continuous branch of the quasi momentum
$k$. The sign of $\I k$ does not change since $(E-W)(S_{-})$ does
not intersect with $\sigma(H_{0})$. If we define
$\kappa_{i}(\varphi)=k(E-W(\varphi))$, $\I\kappa_{i}$ we can
assume that $\I\kappa_{i}>0$ on $S_{-}$.
\item Now, we consider $\Ra\kappa_{i}$. According to section
\ref{qm}, we can choose the branch $\kappa_{i}$ such that
$\kappa_{i}(\varphi_{r}^{-}+i0)\in\{0,\pi\}$. First, we study the
case $\kappa_{i}(\varphi_{r}^{-})=0$; this assumption implies two
possibilities.
\begin{enumerate}\item The point $0$ is a minimum for $W$ and the
band $B$ in $(H_{J})$ is in the form $[E_{4l+1},E_{4l+2}]$. The
points $E_{r}$ and $E_{i}$ satisfy $E_{r}=E_{4l+1}$ and
$E_{i}=E_{4l+2}$. There exists a neighborhood $V$ of
$[\varphi_{r}^{-},0]\cup\sigma$ such that $(E-W)(S_{-}\cap
V)\subset\C_{+}\backslash\R$. Actually, in the neighborhood of
$0$, we have $\I(E-W(\varphi))\geq 0$; according to $(H_{W,g})$,
there exists a neighborhood $V$ of $[\varphi_{r}^{-},0]\cup\sigma$
such that $(E-W)(S_{-}\cap V)$ does not intersect $\R$. By
continuity of the mapping $\varphi\mapsto\I(E-W(\varphi))$, the
sign of $\I(E-W(\varphi))$ remains positive on $S_{-}\cap V$.
There exists a branch $k$ of the quasi-momentum such that $$\I
k(\mathcal{E})>0 \textrm{ for }\I\mathcal{E}>0 \textrm{ and }
k(E_{n}+i0)=0,\ k(E_{p}+i0)=\pi.$$ We define
$\kappa_{i}(\varphi)=k(E-W(\varphi))$.
\item The point $0$ is a maximum for $W$ and the band $B$ is in
the form $[E_{4l+3},E_{4l+4}]$; the points $E_{r}$ and $E_{i}$
satisfy $E_{r}=E_{4l+4}$ and $E_{i}=E_{4l+3}$. Let $k$ be the
branch of the quasi-momentum such that $\I k(\mathcal{E})>0$ for
$\I\mathcal{E}<0$, and $k(E_{n})=0$; then $k(E_{p})=\pi$. We
define $\kappa_{i}(\varphi)=k(E-W(\varphi))$.
\end{enumerate}
The case $\kappa_{i}(\varphi_{r}^{-})=\pi$ is similar.
\end{itemize}
This completes the proof of the lemma.
\end{dem}\\
For the sake of clarity, for all the proofs, we shall consider the
case:
\begin{equation}
\label{premcassc} \kappa_{i}(\varphi_{r}^{-}+i0)=0\textrm{ et
}\kappa_{i}(\varphi_{i}-0)=\pi
\end{equation}
The arguments in the second case are similar and we will not give
the details.
\subsubsection{Other branches of the complex momentum}
\label{compmom} \label{detsc} The properties of the complex
momentum near the branch points $\varphi_{i},\
\overline{\varphi_{i}},\ \varphi_{r}^{\pm}$ are
determined by the behavior of $k$ near $E_{r}$ and $E_{i}$.\\
Now, we describe other branches of $\kappa$, which are obtained
from the branch $\kappa_{i}$ (Lemma \ref{dtepos}) by analytic
continuation. We consider the case \eqref{premcassc}.
\begin{itemize}
\item We denote by $\kappa_{g}$ the continuation of $\kappa_{i}$
to the domain $\{\Ra(\varphi)<\Ra(\varphi_{r}^{-})\}$.
$\kappa_{g}$ satisfies
$$\begin{array}{c}\I(\kappa_{g}(\varphi))>0\textrm{ for }\{\Ra(\varphi)<\Ra(\varphi_{r}^{-})\}\\\Ra(\kappa_{g})(\varphi)\rightarrow 0\textrm{ as }\Ra(\varphi)\rightarrow -\infty. \end{array}$$
$\kappa_{g}$ is the continuation of $\kappa_{i}$ through
$(-\infty,\varphi_{r}^{-}]$.
\item We consider the strip $S_{Y}$ cut along
$(\Sigma\backslash\sigma)\cup(\overline{\Sigma}\backslash\overline{\sigma})\cup(-\infty,\varphi_{r}^{-})\cup(\varphi_{r}^{+},+\infty)$.
We always denote by $\kappa_{i}$ the continuation of $\kappa_{i}$
through $C$.
\item On $\{\Ra(\varphi)>\Ra(\varphi_{r}^{+})\}$, we fix a
continuous branch $\kappa_{d}$ with the conditions:
$$\begin{array}{c}\I(\kappa_{d}(\varphi))>0\textrm{ for }\{\Ra(\varphi)>\Ra(\varphi_{r}^{+})\}\\\Ra(\kappa_{d})(\varphi)\rightarrow 0\textrm{ as }\Ra(\varphi)\rightarrow +\infty. \end{array}$$
$\kappa_{d}$ is the continuation of $\kappa_{i}$ through
$\overline{S_{+}}$.
\end{itemize}
Here, we describe the behavior of the different branches of
$\kappa$.
\begin{equation}
\label{kgki} \forall\varphi\in
S_{-},\quad\kappa_{g}(\varphi)=\kappa_{i}(\varphi)\quad;\quad\forall\varphi\in\overline{S_{-}},\quad\kappa_{g}(\varphi)=-\kappa_{i}(\varphi).
\end{equation}
\begin{equation}
\label{kdki} \forall\varphi\in
S_{+},\quad\kappa_{d}(\varphi)=-\kappa_{i}(\varphi)\quad;\quad\forall\varphi\in\overline{S_{+}},\quad\kappa_{d}(\varphi)=\kappa_{i}(\varphi).
\end{equation}
\psset{unit=1em,linewidth=.05}

\psset{unit=0.5em,linewidth=.05}
\begin{center}
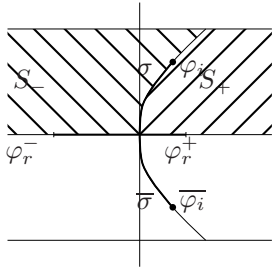
\begin{figure}
\begin{pspicture}(-50,-10)(10,10)
\psline[linewidth=0.05](-10,0)(10,0)\psline[linewidth=0.05](0,-10)(0,10)
\psline(-10,8)(10,8) \psline(-10,-8)(10,-8)
\psline(-6.5,-0.2)(-6.5,0.2)\uput[180](-6.5,-1.2){$\varphi_{r}^{-}$}
\psline(3.5,-0.2)(3.5,0.2)\uput[180](5.5,-1.2){$\varphi_{r}^{+}$}
\psline(-6.5,0)(3.5,0)
\pscurve[linewidth=0.05](0,0)(0.25,2.3)(2.5,5.5)(5,8)\psdots[dotstyle=*](2.5,5.5)\uput[180](6,5){$\varphi_{i}$}\uput[180](2,5){$\sigma$}
\pscurve(0,0)(0.25,2.3)(2.5,5.5)
\pscurve[linewidth=0.05](0,0)(0.25,-2.3)(2.5,-5.5)(5,-8)\psdots[dotstyle=*](2.5,-5.5)\uput[180](6,-5){$\overline{\varphi_{i}}$}\uput[180](2,-5){$\overline{\sigma}$}
\pscurve(0,0)(0.25,-2.3)(2.5,-5.5)
\pscurve[linewidth=0.15](0,0)(0.25,2.3)(2.5,5.5)
\pscurve[linewidth=0.15](0,0)(0.25,-2.3)(2.5,-5.5)
\psline[linewidth=0.15](-6.5,0)(3.5,0)
\pscustom[linestyle=none,fillstyle=vlines,hatchsep=1.5,hatchangle=45]{\psline(-10,0)(0,0)\pscurve(0,0)(0.25,2.3)(2.5,5.5)(5,8)\psline(5,8)(-10,8)}
\pscustom[linestyle=none,fillstyle=vlines,hatchsep=1.5,hatchangle=-45]{\psline(10,0)(0,0)\pscurve(0,0)(0.25,2.3)(2.5,5.5)(5,8)\psline(5,8)(10,8)}
\uput[180](-6,4){$S_{-}$}\uput[180](8,4){$S_{+}$}
\end{pspicture}
\caption{The domains $S_{-}$ and $S_{+}$}\label{smoinsa}
\end{figure}
\end{center}
\subsection{Stokes lines}
\label{stline} This section is devoted to the description of the
Stokes lines under assumptions $(H_{W,g})$ and $(H_{J})$. We
describe the Stokes lines beginning at $\varphi_{r}^{-}$,
$\varphi_{r}^{+}$, $\varphi_{i}$ and $\overline{\varphi_{i}}$.
Since $W$ is real on the real line, the set of the Stokes lines is
symmetric with respect to the real line.\\
First, $\kappa_{i}$ is real on the interval
$[\varphi_{r}^{-},\varphi_{r}^{+}]\subset\R$; therefore,
$[\varphi_{r}^{-},\varphi_{r}^{+}]$ is a part of a Stokes line
beginning at $\varphi_{r}^{-}$. The two other Stokes lines
beginning at $\varphi_{r}^{-}$ are symmetric with respect to the
real line. We denote by $b$ the Stokes line going upward and by
$\bar{b}$ its symmetric. Similarly, we denote by $a$ and $\bar{a}$
the two other Stokes lines beginning at $\varphi_{r}^{+}$; $a$
goes upwards.\\
Consider the Stokes lines beginning at $\varphi_{i}$. The angles
between the Stokes lines at this point are equal to $2\pi/3$. So,
one of the Stokes lines is situated between $\Sigma$ and
$e^{\frac{2i\pi}{3}}\Sigma$. It is locally going to the right of
$\Sigma$; we denote by $d$ this line. Similarly, we denote by $e$
the Stokes line between $\Sigma$ and $e^{-\frac{2i\pi}{3}}\Sigma$.
Finally, we denote by $c$ the third Stokes line beginning at
$\varphi_{i}$; $c$ is going upwards.\\
By symmetry, we denote by $\bar{c}$, $\bar{d}$ and $\bar{e}$ the
Stokes lines beginning at $\overline{\varphi_{i}}$.\\
We describe the behavior of $a$, $b$, $c$, $d$ and $e$ in the
strip $S_{Y}$. We have represented these lines in figure \ref{ls}.
In this figure, we have precised the values of $\kappa$ in the
branch points.
\begin{lem}
\label{stlinea} We assume that $V$, $W$ and $J$ satisfy $(H_{V})$,
$(H_{W})$ and $(H_{J})$. Then, the Stokes lines described in
figure \ref{ls} have the following properties:
\begin{enumerate}
\item $a$ stays vertical; it intersects $\{\I(\varphi)=Y\}$.
\item $b$ stays vertical; it intersects $\{\I(\varphi)=Y\}$.
\item $d$ intersects $a$ above $\varphi_{r}^{+}$; the segment
between $\varphi_{i}$ and this intersection with $a$ is vertical.
\item $e$ intersects $b$ above $\varphi_{r}^{-}$; the segment
between $\varphi_{i}$ and this intersection with $b$ is vertical.
\item $c$ stays vertical; it intersects $\{\I(\varphi)=Y\}$ and does
not intersect $\sigma$.
\item $a$ and $c$ do not intersect one another in the strip $S_{Y}$.
\item $b$ and $c$ do not intersect one another in the strip $S_{Y}$.
\end{enumerate}
\end{lem}
\begin{dem}
First, we note that a Stokes line can become horizontal only at a
point where $\I \kappa=0$, i.e. at a point of the pre-image of a
spectral band. Besides, a Stokes line beginning at
$\varphi_{1}^{\pm}$ (respectively at $\varphi_{2}$ or
$\overline{\varphi_{2}}$) is locally orthogonal to $i\
\overline{\kappa(\varphi)}$ (respectively $i\
\overline{(\pi-\kappa(\varphi))}$).\\
We first prove 1). According $(H_{_{J}})$, the pre-image of the
spectrum is $[\varphi_{r}^{-},\varphi_{r}^{+}]\cup\sigma$. So, $a$
becomes horizontal only if it intersects $\sigma$. Let us prove by
contradiction that it is impossible. Let us assume that $a$
intersects $\sigma$ in $\varphi_{a}$, then:
$$\I\int_{\varphi_{r}^{+}}^{\varphi_{a}}\kappa(u)du=0=\I\int_{0,\textrm{ along }\sigma}^{\varphi_{a}}\kappa(u)du$$
$$=\int_{0}^{\varphi_{a}}(\Ra\kappa(u))d(\I(u))\leq-k_{1}(E-W_{-})\I\varphi_{a}<0$$
which is impossible. Therefore, $a$ stays vertical. Moreover, as
$\varphi\rightarrow\infty,\ \varphi\in S_{Y},\
\I(i\bar{\kappa})\rightarrow 0$. Thus, $a$ admits a vertical
asymptote and intersects $\{\I(\varphi)=Y\}$.\\
Similarly, we prove 2).\\
To prove 3), we consider the Stokes line $d$. If $a$ and $d$ do
not intersect one another, then $d$ intersect either $\sigma$ or
$[0,\varphi_{r}^{+}]$. In this case, we denote by $\varphi_{d}$
the intersection between $d$ and $\sigma$ and we have:
$$\I\int_{\varphi_{d}}^{\varphi_{i}}(\kappa(u)-\pi)du=0=\int_{\varphi_{d}}^{\varphi_{i}}\Ra(\kappa(u)-\pi)d(\I(u))<0$$
Consequently, $d$ and $a$ do not intersect one another. Before its
intersection with $a$, $d$ does not intersect the pre-image of a
spectral band and it stays vertical. We prove similarly the
properties of $e$.\\
We prove now 5). $c$ is going upwards. $c$ does not intersect the
pre-image of a spectral band in
$\{\I\varphi\in]\I\varphi_{i},Y[\}$ and $c$ stays vertical.\\
We prove 6) by contradiction. Let us assume that there is
$\varphi_{a}\in a\cap c$. Then, we compute:
$$\I\int_{0}^{\varphi_{a}}\kappa(u)du=0=\I\int_{\sigma}\kappa(u)du+\I\int_{\varphi_{i}}^{\varphi_{a}}\kappa(u)du$$
First,
$\I\int_{\varphi_{i}}^{\varphi_{a}}\kappa(u)du=\pi\I(\varphi_{a}-\varphi_{i})>0$
and $\I\int_{\sigma}\kappa(u)du=\int_{\sigma}\Ra\kappa(u)d(\I u)>0$.\\
which is impossible. So, $a$ and $c$ do not intersect one another
in $S_{Y}$.
\end{dem}\\
In the following, we choose $\widetilde{Y}\in]\sup\limits_{E\in
J}\I\varphi_{i}(E),Y[$.
\psset{unit=1em,linewidth=.05}

\psset{unit=.7em,linewidth=.1}
\begin{center}
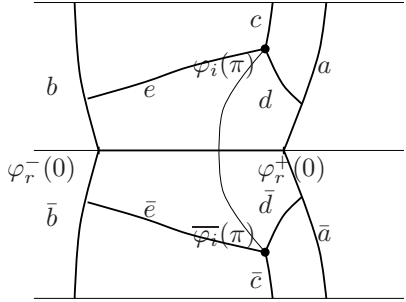
\begin{figure}
\begin{pspicture}(-40,-10)(10,10)
\psline[linewidth=0.005](-10,0)(10,0)
\psline[linewidth=0.05](-10,8)(10,8)
\psline[linewidth=0.05](-10,-8)(10,-8)
\psline(-6.5,-0.2)(-6.5,0.2)\uput[180](-7,-1){$\varphi_{r}^{-}(0)$}
\psline(3.5,-0.2)(3.5,0.2)\uput[180](6.5,-1){$\varphi_{r}^{+}(0)$}
\psline(-6.5,0)(3.5,0)
\pscurve[linewidth=0.005](0,0)(0.25,2.3)(2.5,5.5)\psdots[dotstyle=*](2.5,5.5)\uput[180](2.7,4.6){$\varphi_{i}(\pi)$}
\pscurve[linewidth=0.005](0,0)(0.25,-2.3)(2.5,-5.5)\psdots[dotstyle=*](2.5,-5.5)\uput[180](2.7,-4.6){$\overline{\varphi_{i}}(\pi)$}
\pscurve(3.5,0)(5.4,5.3)(5.8,8)\uput[180](6.8,4.5){$a$}
\pscurve(3.5,0)(5.4,-5.3)(5.8,-8)\uput[180](6.8,-4.5){$\bar{a}$}
\pscurve(-6.5,0)(-7.5,4)(-7.8,8)\uput[180](-8,3.5){$b$}
\pscurve(-6.5,0)(-7.5,-4)(-7.8,-8)\uput[180](-8,-3.5){$\bar{b}$}
\pscurve(2.5,5.5)(2.7,6.5)(2.9,8)\uput[180](3,7){$c$}
 \pscurve(2.5,5.5)(3.5,3.5)(4.5,2.5)\uput[180](3.6,2.8){$d$}
\pscurve(2.5,5.5)(-1.7,4.5)(-4,3.7)(-7.1,2.8)\uput[180](-2.7,3.2){$e$}
\pscurve(2.5,-5.5)(2.7,-6.5)(2.9,-8)\uput[180](3,-7){$\bar{c}$}
\pscurve(2.5,-5.5)(3.5,-3.5)(4.5,-2.5)\uput[180](3.6,-2.8){$\bar{d}$}
\pscurve(2.5,-5.5)(-1.7,-4.5)(-4,-3.7)(-7.1,-2.8)\uput[180](-2.7,-3.2){$\bar{e}$}
\end{pspicture}
\caption{Stokes lines}\label{ls}
\end{figure}
\end{center}
\subsection{Construction of a consistent basis with standard
behavior in the neighborhood of the cross} In this section, we
begin with constructing a canonical line near the cross. To do
that, we follow the methods developed in \cite{FK4}.
\subsubsection{General constructions}
We first recall some general geometric tools presented in
\cite{FK4}, section 4.1.
\begin{itemize}
\item We first introduce the idea of enclosing canonical domain.
\begin{defn}
Let $\gamma\subset D$ be a line canonical with respect to
$\kappa$. Denote its ends by $\xi_{1}$ and $\xi_{2}$. Let a domain
$K\subset D$ be a canonical domain corresponding to the triple
$\kappa,\ \xi_{1}$ and $\xi_{2}$. If $\gamma\subset K$, then $K$
is called a canonical domain {\it enclosing $\gamma$}.
\end{defn}
We have the following property:
\begin{lem}\cite{FK2}\\
\label{enccan} One can always construct a canonical domain
enclosing any given compact canonical curve located in an
arbitrarily small neighborhood of that curve.
\end{lem}
Such canonical domains, whose existence is established using this
lemma are called {\it local}.
\item To construct a canonical domain, we need a canonical line to
start with. To construct such a line, we first build pre-canonical
lines made of some ``elementary'' curves. Let $\gamma\subset D$ be
a vertical curve. We call $\gamma$ {\it pre-canonical} if it is a
finite union of bounded segments of canonical lines and/or lines
of Stokes type. The interest of pre-canonical curves is the
following:
\begin{lem}\cite{FK2}\\
\label{precan} Let $\gamma$ be a pre-canonical curve. Denote the
ends of $\gamma$ by $\xi_{a}$ and $xi_{b}$. Fix $V\subset D$, a
neighborhood of $\gamma$ and $V_{a}\subset D$ a neighborhood of
$\xi_{a}$. Then, there exists a canonical line $\tilde{\gamma}$
connecting the point $\xi_{b}$ to some point in $V_{a}$.
\end{lem}
\end{itemize}
\subsubsection{Constructing a canonical line near the cross}
Here, we mimic the construction of \cite{FK4}, section 4.2. We
assume that assumptions $(H_{V})$, $(H_{W,r})$, $(H_{W,g})$ and
$(H'_{J})$ are satisfied. We now explain the construction of a
canonical line going from $\{\I\xi=-Y\}$ to $\{\I\xi=Y\}$. First,
we consider the curve $\beta$ which is the union of the Stokes
line $\bar{b}$, the segment $[\varphi_{r}^{-},0]$ of the real
line, the closed curve
$\sigma_{+}$ and the Stokes line $c$.\\
We now construct $\alpha$ a pre-canonical line close to the line
$\beta$. We prove:
\begin{prop}
\label{cancurva}
 Fix $\delta>0$. In the
$\delta$-neighborhood of $\beta$, there exists $\alpha$ a
pre-canonical line with respect to the branch $\kappa$ connecting
$\xi_{1}$ to $\xi_{2}$ and having the following properties:
\begin{itemize}
\item at its upper end, $\I\xi_{2}=Y$,
\item at its lower end, $\I\xi_{1}=-Y$,
\item it goes around the branch points of the complex momentum as
the curve shown in figure \ref{cancurve};
\item it contains a canonical line which stays in $S_{-}$, goes
downward from a point in $S_{-}$ to the curve $\sigma$ and then
continues along this curve until it intersects the real line.
\end{itemize}
\end{prop}
\begin{dem}
The proof of this Proposition is completely similar to the proof
of Proposition 4.2 in \cite{FK4}. It consists in breaking down
$\alpha$ in ``elementary'' segments. We do not give the details.
\end{dem}\\
An immediate consequence of Proposition \ref{cancurva} is the
following result:
\begin{prop}
\label{cancurvb} In arbitrarily small neighborhood of the
pre-canonical line $\alpha$, there exists a canonical line
$\gamma$ which has all the properties of the line $\alpha$ listed
in Proposition \ref{cancurva}.
\end{prop}
\psset{unit=.5em,linewidth=.05}

\begin{center}
\begin{figure}
\begin{pspicture}(-40,-10)(10,10)
\psline[linewidth=0.005](-10,0)(10,0)
\psline[linewidth=0.05](-10,8)(10,8)
\psline[linewidth=0.05](-10,-8)(10,-8)
\psline(-6.5,-0.2)(-6.5,0.2)\uput[180](-7.2,-1){$\varphi_{r}^{-}$}
\psline(3.5,-0.2)(3.5,0.2)\uput[180](5.7,-1){$\varphi_{r}^{+}$}
\psline(-6.5,0)(3.5,0)
\pscurve[linewidth=0.005](0,0)(0.25,2.3)(2.5,5.5)\psdots[dotstyle=*](2.5,5.5)\uput[180](3.7,4.6){$\varphi_{i}$}
\pscurve[linewidth=0.005](0,0)(0.25,-2.3)(2.5,-5.5)\psdots[dotstyle=*](2.5,-5.5)\uput[180](3.7,-4.6){$\overline{\varphi_{i}}$}
\pscurve(-6.5,0)(-7.5,-4)(-7.8,-8)
\pscurve[linewidth=0.2](2.4,8)(2.2,6.5)(2,5.5)(0.25,2.3)(0,0)\pscurve[linewidth=0.2](0,0)(-5.8,-1.5)(-7,-4)(-7.3,-8)
\pscurve(2.5,5.5)(2.7,6.5)(2.9,8)
\end{pspicture}
\caption{A canonical curve}\label{cancurve}
\end{figure}
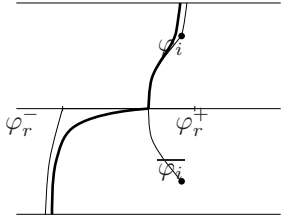
\end{center}
\subsubsection{Some continuation tools}
In this section, we recall some continuation tools; these tools
are developed in \cite{FK4}.
\begin{enumerate}
\item
Now, we present the continuation lemma on compact domains. We
recall that $q$ is defined in \eqref{racq}.
\begin{lem}
\label{lemcontfin} \cite{FK1}
 Let $\varphi_{-}, \varphi_{+},
\varphi_{0}$ be fixed points such that
\begin{itemize}
\item $\I\varphi_{-}=\I\varphi_{+}$;
\item there is no branch point of $\varphi\mapsto\kappa(\varphi)$ on the interval $[\varphi_{-},
\varphi_{+}]$;
\item $\varphi_{0}\in(\varphi_{-} \varphi_{+}),
q(\varphi_{0})\neq 0.$
\end{itemize}
Fix a continuous branch of $\kappa$ on $[\varphi_{-},
\varphi_{+}]$. Let $f(x,\varphi,E,\varepsilon)$, $ f
_{\pm}(x,\varphi,E,\varepsilon)$ be solutions of \eqref{eqp} for
$\varphi\in[\varphi_{-}, \varphi_{+}]$ and $x\in[-X,X]$ satisfying
\eqref{coh} and such that:
\begin{enumerate}
\item $f(x,\varphi,E,\varepsilon)=e^{\frac{i}{\varepsilon}\int_{\varphi_{0}}^{\varphi}\kappa(u)du}(\psi_{+}(x,\varphi,E)+o(1))$
pour $\varphi\in[\varphi_{-}, \varphi_{0}]$ for
$\varphi\in[\varphi_{-}, \varphi_{0}]$ when
$\varepsilon\rightarrow 0$ and the asymptotic is differentiable in
$x$;
\item $f_{\pm}(x,\varphi,E,\varepsilon)=e^{\pm\frac{i}{\varepsilon}\int_{\varphi_{0}}^{\varphi}\kappa(u)du}(\psi_{\pm}(x,\varphi,E)+o(1))$
for $\varphi\in[\varphi_{-}, \varphi_{+}]$ when
$\varepsilon\rightarrow 0$, and the asymptotic is differentiable
in $x$.
\end{enumerate}
Here, $\psi_{\pm}$ are canonical Bloch solutions associated to the
complex momentum $\kappa$.\\
Then,\\
\begin{itemize}
\item if $\I(\kappa(\varphi))>0$ for all $\varphi\in
[\varphi_{-}, \varphi_{+}]$, there exists $C>0$ such that, for
$\varepsilon>0$ small enough,
\begin{equation}
\left|\frac{df}{dx}(x,\varphi,E,\varepsilon)\right|+|f(x,\varphi,E,\varepsilon)|\leq
C
e^{\frac{1}{\varepsilon}\int_{\varphi}^{\varphi_{0}}|\I\kappa(u)|du},\quad\varphi\in
[\varphi_{0}, \varphi_{+}];
\end{equation}
\item if $\I(\kappa(\varphi))<0$ for all
$\varphi\in[\varphi_{-}, \varphi_{+}]$, then
\begin{equation}
f(x,\varphi,E,\varepsilon)=e^{\frac{i}{\varepsilon}\int_{\varphi_{0}}^{\varphi}\kappa(u)du}(\psi_{+}(x,\varphi,E)+o(1)),\quad\varphi\in[\varphi_{0},
\varphi_{+}],
\end{equation}
and the asymptotic is differentiable in $x$.
\end{itemize}
\end{lem}
Intuitively, this lemma means that a function $f$ has the standard
behavior along a horizontal line as long as the leading term of
its asymptotics is growing along that line. For analogous results
with real WKB method, we refer to \cite{Vo}.
\item
The estimate we obtained in Lemma \ref{lemcontfin} can be far from
optimal. The Adjacent Canonical Domain Principle gives a more
precise result:
\begin{prop}\cite{FK3}
\label{adjdom} Assume that a solution $f$ has standard behavior in
either the left hand side or the right hand side of a constant
neighborhood of a vertical curve $\gamma$. Assume that $\gamma$ is
canonical with respect to some branch of the complex momentum.
Then $f$ has standard behavior in any bounded canonical domain
enclosing $\gamma$.
\end{prop}
\item The last tool we shall need in the sequel is the Stokes
Lemma.\\
Notations and assumptions:\\
Assume that $\xi_{0}$ is a branch point of the complex momentum
such that $W'(\xi_{0})\neq 0$. There are three Stokes lines
beginning at $\xi_{0}$. The angles between them at $\xi_{0}$ are
equal to $2\pi/3$. We denote these lines by $\sigma_{1}$,
$\sigma_{2}$ and $\sigma_{3}$, so that $\sigma_{1}$ is vertical at
$\xi_{0}$. Let $V$ be a neighborhood of $\xi_{0}$; assume that $V$
is so small that $\sigma_{1}$, $\sigma_{2}$ and $\sigma_{3}$
divide it into three sectors. We denote them by $S_{1}$, $S_{2}$
and $S_{3}$ so that $S_{1}$ be situated between $\sigma_{1}$ and
$\sigma_{2}$, and the sector $S_{2}$ be between $\sigma_{2}$ and
$\sigma_{3}$ (see figure \ref{lsbp}).\\
We recall now the result:
\begin{lem}\cite{FK4}.\label{stoklemma} Let $V$ be sufficiently
small. Let $f$ be a solution that has standard behavior
$f=e^{\frac{i}{\varepsilon}\int^{\varphi}\kappa(u)du}(\psi_{+}(x,\varphi)+o(1))$
inside the sector $S_{1}\cup\sigma_{2}\cup S_{2}$ of $V$.
Moreover, assume that, in $S_{1}$ near $\sigma_{1}$, one has
$\I\kappa>0$ if $S_{1}$ is to the left of $\sigma_{1}$ and
$\I\kappa<0$ otherwise. Then, $f$ has standard behavior
$f=e^{\frac{i}{\varepsilon}\int^{\varphi}\kappa(u)du}(\psi_{+}(x,\varphi)+o(1))$
inside $V\backslash\sigma_{1}$, the asymptotics being obtained by
analytic continuation from $S_{1}\cup\sigma_{2}\cup S_{2}$ to
$V\backslash\sigma_{1}$.
\end{lem}
\end{enumerate}
\psset{unit=1em,linewidth=.05}

\psset{unit=0.7em,linewidth=.05}
\begin{center}
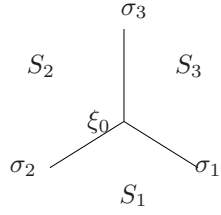
\begin{figure}
\begin{pspicture}(-30,-10)(10,10)
\psline(0,0)(0,5) \psline(0,0)(4,-2.5) \psline(0,0)(-4,-2.5)
\uput[180](0,0){$\xi_{0}$}\uput[180](6,-2.5){$\sigma_{1}$}\uput[180](-4,-2.5){$\sigma_{2}$}\uput[180](2,6){$\sigma_{3}$}
\uput[180](2,-4){$S_{1}$}\uput[180](5,3){$S_{3}$}\uput[180](-3,3){$S_{2}$}
\end{pspicture}
\caption{Stokes lines and Stokes Lemma}\label{lsbp}
\end{figure}
\end{center}
\subsubsection{Construction of a basis with standard asymptotic
behavior near the cross} We prove the existence of a consistent
basis with standard asymptotic behavior near the canonical line
$\alpha$. Let $\alpha$ be the curve described in Proposition
\ref{cancurvb}. According to Lemma \ref{enccan}, we can construct
a local canonical domain $K_{i}$ enclosing $\alpha$.
\begin{prop}
\label{bcki} Assume that $(H_{V})$, $(H_{W,r})$, $(H_{W,g})$ and
$(H_{J})$ are satisfied. Fix $E_{0}\in J$, $X>1$ and
$\tilde{Y}\in]0,\tilde{Y}[$. Then, there exist a complex
neighborhood $\mathcal{U}_{0}$ of $E_{0}$, a real number
$\varepsilon_{0}>0$ and a function $f_{i}$ satisfying the
following properties:
\begin{itemize}
\item The function $(x,\varphi,E,\varepsilon)\mapsto
f_{i}(x,\varphi,E,\varepsilon)$ is defined on $\R\times
S_{\tilde{Y}}\times\mathcal{U}_{0}\times]0,\varepsilon_{0}[$.
\item For any $x\in\R$, for any
$\varepsilon\in]0,\varepsilon_{0}[$, the function
$((\varphi,E)\mapsto f_{i}(x,\varphi,E,\varepsilon))$ is analytic
on $S_{\tilde{Y}}\times\mathcal{U}_{0}$.
\item For $(x,\varphi,E)\in [-X,X]\times
K_{i}\times\mathcal{U}_{0}$, the function $f_{i}$ has the
asymptotic behavior:
\begin{equation}
\label{asca}
f_{i}(x,\varphi,E,\varepsilon)=e^{\frac{i}{\varepsilon}\int_{0}^{\varphi}\kappa(u)
du}\left(\psi_{\pm}(x,\varphi,E)+o(1)\right),\quad\varepsilon\rightarrow
0.
\end{equation}
\item The asymptotics \eqref{asca} are uniform in $(x,\varphi,E)\in [-X,X]\times K_{i}\times\mathcal{U}_{0}$.
\item The asymptotics \eqref{asca} can be differentiated once in $x$.
\item There exists a real number $\sigma_{i}\in\{-1,1\}$ such that the function $f_{i}$ satisfies the relation:
$$w(f_{i},f_{i}^{*})=w(f_{i}(\cdot,\varphi,E,\varepsilon),f_{i}(\cdot,\overline{\varphi},\overline{E},\varepsilon))=\sigma_{i}(k'_{i}w_{i})(E-W(0))$$
\end{itemize}
\end{prop}
The end of this section is devoted to the proof of Proposition
\ref{bcki}. This Proposition mainly follows from Theorem
\ref{finwkbthm}.
\subsubsection{Existence of $f_{i}$}
The domain $K_{i}$ is a local canonical domain. According to
Theorem \ref{finwkbthm}, we can build a function $f_{i}$ such
that, on $K_{i}$, $f_{i}$ has the following asymptotic behavior:
$$f_{i}\sim
e^{\frac{i}{\varepsilon}\int^{\varphi}\kappa_{i}}\psi_{+}.$$ Let
us normalize $f_{i}$ in $0$.
\subsubsection{Computation of the Wronskian $w(f_{i},f_{i}^{*})$}
\label{precsigma} To finish the proof of Proposition \ref{bcki},
it remains to
compute $w(f_{i},f_{i}^{*})$.\\
Let $R_{+}$ be a small enough rectangle to the left of
$\alpha_{+}$, so that $R_{+}\subset K_{i}\cap S_{\tilde{Y}}$. We
define $R=R_{+}\cup R_{-}$; we study the behavior of $f_{i}$ and
$(f_{i})^{*}$ in $R$.
\begin{itemize}
\item First, by construction, in $R_{+}$, the function $f_{i}$
satisfies:
$$ f_{i}\sim
e^{\frac{i}{\varepsilon}\int_{0}^{\varphi}\kappa_{i}}\psi_{+}^{i}.$$
\item To the right of $\alpha_{-}$, the function $f_{i}$ satisfies
$$ f_{i}\sim
e^{\frac{i}{\varepsilon}\int_{0}^{\varphi}\kappa_{i}}\psi_{+}^{i},$$
with $\I\kappa_{i}<0$. According to Lemma \ref{lemcontfin}, we
know that, in $\bar{S_{-}}$, the function $f_{i}$ admits the
asymptotic behavior:
$$ f_{i}\sim
e^{\frac{i}{\varepsilon}\int_{0}^{\varphi}\kappa_{i}}\psi_{+}^{i}.$$
\item Thus, the function $f_{i}$ has the standard asymptotic behavior in
$R$.
\item Now, we study the behavior of $f_{i}^{*}$. To do that, we
start with describing the main objects related to $\kappa_{i}$ in
$R$. Let $k_{i}$ be the branch of the quasi-moment of
\eqref{espc}, analytically continued through $[E_{r},E_{i}]$ and
satisfying:
\begin{equation*}
k_{i}(E_{r})=0\quad\textrm{ and }\quad k_{i}(E_{i})=\pi
\end{equation*}
$k_{i}$ is real on $[E_{r},E_{i}]$. Therefore, $k_{i}$ satisfies:
$$k_{i}(\overline{\E})=\overline{k_{i}(\E)}.$$
The branch $\kappa_{i}$ satisfies
$\kappa_{i}(\varphi)=k_{i}(E-W(\varphi))$. The associated
canonical Bloch solutions $\Psi_{\pm}^{i}$ are such that:
$$\overline{\Psi_{+}^{i}(x,\overline{\varphi})}=\Psi_{-}^{i}(x,\varphi).$$
Therefore, we have in $R$:
\begin{equation}
\label{kappai} \kappa_{i}^{*}(\varphi)=\kappa_{i}(\varphi)\quad
(\Psi_{+}^{i})^{*}(\varphi)=\Psi_{-}^{i}(\varphi)\quad(\omega_{+}^{i})^{*}(\varphi)=\omega_{-}^{i}(\varphi)\quad
\forall\varphi\in R.
\end{equation}
Besides, since $k'_{i}$ is real on the band, there exists a real
number $\sigma_{i}\in\{-1,1\}$ such that:
\begin{equation}
q_{i}^{*}(\varphi)=\sigma_{i}q_{i}(\varphi).
\end{equation}
We shall precise this coefficient in section \ref{deco}.\\
We compute:
$$w(f_{+}^{i}(\cdot,\varphi,E,\varepsilon),(f_{+}^{i})^{*}(\cdot,\varphi,E,\varepsilon))=q_{i}(0)q_{i}^{*}(0)w(\Psi_{+}^{i}(\cdot,0),\Psi_{-}^{i}(\cdot,0))g(\varphi,E,\varepsilon).$$
Since
$\overline{w(\Psi_{+}^{i}(\cdot,0),\Psi_{-}^{i}(\cdot,0))}=-w(\Psi_{+}^{i}(\cdot,0),\Psi_{-}^{i}(\cdot,0))$,
the term $g(\varphi,E,\varepsilon)$ satisfies:
$$g^{*}(\varphi,E,\varepsilon)=\overline{g(\bar{\varphi},\bar{E},\varepsilon)}=g(\varphi,E,\varepsilon),$$
$$g(\varphi,E,\varepsilon)=[1+o(1)].$$
Since the Wronskian is analytic and $\varepsilon$-periodic, this
asymptotic is valid in $S_{\widetilde{Y}}$. \\
Since $g^{*}=g$ and $g=[1+o(1)]$, there exists an analytic
function $(\varphi,E)\mapsto h(\varphi,E,\varepsilon)$ on
$S_{\widetilde{Y}}\times\mathcal{U}$ such that:\\
- $g(\varphi,E,\varepsilon)=h(\varphi,E,\varepsilon)h^{*}(\varphi,E,\varepsilon)$,\\
- $h(\varphi,E,\varepsilon)=[1+o(1)]$.\\
We slightly deform $f_{i}$, i.e., we replace $f_{i}$ by
$\frac{f_{i}}{h(\varphi,E,\varepsilon)}$; the basis
$\{f_{i},f_{i}^{*}\}$ is consistent.
\end{itemize}
This ends the proof of Proposition \ref{bcki}.\\
\section{Consistent Jost solutions of \eqref{eqp}}
\label{scattheory} This section is devoted to the proof of the
following result.
\begin{prop}
\label{propconstinf} We assume that $(H_{V})$, $(H_{W,r})$ and
$(H_{J}^{0})$ are satisfied. Fix $X>1$ and $\lambda>1$. Then,
there exist a complex neighborhood
$\mathcal{V}=\overline{\mathcal{V}}$ of $J$, a real
$\varepsilon_{0}>0$, a constant $C>0$, two complex numbers
$m_{g},\ m_{d}$ and two functions
$(x,\varphi,E,\varepsilon)\mapsto
h_{-}^{g}(x,\varphi,E,\varepsilon)$,
$(x,\varphi,E,\varepsilon)\mapsto
h_{+}^{d}(x,\varphi,E,\varepsilon)$ such that, if we define
$$B_{\varepsilon}^{g}=\left\{\varphi\in S_{Y}\ ;\
\Ra\varphi<-C\varepsilon^{-\frac{\lambda}{s-1}}\right\}\textrm{ et
}B_{\varepsilon}^{d}=\left\{\varphi\in S_{Y}\ ; \
\Ra\varphi>C\varepsilon^{-\frac{\lambda}{s-1}}\right\},$$ then
\begin{itemize}
\item The functions $(x,\varphi,E,\varepsilon)\mapsto
h_{-}^{g}(x,\varphi,E,\varepsilon)$ and
$(x,\varphi,E,\varepsilon)\mapsto
h_{+}^{d}(x,\varphi,E,\varepsilon)$ are clearly defined and
consistent on $\R\times S_{Y}\times
\mathcal{V}\times]0,\varepsilon_{0}[$.
\item For any $x\in[-X,X]$ and $\varepsilon\in]0,\varepsilon_{0}[$, $(\varphi,E)\mapsto
h_{-}^{g}(x,\varphi,E,\varepsilon)$ and $(\varphi,E)\mapsto
h_{+}^{d}(x,\varphi,E,\varepsilon)$ are analytic on $S_{Y}\times
\mathcal{V}$.
\item The function $x\mapsto h_{-}^{g}(x,\varphi,E,\varepsilon)$ (resp. $x\mapsto h_{+}^{d}(x,\varphi,E,\varepsilon)$)
is a basis of $\mathcal{J}_{-}$ (resp. $\mathcal{J}_{+}$).
\item The functions $h_{-}^{g}$ and $h_{+}^{d}$ have the following asymptotic behavior:
\begin{equation}
\label{as1}
h_{-}^{g}(x,\varphi,E,\varepsilon)=e^{\frac{-i}{\varepsilon}\int_{m_{g}}^{\varphi}\kappa(u)du}\psi_{-}(x,\varphi,E)(1+R_{g}(x,\varphi,E,\varepsilon)),
\end{equation}
and
\begin{equation}
\label{asda1}
h_{+}^{d}(x,\varphi,E,\varepsilon)=e^{\frac{i}{\varepsilon}\int_{m_{d}}^{\varphi}\kappa(u)du}\psi_{+}(x,\varphi,E)(1+R_{d}(x,\varphi,E,\varepsilon)),
\end{equation}
where
\begin{itemize}
\item $R_{g}$ and $R_{d}$ satisfy:
$$\exists M>0,\quad\forall\varepsilon\in]0,\varepsilon_{0}[,\quad,\forall
x\in[-X,X],\quad\forall E\in\mathcal{V},\quad\forall\varphi\in
B_{\varepsilon}^{g},\quad|R_{g}(x,\varphi,E,\varepsilon)|\leq\frac{M}{\varepsilon|\Ra\varphi|^{s-1}},$$
$$\exists M>0,\quad\forall\varepsilon\in]0,\varepsilon_{0}[,\quad,\forall
x\in[-X,X],\quad\forall E\in\mathcal{V},\quad\forall\varphi\in
B_{\varepsilon}^{d},\quad|R_{d}(x,\varphi,E,\varepsilon)|\leq\frac{M}{\varepsilon|\Ra\varphi|^{s-1}}.$$
\item The functions $\psi_{+}$ and $\psi_{-}$ are the Bloch canonical solutions
of the periodic equation \eqref{espa} defined in section
\ref{cansolbloch}.
\end{itemize}
\item The asymptotics (\ref{as1}) and (\ref{asda1}) may be differentiated
once in $x$.
\item There exist two real numbers $\sigma_{g}\in\{-1,1\}$, $\sigma_{d}\in\{-1,1\}$, an integer $p$ and two functions $E\mapsto\alpha_{g}(E)$ and $E\mapsto\alpha_{d}(E)$ such that:
\begin{enumerate}
\item
For any $\varepsilon\in]0,\varepsilon_{0}[$, $x\in\R$,
$E\in\mathcal{V}$,et $\varphi\in S_{Y}$ ,we have:
\begin{equation}
\label{starg}
\overline{\alpha_{g}(E)h_{-}^{g}(x,\overline{\varphi},\overline{E},\varepsilon)}=i\sigma_{g}e^{-\frac{i}{\varepsilon}2p\pi
x} \alpha_{g}(E)h_{-}^{g}(x,\varphi,E,\varepsilon)
\end{equation}
\begin{equation}
\label{stard}
\overline{\alpha_{d}(E)h_{+}^{d}(x,\overline{\varphi},\overline{E},\varepsilon)}=i\sigma_{d}e^{\frac{i}{\varepsilon}2p\pi
x} \alpha_{d}(E)h_{+}^{d}(x,\varphi,E,\varepsilon)
\end{equation}
\item The functions $\alpha_{g}$ and $\alpha_{d}$ are analytic
and given by \eqref{renormconstg} and \eqref{renormconstd}. They
do not vanish on $\mathcal{V}$.
\end{enumerate}
\end{itemize}
\end{prop}
We shall construct some consistent Jost solutions of \eqref{eqp}.
To do that, we regard equation \eqref{eqp} as a perturbation of
equation \eqref{esp} with $\E=E$. We adapt the construction of
Jost functions developed in \cite{Fi1, New}. Precisely, we look
for solutions of \eqref{esp} in the form :
$$F_{-}^{g}=e^{-
ik(E)\varphi/\varepsilon}\psi_{-}^{0}(x,E)(1+o(1)),\quad
x\rightarrow -\infty,$$
$$ F_{+}^{d}=e^{ik(E)\varphi/\varepsilon}\psi_{+}^{0}(x,E)(1+o(1)),\quad x\rightarrow +\infty.$$
Since the functions $(x,\varphi,E,\varepsilon)\mapsto e^{\pm
ik(E)\varphi/\varepsilon}\psi_{\pm}(x,E)$ are consistent, they
allow us to construct a consistent resolvent for the periodic
equation. Using this property and the fact that equation
\eqref{eqp} is invariant by the consistency transformation
$(x,\varphi)\mapsto(x-1,\varphi+\varepsilon)$, we obtain the
consistency of the Jost functions.
\subsection{Construction of the Jost functions}
\label{scathyp} We start with constructing $F_{-}^{g}$. The
construction of $F_{+}^{d}$ is similar. Since the parameter $E$
lies in the neighborhood of a gap, $\I k(E)$ is non zero; the
function $F_{-}^{g}$ is therefore exponentially decreasing and
goes to zero as $x$ goes to $-\infty$. Such a solution is called
recessive.
\subsubsection{}
On a small enough complex neighborhood of $J$,
$\mathcal{V}=\overline{\mathcal{V}}$, one can fix a determination
$k$ of the quasi-momentum such that:
$$\I k(E)\geq\beta>0,\quad\forall E\in\mathcal{V}.$$
Fix $m_{g}$ in $S_{Y}$ such that:
\begin{itemize}
\item The point $m_{g}$ is not a branch point of $\kappa$.
\item It satisfies $\I m_{g}>0$, $k'_{E}(m_{g})\neq 0$.
\item The domain $\{\varphi\in S_{Y}\ ;\ \Ra(\varphi-m_{g})<0\textrm{
and }\I(\varphi-m_{g})>0\}$ does not contain any branch point of
$\kappa$.
\end{itemize}
We define $E_{g}=E-W(m_{g})$.
 We denote by $\psi_{\pm}^{0}$
the analytic Bloch solutions of equation \eqref{esp} normalized at
the point $E_{g}$ ( $k'(E_{g}))\neq 0$). These solutions are
constructed in Lemma \ref{anasol}.
$$ \psi_{\pm}^{0}(x,E)=e^{\pm
ik(E)x}p_{\pm}^{0}(x,E)\quad\textrm{with}\quad
p_{\pm}^{0}(x+1,E)=p_{\pm}^{0}(x,E).$$
 We define :
$$\widetilde{\psi_{\pm}}(x,\varphi,E,\varepsilon)=e^{\pm
ik(E)(x+\frac{\varphi}{\varepsilon})}p_{\pm}^{0}(x,E)=e^{\pm
ik(E)\frac{\varphi}{\varepsilon}}\psi_{\pm}^{0}(x,E).$$ We
consider the resolvent $R$ of $H_{0}$ :
$$(Rg)(x)=-\int_{-\infty}^{x}\frac{\psi_{+}^{0}(x,E)\psi_{-}^{0}(x',E)-\psi_{+}^{0}(x',E)\psi_{-}^{0}(x,E)}{(k'w_{0})(E_{g})}g(x')dx'$$
\subsection{}
Since $\widetilde{\psi_{-}}$ goes to zero as $x$ goes to
$-\infty$, we look for a recessive consistent solution
$\widetilde{f}$ of \eqref{eqp} in the form :
\begin{equation}
\label{eqres}
\tilde{f}(x,\varphi,E,\varepsilon)=\widetilde{\psi_{-}}(x,\varphi,E,\varepsilon)+R[W(\varepsilon
x+\varphi)\tilde{f}(x,\varphi,E,\varepsilon)].
\end{equation}
We define
$\tilde{f}(x,\varphi,E,\varepsilon)=e^{-ik(E)(x+\frac{\varphi}{\varepsilon})}f(x,\varphi,E,\varepsilon)$;
equation \eqref{eqres} is transformed into:
\begin{equation}
\label{eqres_a}
f(x,\varphi,E,\varepsilon)=p_{-}^{0}(x,E)+\int_{-\infty}^{x}A(x,x',E)W(\varepsilon
x'+\varphi)f(x',\varphi,E,\varepsilon)dx'
\end{equation}
where the function $A$ satisfies:
\begin{equation}
\label{noy}
A(x,x',E)=\frac{e^{2ik(E)(x-x')}p_{+}^{0}(x,E)p_{-}^{0}(x',E)-p_{+}^{0}(x',E)p_{-}^{0}(x,E)}{(k'w_{0})(E_{g})}.
\end{equation}
Since $\I k(E)\geq \beta>0$ for $E\in\mathcal{V}$, there exists a
constant $C>0$ such that:
\begin{equation}
\forall x>x',\quad\forall E\in\mathcal{V},\quad |A(x,x',E)|\leq C
\end{equation}
\subsubsection{}
Fix $X_{0}\in\R$ and $a>0$. If $I$ is a real interval, we define:
$$R_{I}=\{\varphi\in S_{Y}\ ;\ \Ra\varphi\in I\}.$$
Let $B((-\infty,X_{0}]\times R_{[-a,a]})$ the set of bounded
functions $\{f\ :\ (x,\varphi)\mapsto f(x,\varphi)\}$ on
$(-\infty,X_{0}]\times R_{[-a,a]}$. The set
$B((-\infty,X_{0}]\times R_{[-a,a]})$ equipped with the norm
$$
\|f\|_{\infty}=\sup\limits_{x\in(-\infty,X_{0}],\Ra\varphi\in[-a,a]}|f(x,\varphi)|$$
is a Banach space.\\
We define the integral operator $T_{E}$ by:
$$\begin{array}{ccccc}T_{E}:&B((-\infty,X_{0}]\times R_{[-a,a]})&\rightarrow &B((-\infty,X_{0}]\times R_{[-a,a]})&\\
&f&\mapsto &F
\end{array}$$
\begin{equation}
\label{opea} \textrm{ where }
F(x,\varphi)=\int_{-\infty}^{x}A(x,x',E)W(\varepsilon
x'+\varphi)f(x',\varphi)dx'.
\end{equation}
The operator $T_{E}$ is a bounded operator on
$B((-\infty,X_{0}]\times R_{[-a,a]})$ and satisfies the estimate:
$$\forall x\in(-\infty,X_{0}],\ \forall\varphi\in R_{[-a,a]},\quad |T_{E}(f)(x,\varphi)|\leq
C\|f\|_{\infty}\int_{-\infty}^{x}|W(\varepsilon x'+\Ra
(\varphi)+i\I(\varphi))|dx'.$$
$$\|T_{E}(f)\|_{\infty}\leq\frac{M}{\varepsilon}\sup\limits_{x\in(-\infty,X_{0}],\Ra\varphi\in[-a,a]}\frac{1}{|\varepsilon x+\Ra(\varphi)|^{s-1}}.$$
\subsubsection{} Fix $\lambda>1$. There exists a constant $C>0$ such that:
$$ |X_{0}|>
C\varepsilon^{-\frac{\lambda
s}{s-1}}\Rightarrow\||T_{E}\||<\varepsilon^{s(\lambda-1)}.$$ We
rewrite \eqref{eqres_a} in the form:
\begin{equation}
\label{eqres_b} (1-T_{E})f=p_{-}^{0}(x,E)
\end{equation}
The operator is then invertible on $B((-\infty,X_{0}]\times
R_{[-a,a]})$. We define:
\begin{equation}
\label{eqresc}
F_{-}^{g}(x,\varphi,E,\varepsilon)=(1-T_{E})^{-1}p_{-}^{0}(x,\varphi,E,\varepsilon)
\end{equation}
We now give some properties of $F_{-}^{g}$.
\subsection{Properties of $F_{-}^{g}$}
\subsubsection{Asymptotic behavior in $x$}
Substituting $$(1-T_{E})^{-1}=1-(1-T_{E})^{-1}T_{E}$$ in equation
(\ref{eqresc}), we obtain:
\begin{equation}
\label{eqresd}
F_{-}^{g}(x,\varphi,E,\varepsilon)=e^{-ik(E)\varphi/\varepsilon}\psi_{-}(x,E)(1+R_{g}(x,\varphi,E,\varepsilon)),
\end{equation}
with
$$
|R_{g}(x,\varphi,E,\varepsilon)|\leq\frac{M}{\varepsilon|\varepsilon
x|^{s-1}},$$ for $x\in(-\infty,X_{0}]$ and $\varphi\in
R_{[-a,a]}$.\\
The function $F_{-}^{g}$ is therefore in the Jost subspace
$\mathcal{J}_{-}$ of equation \eqref{eqp}.
\subsubsection{Study of the consistency}
We assume that $a>1$ and $\varepsilon<1$. We now prove that the
function $F_{-}^{g}$
is consistent.\\
We denote by $G$ the function:
$$G\ :\ (x,\varphi,E,\varepsilon)\mapsto
G(x,\varphi,E,\varepsilon)=F_{-}^{g}(x+1,\varphi-\varepsilon,E,\varepsilon)$$
$G$ is defined for $x\in(-\infty,X_{0}-1]$ and $\varphi\in
R_{[-a+1,a-1]}$. Moreover, the function $G$ belongs to $B((-\infty,X_{0}-1]\times R_{[-a+1,a-1]})$.\\
We define the operator:
$$\begin{array}{ccccc}\widetilde{T_{E}}:&B((-\infty,X_{0}-1]\times R_{[-a+1,a-1]})&\rightarrow &B((-\infty,X_{0}-1]\times R_{[-a+1,a-1]})&\\
&f&\mapsto &F
\end{array}$$
\begin{equation}
\label{opeb} \textrm{ where }
F(x,\varphi)=\int_{-\infty}^{x}A(x,x',E)W(\varepsilon
x'+\varphi)f(x',\varphi)dx'.
\end{equation}
Since $B((-\infty,X_{0}]\times R_{[-a,a]})\subset
B((-\infty,X_{0}-1]\times R_{[-a+1,a-1]})$ and according to
equations \eqref{opea} and \eqref{opeb}, the operator
$\widetilde{T_{E}}$ is an extension of the operator $T_{E}$. Let
us denote by $\widetilde{F_{-}^{g}}$ the restriction of
$F_{-}^{g}$ to $(-\infty,X_{0}-1]\times R_{[-a+1,a-1]}$.\\
We compute in $ B((-\infty,X_{0}-1]\times R_{[-a+1,a-1]})$:
$$(\widetilde{T_{E}}(G))(x,\varphi,E,\varepsilon)=(T_{E}(F_{-}^{g}))(x+1,\varphi-\varepsilon,E,\varepsilon)$$
This leads to:
$$((1-\widetilde{T_{E}})(G))(x+1,\varphi-\varepsilon,E,\varepsilon)=((1-T_{E})(F_{-}^{g}))(x+1,\varphi-\varepsilon,E,\varepsilon)=p^{0}_{-}(x+1,E)=p^{0}_{-}(x,E),$$
The functions $\widetilde{F_{-}^{g}}$ and $G$ satisfy the
relation:
$$((1-\widetilde{T_{E}})(G))=((1-\widetilde{T_{E}})(\widetilde{F_{-}^{g}})).$$
For a sufficiently small $\varepsilon_{0}$, the operator
$\widetilde{T_{E}}$ satisfies, for any
$\varepsilon\in]0,\varepsilon_{0}[$:
$$\||\widetilde{T_{E}}\||<\frac{1}{2}.$$
 The operator $(1-\widetilde{T_{E}})$ is invertible in
$B((-\infty,X_{0}-1]\times R_{[-a+1,a-1]})$ and:
$$\widetilde{F_{-}^{g}}=G.$$
For $\varphi\in R_{[-a+1,a-1]}$, the functions $F_{-}^{g}$ and $G$
coincide on $(-\infty,X_{0}-1]$; according to the Cauchy-Lipschitz
Theorem, they coincide for $x\in\R$. Fix $x\in\R$; $F_{-}^{g}$ and
$G$ coincide for $\varphi\in R_{[-a+1,a-1]}$. By analyticity, they
are equal for $\varphi\in S_{Y}$.
\subsubsection{Asymptotic behavior in $\varphi$}
We use now the consistency of $F_{-}^{g}$ to compute its
asymptotics as $\Ra\varphi$ goes to $-\infty$. Fix $X>0$. We study
$F_{-}^{g}$ for $x\in[-X,X]$. The function $F_{-}^{g}$ is
consistent, and:
$$F_{-}^{g}(x,\varphi,E,\varepsilon)=F_{-}^{g}(x+\frac{[\Ra(\varphi)]}{\varepsilon},\varphi-[\Ra(\varphi)],E,\varepsilon)$$
$$=e^{-ik(E)(x+\varphi/\varepsilon)}p_{-}^{0}(x,E)\left(1+O(\frac{1}{\varepsilon|\varepsilon x+[\Ra\varphi]|^{s-1}})\right).$$
As a result, there exists a constant $C$ such that:
$$\Ra\varphi<
-C\varepsilon^{-\frac{\lambda}{s-1}}\Rightarrow
F_{-}^{g}(x,\varphi,E,\varepsilon)=e^{-ik(E)(x+\varphi/\varepsilon)}p_{-}(x,E)(1+\widetilde{R_{g}}(x,\varphi,E,\varepsilon)),$$
where
$$|\widetilde{R_{g}}(x,\varphi,E,\varepsilon)|\leq\frac{M}{\varepsilon|\Ra\varphi|^{s-1}},$$
for $x\in[-X,X]$ and $\Ra\varphi<
-C\varepsilon^{-\frac{\lambda}{s-1}}$.\\
We define $B_{\varepsilon}^{g}=\{\varphi\in S_{Y}\ ;\ \Ra\varphi<
-C\varepsilon^{-\frac{\lambda}{s-1}}\}$.
\subsection{Renormalization of $F_{-}^{g}$}
We now renormalize $F_{-}^{g}$. We define:
\begin{equation}
\label{renorm}
f_{-}^{g}(x,\varphi,E,\varepsilon)=e^{-\frac{i}{\varepsilon}\int_{m_{g}}^{-\infty}[\kappa(u)-k(E)]du}F_{-}^{g}(x,\varphi,E,\varepsilon),
\end{equation}
where the integral $\int_{m_{g}}^{-\infty}[\kappa(u)-k(E)]du$ is
taken in the upper half plane. The function
$E\mapsto\int_{m_{g}}^{-\infty}[\kappa-k(E)]$ is analytic on
$\mathcal{V}$. For $\varphi\in B_{\varepsilon}^{g}$, we have:
$$f_{-}^{g}(x,\varphi,E,\varepsilon)=e^{-\frac{i}{\varepsilon}\int_{m_{g}}^{\varphi}[\kappa(u)-k(E)]du}e^{-\frac{i}{\varepsilon}\int_{\varphi}^{-\infty}[\kappa-k(E)]}e^{-\frac{ik(E)\varphi}{\varepsilon}}\psi_{-}^{0}(x,E)(1+o(1))$$
Since the function $\psi_{-}$ is analytic and since
$W(\varphi)=O(\varepsilon^{\frac{\lambda s}{s-1}})$ for
$\varphi\in B_{\varepsilon}^{g}$, we get:
\begin{equation}
\forall\varphi\in B_{\varepsilon}^{g},\quad
\psi_{-}(x,\varphi,E)=\psi_{-}^{0}(x,E-W(\varphi))=\psi_{-}^{0}(x,E)(1+o(1))
\end{equation}
We finally obtain that, for $x\in[-X,X]$ and $\varphi\in
B_{\varepsilon}^{g}$:
\begin{equation*}
f_{-}^{g}(x,\varphi,E,\varepsilon)=e^{-\frac{i}{\varepsilon}\int_{m_{g}}^{\varphi}\kappa(u)du}\psi_{-}(x,\varphi,E)(1+o(1))
\end{equation*}
\subsubsection{Symmetries}
Let $\gamma$ be a complex path and $f$ be an analytic function on
$\gamma$. We have:
\begin{equation}
\label{symint}
\int_{\gamma}f(z)dz=\overline{\int_{\overline{\gamma}}f^{*}(z)dz}.
\end{equation}
Since $J$ satisfies $(H_{J}^{0})$, according to equation
\eqref{kl}, there exists an integer $p$ such that:
\begin{equation}
\label{entierp} k(E)+k^{*}(E)=2 p \pi.
\end{equation}
We recall that the functions $\omega_{\pm}$ associated to $\kappa$
are defined by equation \eqref{omega}. We consider a path
$\widetilde{\gamma}_{g}$ such that:
\begin{itemize}
\item The path $\widetilde{\gamma}_{g}$ connects $\overline{m_{g}}$ to
$m_{g}$ and is symmetric with respect to the real axis.
\item The path $\widetilde{\gamma}_{g}$ does not contain any
branch point of $\kappa$ and any pole of $\omega_{\pm}$.
\end{itemize}
We fix a continuous determination $q_{g}$ of $\sqrt{k'_{E}}$ on
$\gamma_{g}$. According to relation \eqref{kl}, we have
$(k^{*})'=-k'$, which implies that there exists
$\sigma_{g}\in\{-1,1\}$ such that:
\begin{equation}
\label{symraccarrg} q_{g}^{*}=i\sigma_{g}q_{g}
\end{equation}
The functions $\psi_{\pm}(x,\varphi,E,m_{g})$ satisfy the
relation:
\begin{equation}
\label{symblochnorm}
\psi_{\pm}^{*}(x,\varphi,E,m_{g})=i\sigma_{g}e^{\pm \frac{2i p\pi
x}{\varepsilon}}e^{\int_{\widetilde{\gamma_{g}}}\omega_{\pm}^{g}}\psi_{\pm}^{*}(x,\varphi,E,m_{g}).
\end{equation}
Besides, equations \eqref{symint} and \eqref{symgap} lead to the
following relations:
\begin{equation}
\overline{\int_{\widetilde{\gamma_{g}}}\omega_{+}}=-\int_{\widetilde{\gamma_{g}}}\omega_{+}\
;\
\overline{\int_{\widetilde{\gamma_{g}}}\omega_{-}}=-\int_{\widetilde{\gamma_{g}}}\omega_{-}.
\end{equation}
According to $(H_{W,r})$, $W^{*}=W$. By using \eqref{noy}, we
compute:
$$ A(x,x',E)=\overline{A(x,x',\bar{E})}.$$
The operator $T_{E}$ satisfies:
$$T_{E}(f^{*})=[T_{E}(f)]^{*}.$$
Consequently, according to \eqref{eqres_b} and
\eqref{symblochnorm}, we obtain that, for $E$ in $\mathcal{V}$,
$x$ in $\R$ and $\varphi$ in $B_{\varepsilon}^{g}$,
\begin{equation}
(F_{-}^{g})^{*}(x,\varphi,E,\varepsilon)=\overline{F_{-}^{g}(x,\bar{\varphi},\bar{E},\varepsilon)}=i\sigma_{g}e^{-\frac{i}{\varepsilon}2p\pi
x}e^{\int_{\widetilde{\gamma_{g}}}\omega_{-}^{g}}F_{-}^{g}(x,\varphi,E,\varepsilon).
\end{equation}
This leads to:
$$(h_{-}^{g})^{*}=i\sigma_{g}e^{-\frac{i}{\varepsilon}2p\pi x}\frac{\alpha_{g}(E)}{\alpha_{g}^{*}(E)}h_{-}^{g},$$
where
\begin{equation}
\label{renormconstg}
\alpha_{g}(E)=e^{-\frac{i}{2\varepsilon}\left(\int_{\widetilde{\gamma}_{g}}(\kappa(u)-p\pi)du+p\pi(m_{g}+\overline{m_{g}})\right)}e^{\frac{1}{2}\int_{\widetilde{\gamma}_{g}}\omega_{-}^{g}}
\end{equation}
Similarly, we fix $m_{d}$ in $S_{Y}$ such that:
\begin{itemize}
\item The point $m_{d}$ is not a branch point of $\kappa$.
\item It satisfies $\I m_{d}>0$, $k'_{E}(m_{d})\neq 0$.
\item The domain $\{\varphi\in S_{Y}\ ;\ \Ra(\varphi-m_{d})>0\textrm{
and }\I(\varphi-m_{d})>0\}$ does not contain any branch point of
$\kappa$.
\end{itemize}
We define $E_{d}=E-W(m_{d})$ and we define the function
$h_{+}^{d}$ by:
\begin{equation}
\label{renormd}
h_{+}^{d}(x,\varphi,E,\varepsilon)=e^{\frac{i}{\varepsilon}\int_{m_{d}}^{+\infty}[\kappa(u)-k(E)]du+\frac{ik(E)m_{d}}{\varepsilon}}F_{+}^{d}(x,\varphi,E,\varepsilon)
\end{equation}
where the integral $\int_{m_{d}}^{+\infty}[\kappa(u)-k(E)]du$ is
taken
in the upper half plane.\\
We consider the path $\widetilde{\gamma}_{d}$ such that:
\begin{itemize}
\item The path $\widetilde{\gamma}_{d}$ connects $\overline{m_{d}}$ to
$m_{d}$ and is symmetric with respect to the real axis.
\item The path $\widetilde{\gamma}_{d}$ does not contain any
branch point of $\kappa$ and any pole of $\omega_{\pm}$.
\end{itemize}
We fix a continuous branch $q_{d}$ of $\sqrt{k'_{E}}$ on
$\gamma_{d}$. There exists a real number $\sigma_{d}$ such that:
\begin{equation}
\label{symraccarrd} q_{d}^{*}=i\sigma_{d}q_{d}
\end{equation}
The function $h_{+}^{d}$ satisfies:
$$(h_{+}^{d})^{*}=i\sigma_{d}e^{\frac{i}{\varepsilon}2p\pi x}\frac{\alpha_{d}(E)}{\alpha_{d}^{*}(E)}h_{+}^{d},$$
where
\begin{equation}
\label{renormconstd}
\alpha_{d}(E)=e^{\frac{i}{2\varepsilon}\left(\int_{\widetilde{\gamma}_{d}}(\kappa(u)-p\pi)du+p\pi(m_{d}+\overline{m_{d}})\right)}e^{\frac{1}{2}\int_{\widetilde{\gamma}_{d}}\omega_{+}^{d}}
\end{equation}
We define {\it the transmission coefficient}:
\begin{equation}
\label{transmcoeff}
d(\varphi,E,\varepsilon)=w(\alpha_{g}h_{-}^{g}(\cdot,\varphi,E,\varepsilon),\alpha_{d}h_{+}^{d}(\cdot,\varphi,E,\varepsilon))
\end{equation}
We immediately deduce from Proposition \ref{carvp} and Proposition
\ref{propconstinf} that the eigenvalues of
$H_{\varphi,\varepsilon}$ are characterized by:
\begin{equation}
d(\varphi,E,\varepsilon)=0
\end{equation}
\subsection{Some remarks}
\subsubsection{}
The assumption $(H_{W,r})$ is not optimal. Actually, it suffices
to assume that $W$ is analytic real in $S_{Y}$ and that there
exists a function $f\in L^{1}(\R)$ such that :
$$\forall x\in\R\quad\sup\limits_{y\in[-Y,Y]}|W(x+iy)|\leq f(x).$$
\subsubsection{} In equations \eqref{starg} and \eqref{stard}, we
could have included the numbers $i\sigma_{g}$ and $i\sigma_{d}$
into the functions $\alpha_{g}$ and $\alpha_{d}$, but we prefer
showing the relations between $q_{g}$ and $q_{g}^{*}$, $q_{d}$ and
$q_{d}^{*}$.
\subsubsection{} Note that this construction differs from the
constructions of canonical domains in \cite{FK1}. Indeed, the
domains on which we construct these functions depend on
$\varepsilon$. We shall extend these asymptotics on a fixed strip
in the neighborhood of the real line (section \ref{infwkb}).

\section{WKB Theorem on non compact domains}
\label{infwkb} In this section, we prove a continuation result on
non compact domains of $S_{Y}$. This result is a generalization on
non compact domains of the method developed in \cite{FK1} and
particularly of Lemma \ref{lemcontfin}.\\
We prove that the continuation of asymptotics stay valid on some
half-strips $\{\varphi\in S_{Y}\ ;\ |\Ra\varphi|>A\}$. To do that,
we cover these domains by a countable union of small local
overlapping canonical
domains, called $\delta$-chain (see section \ref{deltcha}).\\
This principle follows the recent developments and improvements of
the WKB method (see \cite{FK3}). The idea is to get over the local
notion of canonical domain in favor of maximal domains. These
domains, constructed as union of local canonical domains are some
domains on which a function keeps the standard behavior (see
\cite{FK3}).
\subsection{Continuation Theorem on non compact domains}
\subsubsection{The main result}
We shall prove the following result:
\begin{thm} Continuation Theorem on non compact domains.\\
\label{infcontle} Fix $\tilde{Y}\in]0,Y[$. Assume that $V$
satisfies $(H_{V})$, that $W$ satisfies $(H_{W,r})$ and that $J$
satisfies $(H_{J}^{0})$. Then, there exist a real
$\varepsilon_{0}>0$, a complex neighborhood $\mathcal{V}$ of $J$
and two real numbers $A_{g}$ and $A_{d}$ such that, if $f$ has the
following properties:
\begin{itemize}
\item The function $f(\cdot,\varphi,E,\varepsilon)$ is a
consistent solution of \eqref{eqp}.
\item The function $(\varphi,E)\mapsto
f(x,\varphi,E,\varepsilon)$ is analytic on $S_{\tilde{Y}}\times
\mathcal{V}$ for any $x\in[-X,X]$ and any
$\varepsilon\in]0,\varepsilon_{0}[$.
\end{itemize}
Then,
\begin{enumerate}
\item There exists $\kappa$ a continuous branch on
$\{\Ra\varphi<A_{g}\}$ such that $\I\kappa>0$. Moreover, for any
$C<B<A_{g}$, if the function $f$ satisfies the asymptotic behavior
\begin{equation}
\label{asprolac}
f(x,\varphi,E,\varepsilon)=e^{-\frac{i}{\varepsilon}\int^{\varphi}\kappa(u)du}(\psi_{-}(x,\varphi,E)+r_{C}(x,\varphi,E,\varepsilon))
\end{equation}
with $\lim\limits_{\varepsilon\rightarrow
0}\sup\limits_{[-X,X]\times
R_{(-\infty,C]}\times\mathcal{V}}\max\{|r_{C}(x,\varphi,E,\varepsilon)|,|\partial_{x}r_{C}(x,\varphi,E,\varepsilon)|\}=0$,\\
then, this behavior stays valid until $B$. Precisely:
\begin{equation}
\label{asprola}
f(x,\varphi,E,\varepsilon)=e^{-\frac{i}{\varepsilon}\int^{\varphi}\kappa(u)du}(\psi_{-}(x,\varphi,E)+r_{B}(x,\varphi,E,\varepsilon))
\end{equation}
with $\lim\limits_{\varepsilon\rightarrow
0}\sup\limits_{[-X,X]\times
R_{(-\infty,B]}\times\mathcal{V}}\max\{|r_{B}(x,\varphi,E,\varepsilon)|,|\partial_{x}r_{B}(x,\varphi,E,\varepsilon)|\}=0$.\\
\item There exists $\kappa$ a continuous branch on
$\{\Ra\varphi>A_{d}\}$ such that $\I\kappa>0$. Moreover, for any
$C>B>A_{d}$, if $f$ satisfies the asymptotic behavior
\begin{equation}
\label{asprolbc}
f(x,\varphi,E,\varepsilon)=e^{\frac{i}{\varepsilon}\int^{\varphi}\kappa(u)du}(\psi_{+}(x,\varphi,E)+r_{C}(x,\varphi,E,\varepsilon))
\end{equation}
with $\lim\limits_{\varepsilon\rightarrow
0}\sup\limits_{[-X,X]\times
R_{[C,+\infty)}\times\mathcal{V}}\max\{|r_{C}(x,\varphi,E,\varepsilon)|,|\partial_{x}r_{C}(x,\varphi,E,\varepsilon)|\}=0$,\\
then this behavior stays valid until $B$. Precisely:
\begin{equation}
\label{asprolb}
f(x,\varphi,E,\varepsilon)=e^{\frac{i}{\varepsilon}\int^{\varphi}\kappa(u)du}(\psi_{+}(x,\varphi,E)+r_{B}(x,\varphi,E,\varepsilon))
\end{equation}
with $\lim\limits_{\varepsilon\rightarrow
0}\sup\limits_{[-X,X]\times
R_{[B,+\infty)}\times\mathcal{V}}\max\{|r_{B}(x,\varphi,E,\varepsilon)|,|\partial_{x}r_{B}(x,\varphi,E,\varepsilon)|\}=0$.
\end{enumerate}
\end{thm}
Theorem \ref{infcontle} and Proposition \ref{propconstinf} clearly
imply Theorem \ref{jostthm}.
\subsubsection{Some remarks}
\label{hypW} We shall prove Theorem \ref{infcontle} as $W$
satisfies
the weaker assumptions:\\
{\bf (H1) $\mathbf{W}$ is an analytic real function in $\mathbf{S_{Y}}$}.\\
{\bf (H2) $\mathbf{\exists\ C>0,\quad \exists\ s>1\textrm{ such that }\forall\ z\in  S_{Y},\quad |W'(z)|\leq\frac{C}{1+|z|^{s}}}$}\\
{\bf (H3) $\mathbf{\exists\ f \in L^{1}(\R)\textrm{ such that }\forall x\in\R\quad \sup\limits_{y\in[-Y,Y]}|W(x+iu)|\leq f(x)}$}\\
The following lemma relates $(H_{W,r})$ and $(H1)$, $(H2)$ and
$(H3)$:
\begin{lem}
\label{hypfaib} Let $W$ satisfy $(H_{W,r})$ on $S_{Y}$. Fix
$\tilde{Y}\in]0,Y[$. Then $W$ satisfies $(H1)$, $(H2)$ and $(H3)$
on $S_{\tilde{Y}}$.
\end{lem}
\begin{dem}
Assume that $W$ satisfy $(H_{W,r})$ on $S_{Y}$. We prove that $W$
satisfies $(H2)$ on $S_{\tilde{Y}}$ by using the following lemma:
\begin{lem}
\label{der} Let $f$ be an analytic function on $S_{Y}$ such that $|f(z)|\leq\frac{C}{1+|z|^{s}}$, $C>0$.\\
Fix $\eta>0$. Then,
$$\forall p\in\N^{*}\quad \exists C_{p}>0/\quad \forall z\in S_{Y-\eta}\quad|f^{(p)}(z)|\leq\frac{C_{p}}{1+|z|^{s}}.$$
\end{lem}
\begin{dem}\\
This result is a consequence of the Cauchy formula. We do not give
the details.
\end{dem}\\
- Clearly, $W$ satisfies $(H1)$ on $S_{Y}$.\\
- $W$ satisfies $(H3)$ with $f(x)=\frac{C}{1+|x|^{s}}$.\\
This completes the proof of Lemma \ref{hypfaib}.
\end{dem}
\subsubsection{}
Let us briefly outline the ideas of the proof. We shall
concentrate on $B_{g}=\{\varphi\in S_{Y};\ \Ra(\varphi)<A_{g}\}$.
There are three steps.\\
First we cover $B_{g}$ with an union of overlapping local compact
canonical domains $K_{m}$.\\
In each canonical domain $K_{m}$, we can construct a consistent
local basis thanks to Theorem \ref{finwkbthm}. To compute the
connection between the consistent bases of $K_{m}$ and $K_{m+n}$,
it suffices to do the product of the $n$ transfer matrices between
the canonical bases of two successive domains. The accuracy of the
rest cannot be better than the sum of the accuracies obtained on
each domain. Theorem \ref{finwkbthm} gives an estimate in $o(1)$;
this accuracy is insufficient when $n$ goes to
infinity.\\
A refinement of the calculation of asymptotics in Theorem
\ref{finwkbthm} is therefore necessary. We prove it by using the
integrability of $W$.
\subsubsection{Branch points}
The following result specifies the location of the branch points
of $\kappa$. We recall that $\Upsilon(E)$  is defined in
\eqref{nupsilon}.
\begin{lem}
Let $\mathcal{V}$ be a complex neighborhood of the interval $J$.
Assume that $W$ satisfies
$$\lim\limits_{x\rightarrow
+\infty}\sup\limits_{y\in[-Y,Y]}|W(x+iy)|=0,$$ then:
\begin{equation*}
\label{bplem} \exists A>0\textrm{ such that }\forall\
E\in\mathcal{V},\quad\varphi\in\Upsilon(E)\cap
S_{Y}\Rightarrow|\Ra(\varphi)|<A.
\end{equation*}
\end{lem}
\begin{dem}
Since $\overline{\mathcal{V}}\cap\partial\sigma(H_{0})=\emptyset$,
there exists $\alpha>0$ such that: $$\forall\
E\in\mathcal{V},\quad\forall\
p\in\N^{*},\quad|E-E_{p}|\geq\alpha.$$ If $\varphi_{p}(E)$
satisfies $E-W(\varphi_{p}(E))=E_{p}$, we get:
$$\forall\
E\in\mathcal{V},\quad\forall\
p\in\N^{*},\quad|W(\varphi_{p}(E))|\geq\alpha.$$ Finally, $\{u\in
S_{Y}\ ;\ |W(u)|\geq\alpha\}$ is a subset of a compact of $S_{Y}$.
This completes the proof of Lemma \ref{bplem}.
\end{dem}
\subsubsection{Uniform asymptotics on a $\delta$-chain}
\label{deltcha} First, we introduce a new definition. We remind
that the width of a complex subset is defined in \eqref{larga}.
\begin{defn}$\delta$-$\textit{chain of strictly canonical domains}$\\
Fix $\widetilde{Y}\in]0,Y[$. Fix $E$. Let $D$ be a simply
connected domain of $S_{\widetilde{Y}}$ containing no branch
points of the complex momentum. We fix on $D$ a continuous branch
$\kappa$ of the complex momentum. Let $\{\tau_{n}\}_{n\in\N}$ be a
sequence of real numbers and $K$ be a compact of $S_{\widetilde{Y}}$.\\
$\{K+\tau_{n}\}_{n\in\N}$ is called a $\delta$-chain for $E$,
$\kappa$ and $D$ if it satisfies the following properties:
\begin{enumerate}
\item $\bigcup\limits_{n=0}^{\infty}(K+\tau_{n})=D.$
\item $\exists\tau>0\textrm{ such that }\forall n\in\N\quad
l((K+\tau_{n})\cap(K+\tau_{n+1}),\widetilde{Y})>\tau.$
\item The domain $K$ is an union of curves $\gamma$ such that, for
any $n$, $\gamma+\tau_{n}$ is a $\delta$-strictly canonical curve
for $\kappa$.
\end{enumerate}
\end{defn}
$K$ is called the fundamental domain of the $\delta$-chain. Now,
we have the intermediate result:
\begin{prop}
\label{infwkbprop} Assume that $V$ satisfies $(H_{V})$ and that
$W$ satisfies $(H_{1})$, $(H_{2})$ and $(H_{3})$. Fix
$\tilde{Y}\in]0,Y[$. Let $\mathcal{V}$ a complex neighborhood of
$J$ and $D\subset S_{\tilde{Y}}$ a domain with the following
properties:
\begin{itemize}
\item $\inf\limits_{p\in\N^{*},E\in\mathcal{V}}
\textrm{dist}\{D,\varphi_{p}(E)\}\geq C,$
\item there exists $\{\tau_{n}\}_{n\in\N}$ such that, for any $E\in\mathcal{V}$,
$\{K+\tau_{n}\}_{n\in\N}$ is a $\delta$-chain for $E$ and $D$.
\end{itemize}
Fix $\varphi_{0}\in D$.\\
Then, there exists $\varepsilon_{0}>0$ such that, for any
$n\in\N$, there exist two functions
$(x,\varphi,E,\varepsilon)\mapsto\psi_{\pm}^{n}(x,\varphi,E,\varepsilon)$
with the following properties:
\begin{itemize}
\item The functions $(x,\varphi,E,\varepsilon)\mapsto\psi_{\pm}^{n}(x,\varphi,E,\varepsilon)$
are defined on
$\R\times(K+\tau_{n})\times\mathcal{V}\times]0,\varepsilon_{0}[$
and form a consistent basis.
\item for any fixed $x\in\R,\
\varepsilon\in]0,\varepsilon_{0}[$, the functions
$(\varphi,E)\mapsto\psi_{\pm}^{n}(x,\varphi,E,\varepsilon)$ are
analytic on $(K+\tau_{n})\times\mathcal{V}$.
\item for $x\in[-X,X]$, $\varphi\in (K+\tau_{n})$ and
$E\in\mathcal{V}$, the functions $\psi_{\pm}^{n}$ have the
asymptotic behavior:
\begin{equation}
\label{asc}
\psi_{\pm}^{n}(x,\varphi,E,\varepsilon)=e^{\pm\frac{i}{\varepsilon}\int_{\varphi_{0}}^{\varphi}\kappa
du}\left(\psi_{\pm}(x,\varphi,E)+\frac{1}{1+|\tau_{n}|^{s}}o(1)\right).
\end{equation}
\item The asymptotics (\ref{asc}) are uniform in $x,\
\tau$, $\varphi\in K+\tau_{n}$ et $E\in\mathcal{V}$.
\item The asymptotics can be differentiated once in $x$.
\end{itemize}
\end{prop}
The proof of Proposition \ref{infwkbprop} mimics this of Theorem
1.1 in \cite{FK1}. We omit the details and we refer to \cite{FK1},
section 4 for an analogous statement.
\subsection{Construction of a $\delta$-chain of strictly canonical
domains} In this section, we shall construct a $\delta$-chain
under assumptions $(H_{1})$, $(H_{2})$ and $(H_{3})$.
\begin{prop}
\label{constdeltachain} Fix $\tilde{Y}\in]0,Y[$. Assume that $V$
satisfies $(H_{V})$, that $W$ satisfies $(H_{1})$, $(H_{2})$ and
$(H_{3})$ and that $J$ satisfies $(H_{J}^{0})$. Then, there exist
a complex neighborhood $\mathcal{V}$ of $J$, two real numbers
$(A_{g},A_{d})\in\R^{2}$, a domain $K\subset S_{\tilde{Y}}$ and
two real sequences $\{\tau_{n}^{1}\}_{n\in\N}$,
$\{\tau_{n}^{2}\}_{n\in\N}$ such that:
\begin{itemize}\item for any $E\in\mathcal{V}$, there exists a
continuous branch $\kappa$ on $\{\varphi\in S_{\tilde{Y}}\ ;\
\Ra\varphi\in(-\infty,A_{g}]\}$ (resp. on $\{\varphi\in
S_{\tilde{Y}}\ ;\ \Ra\varphi\in[A_{d},\infty)\}$),
\item for any $E\in\mathcal{V}$, $\{K+\tau_{n}^{1}\}_{n\in\N}$
(resp. $\{K+\tau_{n}^{2}\}_{n\in\N}$) is a $\delta$-chain for
$\kappa$, $E$ and $\{\Ra\varphi\in(-\infty,A_{g}]\}$ (resp.
$\{\Ra\varphi\in[A_{d},+\infty)$)).
\end{itemize}
\end{prop}
The rest of the section \ref{deltcha} is devoted to the proof of
Proposition \ref{constdeltachain}. This proof is based on
elementary geometrical arguments. We prove the construction for
$\Ra\varphi\in(-\infty,A_{g}]$.
\subsubsection{Construction of $\delta$-strictly canonical
straight-lines} We have defined the canonical lines in section
\ref{lc} and described them in terms of the vector $t(\varphi)$.\\
We set $\alpha=\frac{1}{2}\inf\limits_{E\in J}\I k(E)$ and
$m=2\sup\limits_{E\in J}|\Ra k(E)|$.\\
Since the mapping $(E,\varphi)\mapsto E-W(\varphi)$ is continuous
and since $W(\varphi)$ goes to zero when $\Ra\varphi$ goes to
infinity, there exist a complex neighborhood $\mathcal{V}$ of $J$
and a real number $A_{g}$ such that:
$$\forall E\in\mathcal{V},\quad\forall\varphi\in(-\infty,A_{g}],\quad \Ra k(E-W(\varphi))\in[-m,m],\quad \I k(E-W(\varphi))>\alpha$$
We set $B_{g}=(-\infty,A_{g}]+i[-\tilde{Y},\tilde{Y}]$. The
canonical curves for $\Ra\varphi$ in the neighborhood of $-\infty$
are described by:
\begin{lem}
\label{geom} There exists $\theta_{0}\in]0,\pi/2[$ such that, if
$\gamma$ is a smooth curve in $B_{g}$ satisfying:
\begin{equation}
\label{geoma} \forall\varphi\in\gamma,\quad
\arg[t(\varphi)]\in]\theta_{0},\pi/2-\theta_{0}[.
\end{equation}
then, $\gamma$ is a canonical line for $\kappa$.
\end{lem}
\begin{dem}
For $\arg(u)=\theta$ and
$\cot\theta\in]-\frac{m-\delta}{\alpha},\frac{\pi+m-\delta}{\alpha}[$,
we have:
$$\I(\overline{(\kappa-\delta)}u)>0\quad\textrm{et}\quad\I(\overline{(\pi-\kappa+\delta)}u)>0$$
Consequently, $\cot\theta_{0}=\frac{m-\delta}{\alpha}$ implies
that (\ref{geoma}) is satisfied.
\end{dem}

\subsubsection{The fundamental domain $K$}
Let $\xi_{1}=-i\tilde{Y}$ and $\xi_{2}=i\tilde{Y}$. We denote by
$K$ the lozenge bounded by the straight lines containing $\xi_{1}$
and $\xi_{2}$ whose guiding vectors have the affixes
$e^{i\theta_{0}}$ and $e^{i(\pi-\theta_{0})}$.\\
We set $[-u_{0},u_{0}]=K\cap\{y=0\}$. $K$ is shown in figure
\ref{element}. Fix $x$ such that $K+x\subset B_{g}$; we shall show
that $K+x$ is a $\delta$-strictly canonical domain. According to
Lemma \ref{geom}, it suffices to write $K$ as an union of smooth
curves satisfying (\ref{geoma}).\\
For any $u\in K$, we consider a vertical segment
$[\overline{\xi},\xi]$ containing $u$ and included in $K$ (see
figure \ref{element}). The broken line
$[\xi_{1},\overline{\xi}]\cup[\overline{\xi},\xi]\cup[\xi,\xi_{2}]$
satisfies (\ref{geoma}). The relation (\ref{geoma}) is stable
under small $C^{1}$-perturbation; we slightly deform the line
$[\xi_{1},\overline{\xi}]\cup[\overline{\xi},\xi]\cup[\xi,\xi_{2}]$
to get a smooth curve which satisfies (\ref{geoma}).\\
Consequently, $K$ satisfies the following properties:\\
- $K\cap S_{\tilde{Y}}$ contains a rectangle of width $4\eta>0$.\\
- $l((K-n\eta)\cap(K-(n+1)\eta),\tilde{Y})>\eta$.\\
- $K$ is the union of curves $\gamma$ such that $\gamma-n\eta$ is
$\delta$-strictly canonical for any sufficiently large $n$.
\subsubsection{Conclusion}
To finish the proof, it suffices to adapt the proof of Lemma
\ref{lemcontfin} in section 5.9 of \cite{FK1}, by using
Proposition \ref{constdeltachain} and Proposition
\ref{infwkbprop}. The convergence of the series of general term
$\frac{1}{1+|\tau_{n}|^{s}}$ replaces the compactness. We do not
give the details.
\psset{unit=1em,linewidth=.05}

\psset{unit=1em,linewidth=.05}
\begin{center}
\begin{figure}
\begin{pspicture}(-20,-10)(10,10)
\psline(0,-5)(-3,0)\psline(0,5)(3,0)\psline(0,5)(-3,0)\psline(0,-5)(3,0)\psline(0,-5)(-1,-2)\psline(-1,-2)(-1,2)\psline(-1,2)(0,5)
\psline(-8,4)(8,4)\psline(-8,-4)(8,-4) \psline(-8,0)(8,0)
\uput[180](8,5){$y=\tilde{Y}$}\uput[180](8,-5){$y=-\tilde{Y}$}
\uput[180](-3,-1){$-u_{0}$}\uput[180](5,-1){$u_{0}$}
\uput[180](0.7,-1){$\gamma_{u}$}\uput[180](1,6){$\xi_{2}$}\uput[180](1,-6){$\xi_{1}$}
\end{pspicture}
\caption{The fundamental domain $K$}\label{element}
\end{figure}
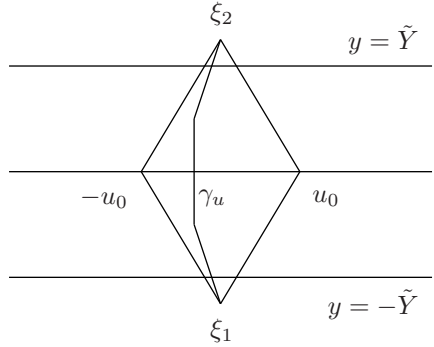
\end{center}
\section{Transmission coefficient. Equation for eigenvalues}
\label{calcmattransf}\label{anares} In Theorem \ref{jostthm}, we
have constructed two functions $h_{-}^{g}$ and $h_{+}^{d}$. We
have defined the transmission coefficient
$d(E,\varphi,\varepsilon)$. We choose $m_{g}=-0+i0$ and
$m_{d}=0+i0$.\\
 In Proposition \ref{bcki}, we have
introduced a consistent basis $(f_{i},f_{i}^{*})$ near the cross.
To compute $d(E,\varphi,\varepsilon)$, we shall project the
functions
$h_{-}^{g}$ and $h_{+}^{d}$ onto the basis $(f_{i},f_{i}^{*})$.\\
\subsection{Preliminaries}
\subsubsection{Introduction. Notations}
Fix $\widetilde{Y}<Y$ and $E_{0}\in J$. We have described in
section \ref{wkbconst} the complex momentum $\kappa$ and the
related geometric objects. We recall that we consider the case
\eqref{premcassc}. We use the notations introduced in section
\ref{wkbconst}. The branch points are called $\varphi_{r}^{\pm}$
and $\varphi_{i},\ \overline{\varphi_{i}}$. We
have described the Stokes lines in section \ref{stline}.\\
We have described in sections \ref{compmom} and \ref{poss} the
different branches $\kappa_{i}$, $\kappa_{g}$ and $\kappa_{d}$.
The branch $\kappa_{g}$, resp. $\kappa_{d}$, is defined and
continuous on the domain $\{\varphi\in S_{Y};\
\Ra\varphi<\varphi_{r}^{-}\}$, resp.
$\{\Ra\varphi>\varphi_{r}^{+}\}$. The branch $\kappa_{i}$ is
defined and continuous on a neighborhood of the cross. The domain
$(E-W)\left(\{\varphi\in S_{Y};\
\Ra\varphi<\varphi_{r}^{-}\}\right)$ is a simply connected domain
which intersects with real axis in only one gap. Thus, we can fix
a determination $k_{g}$ of the quasi-momentum such that:
$$k_{g}(E-W(\varphi))=\kappa_{g}(\varphi).$$
Similarly, we fix the branches $k_{i}$ and $k_{d}$ of the
quasi-momentum such that :
$$ k_{i}(E-W(\varphi))=\kappa_{i}(\varphi),\quad
k_{d}(E-W(\varphi))=\kappa_{d}(\varphi).$$ Finally, we set:
$$ q_{i}(\varphi)=\sqrt{k'_{i}(E-W(\varphi))},\quad q_{g}(\varphi)=\sqrt{k'_{g}(E-W(\varphi))}\quad q_{d}(\varphi)=\sqrt{k'_{d}(E-W(\varphi))}.$$
Let $\varphi_{g}\in\R$ such that $\varphi_{g}<\varphi_{r}^{-}$ and
such that the interval $[\varphi_{g},\varphi_{r}^{-}]$ does not
contain any pole of $\omega_{\pm}$. We define the path
$\gamma_{g}$ in the complex plane by:
$$\gamma_{g}=[-0+i0,\varphi_{g}+i0]\cup[\varphi_{g}-i0,-0-i0].$$
Similarly, fix $\varphi_{d}\in\R$ such that
$\varphi_{d}>\varphi_{r}^{+}$ and such that the interval
$[\varphi_{r}^{+},\varphi_{d}]$ does not contain any pole of
$\omega_{\pm}$. We define the path $\gamma_{d}$ in the complex
plane by:
$$\gamma_{d}=[0+i0,\varphi_{d}+i0]\cup[\varphi_{d}-i0,0-i0].$$
In the following section, we explain the choice of the
determinations $q_{i}$, $q_{g}$ and $q_{d}$.
\subsubsection{The determination $q$}
\label{deco} We recall that there exists a real number
$\sigma_{i}\in\{-1,1\}$ such that:
\begin{equation}
\label{sigmareli} \frac{q_{i}^{*}}{q_{i}}=\sigma_{i}.
\end{equation}
We refer to section \ref{precsigma}.\\
The Wronskian satisfies $w(f_{i},(f_{i})^{*})=\sigma_{i}(w_{0}k_{i}')(E-W(0)).$\\
The number $\sigma_{i}$ depends on the sign of $k'$ along the band
$B$:
\begin{itemize}
\item If the band $B$ can be written $[E_{4p+1},E_{4p+2}]$, then $k'>0$
on $B$ and $\sigma_{i}=1$.
\item If the band $B$ can be written $[E_{4p+3},E_{4p+4}]$, then $k'<0$
on $B$ and $\sigma_{i}=-1$.
\end{itemize}
We fix the branch $q_{g}$ such that $q_{g}=q_{i}$ in $S_{-}$ and
such that $q_{g}$ is analytically continued in $\{\varphi\in
S_{Y};\ \Ra\varphi<\varphi_{r}^{-}\}$. According to relation
\eqref{sigmareli}, the branch $q_{g}$ satisfies:
\begin{equation}
\label{sigmarelg} q_{g}^{*}=i\sigma_{i}q_{g}
\end{equation}
Similarly, we fix $q_{d}$ such that $q_{d}=q_{i}$ in
$\overline{S_{+}}$ and such that $q_{d}$ is analytically continued
in $\{\varphi\in S_{Y};\ \Ra\varphi>\varphi_{r}^{+}\}$. The branch
$q_{d}$ satisfies:
\begin{equation}
\label{sigmareld} q_{d}^{*}=i\sigma_{i}q_{d}
\end{equation}
According to equations \eqref{sigmarelg} and \eqref{sigmareld}, we
have also:
$$\sigma_{g}=\sigma_{i}\quad;\quad\sigma_{d}=\sigma_{i}.$$
We denote by $\widetilde{\Psi_{\pm}}^{g}(x,\E)$,
$\widetilde{\Psi_{\pm}}^{i}(x,\E)$ and
$\widetilde{\Psi_{\pm}}^{d}(x,\E)$ the Bloch solutions described
in section \ref{bloch}. We set: {\small
$$\Psi_{\pm}^{i}(x,\varphi,E)=\widetilde{\Psi_{\pm}}^{i}(x,E-W(\varphi))\
;\
\Psi_{\pm}^{g}(x,\varphi,E)=\widetilde{\Psi_{\pm}}^{g}(x,E-W(\varphi))\
;\
\Psi_{\pm}^{d}(x,\varphi,E)=\widetilde{\Psi_{\pm}}^{d}(x,E-W(\varphi)).$$}
We define the functions $\omega_{\pm}^{g}$, $\omega_{\pm}^{i}$ and
$\omega_{\pm}^{d}$ associated by \eqref{omega} to the branches
$k_{g}$, $k_{i}$ and $k_{d}$.
\subsubsection{Ideas of the method}
The computation is similar to this done in \cite{FK2, FK1, FK4}.
It is based on some elementary principles that we outline now.
\begin{enumerate}
\item Periodicity.\\
The consistency condition \eqref{coh} implies that the Wronskians
are $\varepsilon$-periodic in $\varphi$. To get a total control of
the Wronskians in a horizontal strip, we only need to control them
in some vertical sub-strip of width $\varepsilon$.
\item Analyticity.\\
Since the functions $(\varphi,E)\mapsto
f_{-}^{g}(x,\varphi,E,\varepsilon)$, $(\varphi,E)\mapsto
f_{+}^{d}(x,\varphi,E,\varepsilon)$, $(\varphi,E)\mapsto
f_{\pm}^{i}(x,\varphi,E,\varepsilon)$ are analytic on
$S_{\tilde{Y}}\times \mathcal{U}$, their Wronskians are analytic
in $(\varphi,E)\in S_{\tilde{Y}}\times \mathcal{U}$. This allows
us to expand them into exponentially converging series. \\
Let $w(\varphi,E,\varepsilon)$ be an analytic function in
$(\varphi,E)$ which is $\varepsilon$-periodic in $\varphi$. We
set:
$$
w(\varphi,E,\varepsilon)=\sum\limits_{k\in\Z}w_{k}(E,\varepsilon)e^{\frac{2i\pi\varphi}{\varepsilon}}$$
The Cauchy formula gives an estimate of the Fourier coefficients:
\begin{equation}
\label{fourgen}
w_{k}(E,\varepsilon)=\frac{1}{\varepsilon}\int_{\varphi_{0}}^{\varphi_{0}+\varepsilon}w(\varphi,E,\varepsilon)e^{-\frac{2i
k\pi\varphi}{\varepsilon}}d\varphi,\quad\forall
k\in\N,\quad\forall\varphi_{0}\in S_{\tilde{Y}}.
\end{equation}
By moving $\I\varphi_{0}$ in $[-\tilde{Y},\tilde{Y}]$, we get a
control of positive and negative coefficients.
\end{enumerate}
\subsection{Asymptotic expansion of $d(\varphi,E,\varepsilon)$}
In this section, we shall establish the following result.
\begin{prop}
\label{colina} For any $E_{0}$ in $J$, there exist a complex
neighborhood $\mathcal{U}_{0}$ of $E_{0}$ and two functions
$(\varphi,E,\varepsilon)\mapsto b_{g}^{-}(\varphi,E,\varepsilon)$
and $(\varphi,E,\varepsilon)\mapsto
b_{d}^{+}(\varphi,E,\varepsilon)$ such that:
\begin{itemize}
\item The coefficient $d$ defined in \eqref{transmcoeff} can be written:
\begin{equation}
d(\varphi,E,\varepsilon)=i\sigma_{i}w(f_{i},(f_{i})^{*})[b_{g}^{-}(b_{d}^{+})^{*}-(b_{g}^{-})^{*}b_{d}^{+}].
\end{equation}
\item The functions $(\varphi,E)\mapsto b_{g}^{-}(\varphi,E,\varepsilon)$ and $(\varphi,E)\mapsto
b_{d}^{+}(\varphi,E,\varepsilon)$ are analytic on $S_{Y}\times
\mathcal{U}_{0}$.
\item The functions $\varphi\mapsto b_{g}^{-}(\varphi,E,\varepsilon)$ and $\varphi\mapsto
b_{d}^{+}(\varphi,E,\varepsilon)$ are $\varepsilon$-periodic and
admit the following Fourier asymptotic expansion, when
$\varepsilon\rightarrow 0$:
\begin{equation}
\label{bscoeffaaa}
b_{g}^{-}(\varphi,E,\varepsilon)=\sum\limits_{k\in\Z}(b_{g}^{-})_{k}(E,\varepsilon)e^{\frac{2ik\pi\varphi}{\varepsilon}},
\end{equation}
with
\begin{equation}
\label{bscoeffaa}
(b_{g}^{-})_{0}(E,\varepsilon)=\sigma_{i}e^{-\frac{i}{\varepsilon}\int_{0}^{\varphi_{r}^{-}}\kappa_{i}}e^{\frac{1}{2}\int_{0}^{\varphi_{r}^{-}}(\omega_{+}^{i}-\omega_{-}^{i})}[1+o(1)],
\end{equation}
and \begin{equation} \label{bscoeffbb}\forall k \neq 0,\quad
|(b_{g}^{-})_{k}(E,\varepsilon)|<C
e^{-\alpha/\varepsilon}e^{\frac{-2|k|\pi Y_{0}}{\varepsilon}},
\end{equation}
\begin{equation}
\label{bscoeffbbb}
b_{d}^{+}(\varphi,E,\varepsilon)=\sum\limits_{k\in\Z}(b_{d}^{+})_{k}(E,\varepsilon)e^{\frac{2ik\pi\varphi}{\varepsilon}},
\end{equation}
with
\begin{equation} \label{bscoeffcc}(b_{d}^{+})_{0}(E,\varepsilon)=i\sigma_{i}e^{\frac{i}{\varepsilon}\int_{0}^{\varphi_{r}^{+}}\kappa_{i}}e^{\frac{1}{2}\int_{0}^{\varphi_{r}^{+}}(\omega_{+}^{i}-\omega_{-}^{i})}[1+o(1)],\end{equation}
\begin{equation} \label{bscoeffdd}(b_{d}^{+})_{1}(E,\varepsilon)=-i\sigma_{i}e^{\frac{i}{\varepsilon}\int_{\varphi_{r}^{+}}^{0}\kappa_{i}}e^{\frac{2i}{\varepsilon}\int_{0}^{\overline{\varphi_{i}}}(\kappa_{i}-\pi)}e^{\frac{1}{2}\int_{\varphi_{r}^{+}}^{0}\omega_{+}^{i}-\omega_{-}^{i}}e^{\int_{0}^{\overline{\varphi_{i}}}(\omega_{+}^{i}-\omega_{-}^{i})}[1+o(1)],\end{equation} et
\begin{equation} \label{bscoeffee}\forall k>1,\quad |(b_{d}^{+})_{k}(\varphi,E,\varepsilon)|<C |(b_{d}^{+})_{1}(E,\varepsilon)|e^{-\alpha/\varepsilon}e^{\frac{-2|k-1|\pi Y_{0}}{\varepsilon}},\end{equation}
\begin{equation} \label{bscoeffff}\forall k<0,\quad |(b_{d}^{+})_{k}(\varphi,E,\varepsilon)|<C e^{-\alpha/\varepsilon}e^{\frac{-2|k|\pi Y_{0}}{\varepsilon}},\end{equation}
\end{itemize}
\end{prop}
The rest of the section is devoted to the proof of Proposition
\ref{colina}.\\
Fix $E_{0}\in J$. According to the choice of $\kappa_{g}$ and
$\kappa_{d}$ (sections \ref{compmom} and \ref{poss}), there exist
two analytic functions $\alpha_{g}(E)$ and $\alpha_{d}(E)$ such
that:
$$(\alpha_{g}h_{-}^{g})^{*}=i\sigma_{g}\alpha_{g}h_{-}^{g},$$
$$(\alpha_{d}h_{+}^{d})^{*}=i\sigma_{d}\alpha_{d}h_{+}^{d}.$$
Now, we use the function $f_{i}$ constructed in Proposition
\ref{bcki}. There exists a neighborhood $\mathcal{U}_{0}$ of
$E_{0}$ such that we can write:
$$\alpha_{g}h_{-}^{g}=-i\sigma_{g}(b_{g}^{-})^{*}f_{i}+b_{g}^{-}(f_{i})^{*},$$
$$\alpha_{d}h_{+}^{d}=-i\sigma_{d}(b_{d}^{+})^{*}f_{i}+b_{d}^{+}(f_{i})^{*}.$$
The coefficients $\alpha_{g}$ and $\alpha_{d}$ are defined in
equations \eqref{renormconstg} and \eqref{renormconstd}. We
compute:
$$\int_{\gamma_{g}}\omega_{-}^{g}=\int_{0}^{\varphi_{r}^{-}}(\omega_{+}^{i}-\omega_{-}^{i}),$$
$$\int_{\gamma_{d}}\omega_{+}^{d}=\int_{0}^{\varphi_{r}^{+}}(\omega_{+}^{i}-\omega_{-}^{i}).$$
This leads to:
$$\alpha_{g}(E)=e^{-\frac{i}{\varepsilon}\int_{0}^{\varphi_{r}^{-}}\kappa_{i}}e^{\frac{1}{2}\int_{0}^{\varphi_{r}^{-}}(\omega_{+}^{i}-\omega_{-}^{i})},$$
$$\alpha_{d}(E)=e^{\frac{i}{\varepsilon}\int_{0}^{\varphi_{r}^{+}}\kappa_{i}}e^{\frac{1}{2}\int_{0}^{\varphi_{r}^{+}}(\omega_{+}^{i}-\omega_{-}^{i})}.$$
The coefficients $b_{g}^{-}$ and $b_{d}^{+}$ satisfy:
$$b_{g}^{-}=\alpha_{g}a_{g}^{-},\quad
b_{d}^{+}=\alpha_{d}a_{d}^{+},$$ where the coefficients
$a_{g}^{-}$ and $a_{d}^{+}$ are given by:
\begin{equation}
\label{asag}
a_{g}^{-}=\frac{w(f_{i},h_{-}^{g})}{w(f_{i},f_{i}^{*})},
\end{equation}
and:
\begin{equation}
\label{asad}
a_{d}^{+}=\frac{w(f_{i},h_{+}^{d})}{w(f_{i},f_{i}^{*})}.
\end{equation}
We compute, for $E\in\mathcal{U}_{0}$:
$$d(\varphi,E,\varepsilon)=w(\alpha_{g}h_{-}^{g},\alpha_{d}h_{+}^{d})$$
$$=\left[b_{g}^{-}(b_{d}^{+})^{*}-b_{d}^{+}(b_{g}^{-})^{*}\right]i\sigma_{i}w(f_{i},(f_{i})^{*}).$$
\subsubsection{Continuation diagram of $f_{i}$}
First, we describe the asymptotic behavior of the function $f_{i}$
in some domains of the complex plane.
\begin{lem}
\label{contdiag} We suppose that the assumptions of Proposition
\ref{bcki} are satisfied. Fix $\tilde{Y}<Y$. Fix
$\varphi_{g}<\varphi_{r}^{-}$ and $\varphi_{d}>\varphi_{r}^{+}$.
There exists $y_{0}\in]0,\I\varphi_{i}[$ such that the function
$f_{i}$ has the following asymptotic behavior:
\begin{itemize} \item For $\varphi\in\{\varphi\in S_{\tilde{Y}};\
\Ra\varphi\in[\varphi_{g},\varphi_{r}^{-}]\}$, $f_{i}$ has the
standard asymptotic behavior:
$$f_{i}=q_{g}e^{\frac{i}{\varepsilon}\int_{0}^{\varphi_{r}^{-}}\kappa_{i}}e^{\frac{i}{\varepsilon}\int_{\varphi_{r}^{-}}^{\varphi}\kappa_{g}}e^{\int_{0}^{\varphi_{r}^{-}}\omega_{+}^{i}}e^{\int_{\varphi_{r}^{-}}^{\varphi}\omega_{+}^{g}}\left(\Psi_{+}^{g}+o(1)\right).$$
\item For $\varphi\in\{\varphi\in S_{\tilde{Y}};\
\Ra\varphi\in[\varphi_{r}^{+},\varphi_{d}];\ \I\varphi>-y_{0}\}$,
$f_{i}$ has the standard asymptotic behavior:
$$f_{i}=iq_{d}e^{\frac{i}{\varepsilon}\int_{0}^{\varphi_{r}^{+}}\kappa_{i}}e^{-\frac{i}{\varepsilon}\int_{\varphi_{r}^{+}}^{\varphi}\kappa_{d}}e^{\int_{0}^{\varphi_{r}^{+}}\omega_{+}^{i}}e^{\int_{\varphi_{r}^{+}}^{\varphi}\omega_{-}^{d}}\left(\Psi_{-}^{d}+o(1)\right).$$
\item For $\varphi\in\{\varphi\in S_{\tilde{Y}};\
\Ra\varphi\in[\varphi_{r}^{+},\varphi_{d}];\ \I\varphi<-y_{0}\}$,
$f_{i}$ has the standard asymptotic behavior:
$$f_{i}=-iq_{d}e^{\frac{i}{\varepsilon}\int_{0}^{\varphi_{i}^{-}}\kappa_{i}}e^{\frac{i}{\varepsilon}\int_{\varphi_{i}^{-}}^{\varphi}(2\pi-\kappa_{d})}e^{\int_{0}^{\varphi_{i}^{-}}\omega_{+}^{i}}e^{\int_{\varphi_{i}^{-}}^{\varphi}\omega_{-}^{d}}\left(\Psi_{-}^{d}+o(1)\right).$$
\end{itemize}
\end{lem}
\begin{dem}
This lemma is similar to the continuation diagram presented in
section 6 of \cite{FK4}. Thus, we give only the main ideas of the
study  and refer to this paper for the details. The continuation
diagram is represented in figure \ref{contdiagfig}. In this
figure, the straight arrows indicate the use of continuation lemma
(Lemma \ref{lemcontfin}), the circular arrows the use of the
Stokes lemma (Lemma \ref{stoklemma}) and the hatched zones the use
of the Adjacent Canonical Domain Principle (Lemma \ref{adjdom}).
To complete the proof, it remains to explain the connections
between the different objects of the WKB method.
\begin{itemize}
\item According to the definitions given in section \ref{compmom},
the branches $\kappa_{i}$ and $\kappa_{g}$ are equal in $S_{-}$
and, for all $\varphi\in S_{-}$, we have:
 \begin{equation}
\label{bda}
\kappa_{i}(\varphi+0)=\kappa_{g}(\varphi-0),\quad\Psi_{\pm}^{i}(\varphi+0)=\Psi_{\pm}^{g}(\varphi-0),\quad
 \omega_{\pm}^{i}(\varphi+0)=\omega_{\pm}^{g}(\varphi-0).
 \end{equation}
 Besides, it remains to link $q_{i}$ and $q_{g}$. Section
 \ref{deco} implies:
 $$\forall\varphi\in S_{-},\quad q_{i}(\varphi)=q_{g}(\varphi).$$
 \item Similarly, we have, for all $\varphi\in \overline{S_{+}}$:
 \begin{equation}
\label{bdd} \kappa_{i}(\varphi-0)=-\kappa_{d}(\varphi+0),
\quad\Psi_{\pm}^{i}(x,\varphi-0)=\Psi_{\mp}^{d}(x,\varphi+0),
\end{equation}
\begin{equation*}
\omega_{\pm}^{i}(\varphi-0)=\omega_{\mp}^{d}(\varphi+0),\quad
q_{d}(\varphi+0)=-i\ q_{i}(\varphi-0)
\end{equation*}
\item We study finally the link between $\kappa_{i}$ and $\kappa_{d}$ along the Stokes line $\bar{c}$ beginning at $\overline{\varphi_{i}}$.
We consider the quasi-momenta $k_{d}$ and $k_{i}$ associated to
$\kappa_{d}$ and $\kappa_{i}$. Equation \eqref{kl} for $k_{d}$ and
$k_{i}$, on either side of $[E_{2},E_{3}]$, implies that
$\kappa_{d}$ and $\kappa_{i}$ satisfy the following relations, for
$\varphi\in c$,
\begin{equation}
\label{bdf}
\kappa_{d}(\varphi+0)=2\pi-\kappa_{i}(\varphi-0),\quad\Psi_{\pm}^{d}(x,\varphi+0)=\Psi_{\mp}^{i}(x,\varphi-0)
\end{equation}
\begin{equation*}
\omega_{\pm}^{d}(\varphi+0)=\omega_{\mp}^{i}(\varphi-0)\quad
q_{d}(\varphi+0)=iq_{i}(\varphi-0)
\end{equation*}
\end{itemize}
\end{dem}
\psset{unit=1em,linewidth=.05}

\psset{unit=0.5em,linewidth=.1}
\begin{center}
\begin{figure}
\begin{pspicture}(-30,-10)(30,10) \psline(-10,-8)(10,-8)
\psline(-10,8)(10,8)
\psline(-6.5,-0.2)(-6.5,0.2)\uput[180](-7.7,-1){$\varphi_{r}^{-}$}
\psline(3.5,-0.2)(3.5,0.2)\uput[180](7,-1){$\varphi_{r}^{+}$}
\psline[linewidth=0.01](-6.5,0)(3.5,0)
\psdots[dotstyle=*](2.5,5.5)\uput[180](2,4.5){$\varphi_{i}$}
\psdots[dotstyle=*](2.5,-5.5)\uput[180](2,-5.5){$\overline{\varphi_{i}}$}
\pscurve(3.5,0)(4.1,-1.8)(4.5,-2.5)
\pscurve(2.5,-5.5)(3.5,-3.5)(4.5,-2.5) \psline(4.5,-2.5)(10,-2.5)
\psline{->}(2.5,6.5)(0.5,6.5)\psline{->}(-6.5,3)(-8.5,3)\psline{->}(2.5,3)(4.5,3)\psline{->}(1.5,-7.5)(3.5,-7.5)\psline{->}(-6.5,-2)(-4.5,-2)\psline{->}(3.5,-2)(1.5,-2)\psline{->}(2.5,-4)(0.5,-4)
\pscurve{->}(3.5,-1)(2.5,0)(3.5,1)(4.5,0)(4,-1)\pscurve{->}(-6.5,-1)(-5.5,0)(-6.5,1)(-7.5,0)(-7,-1)
\pscurve{->}(2.5,6.2)(1.5,5.5)(2.5,4.5)(3.5,5.5)(3,6)\pscurve{->}(2.5,-4.5)(1.5,-5.5)(2.5,-6.5)(3.5,-5.5)(3,-5)
\psline{->}(5,-5)(7,-5)\psline{->}(6.5,-1.5)(8.5,-1.5)\psline{->}(5,6)(7,6)\psline{->}(-8,-4)(-10,-4)
\pscurve(2.5,5.5)(3.5,6.5)(4.8,8)
\pscurve(-6.5,0)(-7.5,-4)(-7.8,-8)
\pscustom[linestyle=none,fillstyle=vlines,hatchsep=.8,hatchangle=45]{\psline(2,5.5)(2.5,5.5)\pscurve(2.5,5.5)(3.5,6.5)(4.8,8)\psline(4.8,8)(4.3,8)}\pscurve(4.3,8)(3,6.5)(2,5.5)
\psline(2,5.5)(2.5,5.5)\psline(4.8,8)(4.3,8)
\pscustom[linestyle=none,fillstyle=vlines,hatchsep=.8,hatchangle=45]{\pscurve(-6.5,0)(-7.5,-4)(-7.8,-8)\psline(-6,0)(-6.5,0)}
\pscustom[linestyle=none,fillstyle=vlines,hatchsep=.8,hatchangle=45]{\psline(-7.8,-8)(-7.3,-8)\pscurve(-7.3,-8)(-7,-4)(-6,0)}
\pscurve(0,0)(-2,-0.2)(-4,-0.4)(-6,-0.8)(-7,-4)(-7.3,-8)
\pscurve(0,0)(0.25,-2.3)(2.5,-5.5)
\pscustom[linestyle=none,fillstyle=vlines,hatchsep=.8,hatchangle=45]{\pscurve(-6,-0.8)(-4,-0.4)(-2,-0.2)(0,0)\psline(0,0)(-5,0)}
\pscustom[linestyle=none,fillstyle=vlines,hatchsep=.8,hatchangle=45]{\pscurve(2.5,-5.5)(0.25,-2.3)(0,0)\psline(0,0)(3.5,0)}
\pscustom[linestyle=none,fillstyle=vlines,hatchsep=.8,hatchangle=45]{\pscurve(3.5,0)(4.1,-1.8)(4.5,-2.5)\pscurve(4.5,-2.5)(3.5,-3.5)(2.5,-5.5)}
\end{pspicture}
\caption{Continuation Diagram}\label{contdiagfig}
\end{figure}
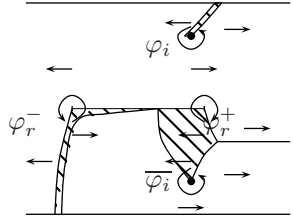
\end{center}
\subsubsection{Computation of $b_{g}^{-}$ and $b_{d}^{+}$}
Now, we compute the coefficients $b_{g}^{-}$ and $b_{d}^{+}$ given
by \eqref{asag} and \eqref{asad}.\\
According to Theorem \ref{infcontle}, we know that the asymptotic
behavior of the function $h_{-}^{g}$ remains valid in the domain
$\{\varphi\in S_{Y};\
\Ra(\varphi)\in[\varphi_{g},\varphi_{r}^{-}]\}$. Lemma
\ref{contdiag} gives the asymptotic behavior of $f_{i}$ in this
domain and we get :
\begin{equation} \forall \varphi\in S_{Y},\quad
a_{g}^{-}(\varphi,E,\varepsilon)=\sigma_{i}[1+o(1)].
\end{equation}
Fix $Y_{0}\in]0,Y[$. In the strip $S_{Y_{0}}$, we write:
\begin{equation}
a_{g}^{-}(\varphi,E,\varepsilon)=\sum\limits_{n\in
\Z}\alpha_{n}e^{\frac{2i\pi n\varphi}{\varepsilon}}
\end{equation}
The coefficients $\alpha_{n}$ satisfy:
\begin{equation}
\label{four}
\alpha_{n}=\frac{1}{\varepsilon}\int_{\varphi_{0}}^{\varphi_{0}+\varepsilon}a_{g}^{-}(\varphi,E,\varepsilon)e^{-2i\pi
n\frac{\varphi}{\varepsilon}}d\varphi,\quad \forall
n\in\N,\quad\forall\varphi_{0}\in\{-Y_{0}\leq\I\varphi\leq
Y_{0}\}.
\end{equation}
Fix $n>0$. We estimate $|\alpha_{n}|$. We use formula (\ref{four})
for $\I\varphi_{0}=-(Y-\delta) $, and we get:
$$ |\alpha_{n}|\leq C e^{-2\pi
n(Y-\delta)/\varepsilon}e^{\frac{C\delta}{\varepsilon}}.$$ We
treat similarly the case $n<0$ with $\I\varphi_{0}=(Y-\delta)$ and
we obtain:
$$ |\alpha_{n}|\leq C e^{2\pi
n(Y-\delta)/\varepsilon}e^{\frac{C\delta}{\varepsilon}}.$$
Besides, we have:
$$\alpha_{0}=\sigma_{i}[1+o(1)].$$
We fix $\delta<\frac{2\pi(Y-Y_{0})}{C+2\pi}$. For a constant $C$
such that $\alpha<2\pi(Y-Y_{0})-\delta(C+2\pi)$, we obtain the
estimates \eqref{bscoeffaa} and \eqref{bscoeffbb}.\\
The arguments for the coefficients $a_{d}^{+}$ and $b_{d}^{+}$ are
similar.
\subsection{Proof of Lemma \ref{phaseactint}}
\label{phaseactintdem} Now, we want to express the coefficient $d$
in a more understandable form. We begin with proving Lemma
\ref{phaseactint}. We recall that we denote by $\varphi_{r}^{\pm}$
and $\varphi_{i},\ \overline{\varphi_{i}}$ the branch points of
the complex momentum, and by $E_{r}$ and $E_{i}$ the related ends
of $\sigma(H_{0})$. We shall prove the lemma in the case
\eqref{premcassc}. Let $\kappa_{i}$ be the branch
described in section \ref{poss}. $\kappa_{i}$ satisfies \eqref{kappai}.\\
We shall prove Lemma \ref{phaseactint} for the branch
$\widetilde{\kappa_{i}}=\kappa_{i}$.
\begin{itemize}
\item First, we express $\Phi$, $S$ and $\Phi_{d}$ as integrals of
the complex momentum along complex paths. Let $\gamma$ be an
oriented curve, we call $\gamma^{\dagger}$ the curve oriented in
the opposite direction. Fix $\varphi_{d}\in\R$ and
$\varphi_{g}\in\R$ such that:
$$\varphi_{d}>\varphi_{r}^{+}\ ;\
\varphi_{g}<\varphi_{r}^{-}.$$ We define the complex paths
$\gamma_{\Phi}$, $\gamma_{S}$ and $\gamma_{g,d}$:
$$\gamma_{\Phi}=[\varphi_{r}^{-}+i0,\varphi_{r}^{+}+i0]\cup[\varphi_{r}^{+}-i0,\varphi_{r}^{-}-i0],$$
$$\gamma_{S}=(\sigma+0)\cup(\sigma^{\dagger}-0),$$
$$\gamma_{g,d}=[\varphi_{g}+i0,0+i0]\cup(\sigma_{+}-0)\cup(\sigma_{+}^{\dagger}+0)\cup[0+i0,\varphi_{d}+i0].$$
These paths are represented in figure \ref{uscoeff}. We have the
following result:
\begin{lem}
\label{contour} The coefficients $\Phi$, $\Phi_{d}$ and $S$ can be
written:
$$\Phi=\frac{1}{2}\oint_{\gamma_{\Phi}}\kappa(u)du,$$
$$S=\frac{1}{2i}\oint_{\gamma_{S}}\kappa(u)du,$$
$$\Phi_{d}=\frac{1}{2}\left(\int_{\gamma_{g,d}}(\kappa(u)-\pi)du+\int_{\overline{\gamma_{g,d}}}(\widetilde{\kappa}(u)-\pi)du\right)+\pi(\varphi_{g}-\varphi_{d}).$$
where $\kappa=\kappa_{i}$ in $S_{-}$ and $\kappa$ is analytically
continued along each path; $\widetilde{\kappa}=\kappa_{i}$ in
$\overline{S_{-}}$ and $\widetilde{\kappa}$ is analytically
continued along $\overline{\gamma_{g,d}}$.
\end{lem}
\begin{dem}
\begin{itemize}
\item First, let us justify the fact that integrals along $\gamma_{\Phi}$ and
$\gamma_{S}$ can be considered along closed curves. It suffices to
show that $\kappa$ can be analytically continued along
$\gamma_{\Phi}$ and $\gamma_{S}$.\\
We consider the curve $\gamma_{\Phi}$. We have taken the cut of
$\gamma_{\Phi}$ in $\varphi_{r}^{-}$. We show that $\kappa$ has
the same values on each side of the cut. $\kappa=\kappa_{i}$ on
$[\varphi_{r}^{-}+i0,\varphi_{r}^{+}+i0]$, since $\kappa$ is
continuous to the right of $\varphi_{r}^{+}$, we obtain that
$\kappa=-\kappa_{i}$ on $[\varphi_{r}^{-}-i0,\varphi_{r}^{+}-i0]$.
In addition,
$\kappa(\varphi_{r}^{-}+i0)=0=\kappa(\varphi_{r}^{-}-i0)$, which
proves that the integral can be taken on the closed curve $\gamma_{\Phi}$ .\\
The arguments for $\gamma_{S}$ are similar.
\item We compute:
$$\frac{1}{2}\oint_{\gamma_{\Phi}}\kappa(u)du=\int_{\varphi_{r}^{-}}^{\varphi_{r}^{+}}\kappa_{i}(u)du=\Phi(E).$$
Similarly, for the coefficient $S(E)$,
$$\frac{1}{2i}\oint_{\gamma_{S}}\kappa(u)du=\frac{1}{i}[\int_{\sigma_{+}}(\pi-\kappa_{i}(u))du+\int_{\sigma_{-}}(\pi-\kappa_{i}(u))du].$$
\item It remains to study $\Phi_{d}$. We introduce the branch
$\kappa_{i}$ and we cut $\gamma_{g,d}$ in elementary segments:
$$\int_{\gamma_{g,d}}(\kappa(u)-\pi)du+\int_{\overline{\gamma_{g,d}}}(\widetilde{\kappa}(u)-\pi)du$$
$$=2\int_{\varphi_{g}}^{0}(\kappa_{i}(u)-\pi)du+2\int_{\sigma_{+}}(\kappa_{i}(u)-\pi)du+\int_{\varphi_{d}}^{0}(\kappa_{i}(u)-\pi)du-2\int_{\sigma_{-}}(\kappa_{i}(u)-\pi)du$$
$$=2\int_{\varphi_{r}^{-}}^{0}(\kappa_{i}(u)-\pi)du+2\int_{\sigma_{+}}(\kappa_{i}(u)-\pi)du+\int_{\varphi_{r}^{+}}^{0}(\kappa_{i}(u)-\pi)du-2\int_{\sigma_{-}}(\kappa_{i}(u)-\pi)du-2\pi(\varphi_{g}-\varphi_{r}^{-})+2\pi(\varphi_{d}-\varphi_{r}^{+})$$
$$=2\Phi_{d}(E)+2\pi(\varphi_{d}-\varphi_{g})$$
\end{itemize}
This ends the proof of Lemma \ref{contour}.
\end{dem}
\item We use Lemma \ref{contour} to prove the analyticity of $\Phi$, $S$ and $\Phi_{d}$.\\
First, we consider $\Phi$. We can deform $\gamma_{\Phi}$ to a
closed curve going around $[\varphi_{r}^{-},\varphi_{r}^{+}]$ and
staying at a nonzero distance from this interval. Besides,
$\kappa$ is analytic in $E$ on the integration contour when $E$ is
close enough to $J$. The analysis of the coefficient $S$ is done
in the same way. To prove that $\Phi_{d}$ is analytic, we deform
the curves $\gamma_{g,d}$ and $\overline{\gamma_{g,d}}$ to stay at
a nonzero distance of the cross.
\item Fix $E\in J$. On the interval $[\varphi_{r}^{-},\varphi_{r}^{+}]$, the
branch $\kappa_{i}$ satisfies $\kappa_{i}\in[0,\pi]$. Thus, the
function $\Phi(E)$ is real positive on $J$.\\
Now, we give a simplified expression of $S$:
\begin{equation}
\label{exps}
S(E)=-i\left[\int_{\sigma_{+}}(\pi-\kappa_{i}(u))du+\int_{\sigma_{-}}(\pi-\kappa_{i}(u))du\right]=2\I\int_{\sigma_{+}}(\pi-\kappa_{i}(u))du
\end{equation}
On $\sigma$, the branch $\kappa_{i}$ satisfies
$\kappa_{i}\in[0,\pi]$. According to \eqref{action} and
\eqref{exps}, we obtain that $0< S(E)\leq
2\pi\I\varphi_{i}(E)$.\\
Finally, we have
$\int_{\sigma_{-}}(\kappa_{i}(u)-\pi)du=-\overline{\int_{\sigma_{+}}(\kappa_{i}(u)-\pi)du}$.
Consequently, the coefficient $\Phi_{d}(E)$ is real.
\item Now, we compute $S'$ and $\Phi'$ on $J$. Let $k$ be the
branch of the Bloch momentum continuous through $[E_{r},E_{i}]$,
then $\kappa(\varphi)=k(E-W(\varphi))$ and
$$\Phi'(E)=\int_{\varphi_{r}^{-}}^{\varphi_{r}^{+}}k'(E-W(u))du+k(E-W(\varphi_{r}^{+}))-k(E-W(\varphi_{r}^{-}))=\int_{\varphi_{r}^{-}}^{\varphi_{r}^{+}}k'(E-W(u))du.$$
We recall that $k$ has some branch points of square root type at
the ends of spectral bands (see section \ref{qm2}); consequently,
the integral
$\int_{\varphi_{r}^{-}}^{\varphi_{r}^{+}}k'(E-W(u))du$ is
convergent. In the interval $[E_{r},E_{i}]$, $k'(\E)>0$ and
$(E_{i}-E_{r})\Phi'$ takes positive values on $J$.\\
The analysis of $S'$ is similar.
\item We complete this section with the following formulas:
\begin{equation}
\label{phida}
\Phi_{d}(E)+iS(E)=\int_{\varphi_{r}^{-}}^{0}\kappa_{i}(u)du-2\int_{\sigma_{-}}(\kappa_{i}-\pi)(u)du+\int_{\varphi_{r}^{+}}^{0}\kappa_{i}(u)du
\end{equation}
\begin{equation}
\label{phidb}
-\Phi_{d}(E)+iS(E)=-\int_{\varphi_{r}^{-}}^{0}\kappa_{i}(u)du-2\int_{\sigma_{+}}(\kappa_{i}-\pi)(u)du-\int_{\varphi_{r}^{+}}^{0}\kappa_{i}(u)du
\end{equation}
When $\kappa(\varphi_{r}^{-})=\pi$, the proof is analogous for the
branch $\widetilde{\kappa_{i}}=2\pi-\kappa_{i}$.\\
\end{itemize}
\psset{unit=1em,linewidth=.05}

\psset{unit=0.5em,linewidth=.1}
\begin{center}
\begin{figure}
\begin{pspicture}(-20,-10)(35,10)
\psline[linewidth=0.01](-10,0)(10,0)
\psline[linewidth=0.01](-10,8)(10,8)
\psline[linewidth=0.01](-10,-8)(10,-8)
\psline[linewidth=0.05](-6.5,-0.2)(-6.5,0.2)\uput[180](-6.5,-1.2){$\varphi_{r}^{-}$}
\psline[linewidth=0.05](3.5,-0.2)(3.5,0.2)\uput[180](3.5,-1.2){$\varphi_{r}^{+}$}
\psline[linewidth=0.05](-6.5,0)(3.5,0)
\pscurve[linewidth=0.05](0,0)(0.25,2.3)(2.5,5.5)\psdots[dotstyle=*](2.5,5.5)\uput[180](3.7,5.5){$\varphi_{i}$}
\pscurve[linewidth=0.05](0,0)(0.25,-2.3)(2.5,-5.5)\psdots[dotstyle=*](2.5,-5.5)\uput[180](3.7,-5.5){$\overline{\varphi_{i}}$}
\pscurve{->}(-1,0)(-0.75,-2.3)(1.5,-5.5)(2.5,-6.5)(3.5,-5.5)(1.6,-2.3)(1,0)(1.6,2.3)(3.5,5.5)(2.5,6.5)(1.5,5.5)(-0.75,2.3)(-1,0)
\uput[180](3.5,7.3){$\gamma_{S}$}
\pscurve[linewidth=0.05]{->}(-1.5,-0.8)(-7,0)(-1.5,0.8)(4,0)(-1.5,-0.8)(-7,0)(-1.5,0.8)
\pscurve[linewidth=0.05](-1.5,0.8)(4,0)(-1.5,-0.8)(-7,0)(-1.5,0.8)(4,0)(-1.5,-0.8)
\uput[180](6,0.2){$\gamma_{\Phi}$}
\psline[linewidth=0.01](15,0)(35,0)
\psline[linewidth=0.01](15,8)(35,8)
\psline[linewidth=0.01](15,-8)(35,-8)
\psdots[dotstyle=*](18.5,0)\uput[180](18.5,-1.2){$\varphi_{r}^{-}$}
\psdots[dotstyle=*](28.5,0)\uput[180](28.5,-1.2){$\varphi_{r}^{+}$}
\psline[linewidth=0.05](18.5,0)(28.5,0)
\pscurve[linewidth=0.05](25,0)(25.25,2.3)(27.5,5.5)\psdots[dotstyle=*](27.5,5.5)\uput[180](28.7,5.5){$\varphi_{i}$}
\pscurve[linewidth=0.05](25,0)(25.25,-2.3)(27.5,-5.5)\psdots[dotstyle=*](27.5,-5.5)\uput[180](28.7,-5.5){$\overline{\varphi_{i}}$}
\pscurve(35,0)(30,-0.3)(25.5,-0.5)\pscurve(25.5,-0.5)(27.5,-6.5)(24.5,-0.5)\pscurve{->}(12,-0)(17,-0.3)(24.5,-0.5)
\pscurve(35,0)(30,0.3)(25.5,0.5)\pscurve(25.5,0.5)(27.5,6.5)(24.5,0.5)\pscurve{->}(12,0)(17,0.3)(24.5,0.5)
\uput[180](35,1){$\varphi_{d}$}\uput[180](15,1){$\varphi_{g}$}\psdots[dotstyle=*](35,0)\psdots[dotstyle=*](12,0)
\uput[180](24,3){$\gamma_{g,d}$}\uput[180](24,-3){$\overline{\gamma_{g,d}}$}
\end{pspicture}
\caption{Some complex paths}\label{uscoeff}
\end{figure}
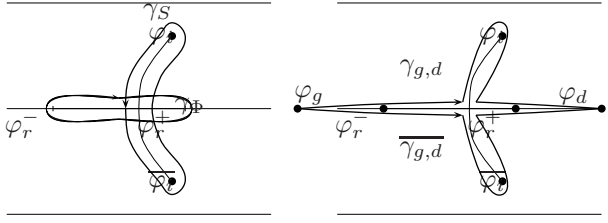
\end{center}
\subsubsection{Further computations}
\label{comp} We recall that the functions $\omega_{+}^{i}$ and
$\omega_{-}^{i}$ are defined in \eqref{omega}. We consider the
integrals of $\omega_{+}^{i}$ and $\omega_{-}^{i}$ along some
paths of the complex plane. We have the following relations:
\begin{lem}
\label{coeffut} The integrals of $\omega_{+}^{i}$ and
$\omega_{-}^{i}$ satisfy:
\begin{equation}
\label{omegaa} \forall E\in
J,\quad\int_{[\varphi_{r}^{-},\varphi_{r}^{+}]}\omega_{+}^{i}(u,E)du=0,\quad\int_{[\varphi_{r}^{-},\varphi_{r}^{+}]}\omega_{-}^{i}(u,E)du=0
\end{equation}
\begin{equation}
\label{omegab} \forall E\in
J,\quad\int_{\sigma}\omega_{+}^{i}(u,E)du=0,\quad\int_{\sigma}\omega_{-}^{i}(u,E)du=0\end{equation}
There exists a real number $\rho$ such that:
\begin{equation}
\label{omegac} \forall E\in
J,\quad\int_{[\varphi_{r}^{+},0]\cup\sigma_{+}}(\omega_{+}^{i}(u,E)-\omega_{-}^{i}(u,E))du-\int_{\sigma_{-}\cup[0,\varphi_{r}^{-}]}(\omega_{+}^{i}(u,E)-\omega_{-}^{i}(u,E))du=i\rho
\end{equation}
\end{lem}
\begin{dem}
We consider the case \eqref{premcassc}.
\begin{itemize}
\item We first prove \eqref{omegaa}. According to \eqref{omega},
we compute:
$$\int_{[\varphi_{r}^{-},\varphi_{r}^{+}]}\omega_{+}^{i}(u,E)du=-\int_{[\varphi_{r}^{-},\varphi_{r}^{+}]}g_{+}^{i}(E-W(u))W'(u)du=\int_{E-W([\varphi_{r}^{-},\varphi_{r}^{+}])}g_{+}^{i}(e)de=0$$
Indeed, for $E\in J$, the subset
$E-W([\varphi_{r}^{-},\varphi_{r}^{+}])$ is a complex path of
energies connecting $E_{r}$ to $E_{r}$ and containing
$(E-W(0))\in]E_{1},E_{2}[)$. We have shown this path in figure
\ref{imagchem}A. Particularly,
$E-W([\varphi_{r}^{-},\varphi_{r}^{+}])$ is a closed path and does
not surround any pole of the meromorphic function $g_{+}^{i}$.
Consequently, the integral is zero. We prove similarly that
$$\int_{[\varphi_{r}^{-},\varphi_{r}^{+}]}\omega_{-}^{i}(u,E)du=0.$$
\item We consider now \eqref{omegab}. We write:
$$\int_{\sigma}\omega_{+}^{i}(u,E)du=-\int_{E-W(\sigma)}g_{+}^{i}(e)de$$
The image of the path $\sigma$ is shown in figure \ref{imagchem}B.
We deal with $\omega_{-}^{i}$ similarly.
\item Finally, we compute:
$$\int_{[\varphi_{r}^{+},0]\cup\sigma_{+}}(\omega_{+}^{i}(u,E)-\omega_{-}^{i}(u,E))du-\int_{\sigma_{-}\cup[0,\varphi_{r}^{-}]}(\omega_{+}^{i}(u,E)-\omega_{-}^{i}(u,E))du$$ $$=\int_{E-W([\varphi_{r}^{+},0]\cup\sigma_{+})}(g_{+}^{i}(e)-g_{-}^{i}(e))de-\int_{E-W(\sigma_{-}\cup[0,\varphi_{r}^{-}])}(g_{+}^{i}(e)-g_{-}^{i}(e))de $$
The images $E-W([\varphi_{r}^{+},0]\cup\sigma_{+})$ and
$E-W(\sigma_{-}\cup[0,\varphi_{r}^{-}])$ are two paths of energies
connecting $E_{r}$ to $E_{i}$ (see figure \ref{imagchem}C). By
analyticity of $(g_{+}^{i}-g_{-}^{i})$ in the domain
$\Ra(e)\in]E_{r},E_{i}[$ , we obtain that:
\begin{equation}
\label{coeffrho}
\int_{[\varphi_{r}^{+},0]\cup\sigma_{+}}(\omega_{+}^{i}(u,E)-\omega_{-}^{i}(u,E))du-\int_{\sigma_{-}\cup[0,\varphi_{r}^{-}]}(\omega_{+}^{i}(u,E)-\omega_{-}^{i}(u,E))du=2\int_{E_{r}}^{E_{i}}(g_{+}^{i}-g_{-}^{i})(e)de
\end{equation}
It remains to show that this coefficient is purely imaginary. To
do that, we point out that $(g_{-}^{i})^{*}=g_{+}^{i}$, according
to \eqref{symband}. Equation \eqref{coeffrho} becomes:
$$\int_{[\varphi_{r}^{+},0]\cup\sigma_{+}}(\omega_{+}^{i}-\omega_{-}^{i})(u,E)du-\int_{\sigma_{-}\cup[0,\varphi_{r}^{-}]}(\omega_{+}^{i}-\omega_{-}^{i})(u,E)du$$ $$=2\left(\int_{E_{r}}^{E_{i}}g_{+}^{i}(e)de-\int_{E_{r}}^{E_{i}}(g_{+}^{i})^{*}(e)de\right)=2\left(\int_{E_{r}}^{E_{i}}g_{+}^{i}(e)de-\overline{\int_{E_{r}}^{E_{i}}g_{+}^{i}(e)de}\right)$$
\end{itemize}
This ends the proof of Lemma \ref{coeffut}.
\end{dem}
\psset{unit=1em,linewidth=.05}

\psset{unit=0.5em,linewidth=.1}
\begin{center}
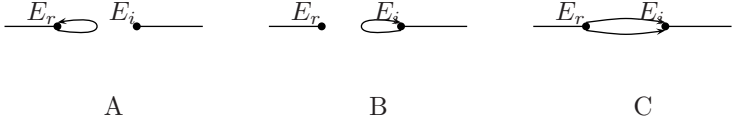
\begin{figure}
\begin{pspicture}(-30,-6)(30,4)
\psline(-30,0)(-26,0)\psline(-20,0)(-15,0)\psdots[dotstyle=*](-20,0)\psdots[dotstyle=*](-26,0)
\pscurve{->}(-26,-0.3)(-23,0)(-26,0.3)\uput[180](-25,1){$E_{r}$}\uput[180](-19,1){$E_{i}$}
\uput[180](-20,-6){A}
\psline(-10,0)(-6,0)\psline(0,0)(5,0)\psdots[dotstyle=*](0,0)\psdots[dotstyle=*](-6,0)
\pscurve{->}(0,-0.3)(-3,0)(0,0.3)\uput[180](-5,1){$E_{r}$}\uput[180](1,1){$E_{i}$}
\uput[180](0,-6){B}
\psline(10,0)(14,0)\psline(20,0)(25,0)\psdots[dotstyle=*](20,0)\psdots[dotstyle=*](14,0)
\pscurve{->}(14,-0.3)(17,-0.6)(20,-0.3)\pscurve{->}(14,0.3)(17,0.6)(20,0.3)\uput[180](15,1){$E_{r}$}\uput[180](21,1){$E_{i}$}
\uput[180](20,-6){C}
\end{pspicture}
\caption{Paths in the complex plane of energy}\label{imagchem}
\end{figure}
\end{center}
\subsection{Equation for the eigenvalues}
\label{demeigeneq} The following result gives a characterization
of the eigenvalues of $H_{\varphi,\varepsilon}$.
\begin{prop} \label{eigeneq2}
We assume that $(H_{V})$, $(H_{W,r})$, $(H_{W,g })$ and $(H_{J})$
are satisfied.\\
There exist $\varepsilon_{0}>0$, a neighborhood
$\mathcal{V}=\overline{\mathcal{V}}$ of $J$, two functions
$(E,\varepsilon)\mapsto\widetilde{\Phi}(E,\varepsilon)$ and
$(E,\varepsilon)\mapsto\widetilde{\Phi_{d}}(E,\varepsilon)$
defined on $\mathcal{V}\times]0,\varepsilon_{0}[$ and two
functions $(\varphi,E,\varepsilon)\mapsto
F(\varphi,E,\varepsilon)$ and $(\varphi,E,\varepsilon)\mapsto
R_{2}(\varphi,E,\varepsilon)$ defined on
$\R\times\mathcal{V}\times]0,\varepsilon_{0}[$ such that:
\begin{enumerate}\item $E$ is an eigenvalue of $H_{\varphi,\varepsilon}$ if and only if:
$$ F(\varphi,E,\varepsilon)=0$$
\item The function $F$ satisfies:
$$\forall\varphi\in \R,\ \forall E\in\mathcal{V},\ \forall\varepsilon\in]0,\varepsilon_{0}[,\quad F^{*}(\varphi,E,\varepsilon)=\overline{F(\overline{\varphi},\overline{E},\varepsilon)}=F(\varphi,E,\varepsilon).$$
\item The function $\varphi\mapsto F(\varphi,E,\varepsilon)$ is $\varepsilon$-periodic and its Fourier expansion is written:
\begin{equation}
\label{decsfoud}
F(\varphi,E,\varepsilon)=\cos\left(\frac{\widetilde{\Phi}(E)}{\varepsilon}\right)+e^{-S(E)/\varepsilon}\cos\left(\frac{\widetilde{\Phi_{d}}(E)}{\varepsilon}+\frac{2\pi\varphi}{\varepsilon}+\rho\right)+e^{-S(E)/\varepsilon}R_{2}(\varphi,E,\varepsilon)
\end{equation}
\item The functions $\widetilde{\Phi}$, $\widetilde{\Phi_{d}}$
satisfy the following properties for any
$\varepsilon\in]0,\varepsilon_{0}[$:
\begin{itemize}\item $E\mapsto\widetilde{\Phi}(E,\varepsilon)$ and
$E\mapsto\widetilde{\Phi_{d}}(E,\varepsilon)$ are analytic on
$\mathcal{V}$.
\item $\widetilde{\Phi}(E,\varepsilon)=\Phi(E)+o(\varepsilon)$ and
$\widetilde{\Phi_{d}}(E,\varepsilon)=\Phi_{d}(E)+o(\varepsilon)$
uniformly for $E\in\mathcal{V}$.
\end{itemize}
\item For any $\varepsilon\in]0,\varepsilon_{0}[$, the function $(\varphi,E)\mapsto
R_{2}(\varphi,E,\varepsilon)$ is analytic on
$\R\times\mathcal{V}$. Besides, there exists a constant $\alpha>0$
such that,for all $\varepsilon\in]0,\varepsilon_{0}[$, and all $E$
in $\mathcal{V}$, the function $R_{2}$ satisfies the following
properties:
$$ \int_{0}^{\varepsilon}R_{2}(u,E,\varepsilon)du=0,\quad\int_{0}^{\varepsilon}R_{2}(u,E,\varepsilon)e^{\frac{ 2i\pi u}{\varepsilon}}du=0,\quad\int_{0}^{\varepsilon}R_{2}(u,E,\varepsilon)e^{\frac{-2i\pi u}{\varepsilon}}du=0,$$
$$\sup\limits_{\varphi\in\R,E\in\mathcal{V}}|R_{2}(\varphi,E,\varepsilon)|\leq e^{-\frac{\alpha}{\varepsilon}}$$
\end{enumerate}
The functions $\Phi$, $\Phi_{d}$, $S$ are defined in Lemma
\ref{phaseactint}. $\rho$ is a real number defined in
\eqref{omegac}.
\end{prop}
Now, we prove Proposition \ref{eigeneq2}.
\begin{itemize}\item Now, it suffices to compute the Fourier
expansion of:
$$b_{g}^{-}(b_{d}^{+})^{*}(\varphi,E,\varepsilon)=\sum\limits_{n\in\Z}\gamma_{n}(E,\varepsilon)e^{\frac{2i
n\pi\varphi }{\varepsilon}}.$$ By using the asymptotic expansion
of the coefficients $b_{g}^{-}$ and $b_{d}^{+}$ given in Lemma
\ref{colina}, we prove that:
$$\gamma_{0}=-ie^{-\frac{i}{\varepsilon}\int_{\varphi_{r}^{-}}^{\varphi_{r}^{+}}\kappa_{i}}
[1+o(1)].$$
$$\gamma_{1}=+ie^{-\frac{i}{\varepsilon}(\int_{\varphi_{r}^{+}}^{0}\kappa_{i}+\int_{\varphi_{r}^{-}}^{0}\kappa_{i})}e^{\frac{2i}{\varepsilon}\int_{0}^{\overline{\varphi_{i}}}(\kappa_{i}-\pi)}e^{\int_{0}^{\varphi_{r}^{+}}(\omega_{+}^{i}-\omega_{-}^{i})}e^{\int_{0}^{\varphi_{i}}(\omega_{-}^{i}-\omega_{+}^{i})}[1+o(1)].$$
$$\left|\sum\limits_{n\in\Z\backslash\{0,1\}}\gamma_{n}e^{\frac{2i
n\pi\varphi
}{\varepsilon}}\right|=O(e^{-\alpha/\varepsilon})\quad\textrm{
pour }\varphi\in S_{Y_{0}}.$$ Actually,
$$\gamma_{0}=\alpha_{0}\beta_{0}+\sum\limits_{n\neq 0}\alpha_{n}\beta_{-n}=-i
e^{-\frac{i}{\varepsilon}\int_{\varphi_{r}^{-}}^{\varphi_{r}^{+}}\kappa_{i}}e^{\frac{1}{2}\left[\int_{0}^{\varphi_{r}^{-}}(\omega_{+}^{i}-\omega_{-}^{i})+\int_{0}^{\varphi_{r}^{-}}(\omega_{-}^{i}-\omega_{+}^{i})\right]}[1+o(1)].$$
According to \eqref{omegaa}, we simplify:
$$\left[\int_{0}^{\varphi_{r}^{-}}(\omega_{+}^{i}-\omega_{-}^{i})+\int_{0}^{\varphi_{r}^{-}}(\omega_{-}^{i}-\omega_{+}^{i})\right]=0.$$
According to \eqref{phi},
$\int_{\varphi_{r}^{-}}^{\varphi_{r}^{+}}\kappa_{i}=\Phi(E)$.
Consequently,
$$\gamma_{0}=-i e^{\frac{i\Phi(E)}{\varepsilon}}[1+o(1)].$$ We compute:
$$\gamma_{1}=\alpha_{0}\beta_{1}+\sum\limits_{n\neq 1}\alpha_{n}\beta_{1-n}$$
We start with computing $\alpha_{0}\beta_{1}$. To do that, we
deduce from equation \eqref{phida} that:
$$\int_{\varphi_{r}^{+}}^{0}\kappa_{i}(\varphi)d\varphi+\int_{\varphi_{r}^{-}}^{0}\kappa_{i}(\varphi)d\varphi-\int_{\sigma_{-}}(\kappa_{i}(\varphi)-\pi)d\varphi=\Phi_{d}(E)+iS(E).$$
$$\alpha_{0}\beta_{1}=ie^{-\frac{i}{\varepsilon}(\int_{\varphi_{r}^{+}}^{0}\kappa_{i}+\int_{\varphi_{r}^{-}}^{0}\kappa_{i})}e^{\frac{2i}{\varepsilon}\int_{0}^{\overline{\varphi_{i}}}(\kappa_{i}-\pi)}e^{\int_{0}^{\varphi_{r}^{+}}\omega_{+}^{i}+\int_{\varphi_{r}^{-}}^{0}\omega_{-}^{i}-\int_{\sigma_{+}}(\omega_{+}^{i}-\omega_{-}^{i})}[1+o(1)]+O(e^{\frac{-\alpha}{\varepsilon}})e^{-S(E)/\varepsilon}.$$
Equation \eqref{phida} leads to:
$$\int_{\varphi_{r}^{+}}^{0}\kappa_{i}(\varphi)d\varphi+\int_{\varphi_{r}^{-}}^{0}\kappa_{i}(\varphi)d\varphi-\int_{\sigma_{-}}(\kappa_{i}(\varphi)-\pi)d\varphi=\Phi_{d}(E)+iS(E).$$
Besides, according to Lemma \ref{coeffut}, we have:
$$\int_{0}^{\varphi_{r}^{+}}\omega_{+}^{i}+\int_{\varphi_{r}^{-}}^{0}\omega_{-}^{i}-\int_{\sigma_{+}}(\omega_{+}^{i}-\omega_{-}^{i})=i\rho.$$
and:
$$\alpha_{0}\beta_{1}=ie^{-S/\varepsilon}e^{-i\Phi_{d}/\varepsilon}e^{i\rho}[1+o(1)].$$
Since $S(E)\leq 2\pi\I\varphi_{i}(E)$, we estimate the remainder
in the expansion:
$$|\sum\limits_{n\neq
0}\alpha_{n}\beta_{-n}|=o(e^{-S/\varepsilon}).$$ Finally, for
$p\neq 0,1$, we estimate:
$$\gamma_{p}=\sum\limits_{n\in\Z}\alpha_{n}\beta_{p-n}.$$
For $p>1$, we have:
$$|\gamma_{p}|=e^{-S/\varepsilon}e^{-\alpha/\varepsilon}O(e^{-\frac{2\pi Y_{0}(p-1)}{\varepsilon}}).$$
Similarly, we estimate for $p<0$,
$$|\gamma_{p}|=e^{-S/\varepsilon}e^{-\alpha/\varepsilon}O(e^{-\frac{2\pi Y_{0}(|p|-1)}{\varepsilon}}).$$
\item Now, we consider $\varphi\in\R$. We compute the Fourier asymptotic expansion of the coefficient $d(E,\varphi,\varepsilon)$
in a neighborhood $\mathcal{U}_{0}$ of $E_{0}$:
$$d(\varphi,E,\varepsilon)=iw(f_{i},\sigma_{i}(f_{i})^{*})\left(\lambda_{0}(E,\varepsilon)+\sum\limits_{n\in\N^{*}}(\lambda_{n}(E,\varepsilon)e^{\frac{2i n\pi\varphi}{\varepsilon}}+(\lambda_{n})^{*}(E,\varepsilon)e^{\frac{-2i n\pi\varphi}{\varepsilon}})\right)$$
$$=i(w_{0}k'_{i})(E-W(0))\sum\limits_{n\in\N}u_{n}(\varphi,E,\varepsilon).$$
where
$u_{n}(\varphi,E,\varepsilon)=\lambda_{n}(E,\varepsilon)e^{\frac{2i
n\pi\varphi}{\varepsilon}}+(\lambda_{n})^{*}(E,\varepsilon)e^{\frac{-2i
n\pi\varphi}{\varepsilon}}$, pour $n\in\N^{*}$, et $u_{0}(\varphi,E,\varepsilon)=\lambda_{0}(E,\varepsilon)$.\\
We have:
$$u_{0}(\varphi,E,\varepsilon)=\gamma_{0}(E,\varepsilon)-\gamma_{0}^{*}(E,\varepsilon)=-ie^{i\frac{\Phi}{\varepsilon}}g(E,\varepsilon)-ie^{-i\frac{\Phi}{\varepsilon}}g^{*}(E,\varepsilon).$$
where $g(E,\varepsilon)=1+o(1)$.\\
We define
$g(E,\varepsilon)=r_{g}(E,\varepsilon)e^{i\theta_{g}(E,\varepsilon)}$
where the functions $E\mapsto r_{g}(E,\varepsilon)$ and $E\mapsto
\theta_{g}(E,\varepsilon)$ are analytic and satisfy $$
r_{g}^{*}=r_{g},\quad r_{g}=1+o(1)\quad
\theta_{g}^{*}=\theta_{g},\quad\theta_{g}=o(1).$$ We simplify:
$$u_{0}(\varphi,E,\varepsilon)=-i r_{g}(E,\varepsilon)\cos\left(\frac{\Phi(E)}{\varepsilon}+\theta_{g}(E,\varepsilon)\right).$$
Similarly, we compute:
$$u_{1}(\varphi,E,\varepsilon)=i
r_{h}(E,\varepsilon)e^{-S(E)/\varepsilon}\cos\left(\frac{\Phi_{d}+2\pi\varphi}{\varepsilon}+\rho+\theta_{h}(E,\varepsilon)\right).$$
where the functions $E\mapsto r_{h}(E,\varepsilon)$ and $E\mapsto
\theta_{h}(E,\varepsilon)$ are analytic and satisfy $$
r_{h}^{*}=r_{h},\quad r_{h}=1+o(1)\quad
\theta_{h}^{*}=\theta_{h},\quad\theta_{h}=o(1).$$ In addition, we
have the following estimate of the remainder:
$$\left|\sum\limits_{p\geq 2}u_{p}(\varphi,E,\varepsilon)\right|\leq
Ce^{\frac{-S(E)}{\varepsilon}}e^{\frac{-\alpha}{\varepsilon}}\quad\textrm{
pour }\varphi\in \R.$$
\item We have proved that, for $E$
in a neighborhood of $E_{0}$, the Fourier expansion of
$d(E,\varphi,\varepsilon)$ can be written:
\begin{equation}
\label{decsfou}
\frac{d(\varphi,E,\varepsilon)}{i(w_{0}k'_{i})(E-W(0))}=-i[1+o(1)]\cos\left(\frac{\Phi(E)}{\varepsilon}+o(1)\right)
\end{equation}
$$+i[1+o(1)]e^{\frac{-S(E)}{\varepsilon}}\cos\left(\frac{\Phi_{d}+2\pi\varphi}{\varepsilon}+\rho+o(1)\right)+e^{\frac{-S(E)}{\varepsilon}}O(e^{\frac{-\alpha}{\varepsilon}}).$$
The compactness of $J$ implies that there exists a finite number
of intervals $\{J_{k}\}_{k\in\{1\cdots p \}}$ such that:
\begin{enumerate}
\item $J\subset\bigcup\limits_{k\in\{1\cdots p \}}J_{k}$ \item
For any $k\in\{1,\cdots ,p-1 \}$, the intervals $J_{k}$ and
$J_{k+1}$ overlap.\item For any $k\in\{1,\cdots, p \}$, there
exists a complex neighborhood $\mathcal{U}_{k}$ of $J_{k}$ such
that the expansion (\ref{decsfou}) is satisfied on
$\mathcal{U}_{k}$.
\end{enumerate}
We shall prove that we can define some functions
$\widetilde{\Phi}$ and $\widetilde{\Phi}_{d}$ on the whole
neighborhood $\mathcal{V}=\bigcup\limits_{k\in\{1\cdots p
\}}\mathcal{U}_{k}$. To do that, we shall ``stick'' the expansions obtained on each interval.\\
The coefficient $u_{0}$ is written:
$$\forall E\in J_{k},\quad u_{0}(E,\varepsilon)=r_{0,k}(E,\varepsilon)\cos\left(\frac{\Phi(E)}{\varepsilon}+\theta_{0,k}(E,\varepsilon)\right)$$
$$\forall E\in J_{k+1},\quad u_{0}(E,\varepsilon)=r_{0,k+1}(E,\varepsilon)\cos\left(\frac{\Phi(E)}{\varepsilon}+\theta_{0,k+1}(E,\varepsilon)\right)$$
where $r_{0,k}(E,\varepsilon)=1+o(1)$ and
$\theta_{0,k}(E,\varepsilon)=o(1)$ (resp.
$r_{0,k+1}(E,\varepsilon)=1+o(1)$ and
$\theta_{0,k+1}(E,\varepsilon)=o(1)$) for $E\in J_{k}$ (resp.
$E\in J_{k+1}$).We get that:
$$r_{0,k}(E,\varepsilon)=r_{0,k+1}(E,\varepsilon)=r_{0}(E,\varepsilon)\textrm{ et } \theta_{0,k}(E,\varepsilon)=\theta_{0,k+1}(E,\varepsilon)=\theta_{0}(E,\varepsilon)\textrm{ for
}E\in J_{k}\cap J_{k+1}$$ The function $\widetilde{\Phi}$ defined
by its restrictions to each $\mathcal{U}_{k}$ is analytic on $\mathcal{V}$.\\
The case of $\widetilde{\Phi_{d}}$ is treated similarly.
\end{itemize}
Defining
$$F(\varphi,E,\varepsilon)=\frac{d(\varphi,E,\varepsilon)}{i(w_{0}k'_{i})(E-W(0))r_{0}(E,\varepsilon)},$$
we finish the proof of Proposition \ref{eigeneq2}.
\subsection{Localization of the eigenvalues}
In this section, we deduce Theorem \ref{eigenloc} from Proposition \ref{eigeneq2}. \\
We solve equation $F(\varphi,E,\varepsilon)=0$, where $F$ is
described in (\ref{decsfoud}).
\subsubsection{Energy levels $E^{(l)}(\varepsilon)$}
\label{koe} For $E\in \mathcal{V}$, we start with solving:
\begin{equation}
\label{phasemodif}
\cos\frac{\widetilde{\Phi}(E,\varepsilon)}{\varepsilon}=0
\end{equation}
$E\mapsto\widetilde{\Phi}(E,\varepsilon)$ is a real analytic
function. For a sufficiently small $\varepsilon_{0}$, by Lemma
\ref{phaseactint}, there exists a constant $m>0$ such that:
\begin{equation} \label{phasemodifdiff}
\forall
E\in\mathcal{V},\quad\forall\varepsilon\in]0,\varepsilon_{0}[,\quad
|\widetilde{\Phi}'(E,\varepsilon)|\geq m
\end{equation}
Consequently, equation \eqref{phasemodif} has a finite number of
zeros in $J$. We denote them by $E^{(l)}(\varepsilon)$, for
$l\in\{L_{-}(\varepsilon),\dots,L_{+}(\varepsilon)\}$. They are
given by:
\begin{equation}
\label{zeroa}
\frac{\widetilde{\Phi}(E^{(l)}(\varepsilon),\varepsilon)}{\varepsilon}=l\pi+\frac{\pi}{2},\quad\quad\forall
l\in\{L_{-}(\varepsilon),\ldots,L_{+}(\varepsilon)\}.
\end{equation}
and satisfy:
\begin{equation}
\label{ecart}
E^{(l+1)}(\varepsilon)-E^{(l)}(\varepsilon)=\frac{1}{\widetilde{\Phi}'(E^{(l)}(\varepsilon))}\pi\varepsilon+o(\varepsilon).
\end{equation}
The distances between two consecutive zeros are of order
$\varepsilon$. Precisely, by combining \eqref{phasemodifdiff} with
\eqref{ecart}, we obtain that there exists a constant $c>0$ such
that:
\begin{equation}
\label{ecarta}
\frac{1}{c}\varepsilon<|E^{(l+1)}(\varepsilon)-E^{(l)}(\varepsilon)|<c\varepsilon,\quad
\forall l\in \{L_{-}(\varepsilon),\ldots,L_{+}(\varepsilon)-1\}
\end{equation}
First, we prove that the zeros of $F$ are in an exponentially
small neighborhood of the points $E^{(l)}(\varepsilon)$.
\subsubsection{First order approximation}
We give a first order approximation of the zeros of $F$.\\
We set
$$a_{0}(E,\varepsilon)=\cos\frac{\widetilde{\Phi}(E,\varepsilon)}{\varepsilon}.$$
We can assume that the neighborhood $\mathcal{V}$ is sufficiently
small and such that, for any $E\in\mathcal{V}$,
$$\Ra(S(E))>\beta>0.$$ Then, there exists a positive constant $A$
such that
$$|F(\varphi,E,\varepsilon)-a_{0}(E,\varepsilon)|<Ae^{-\beta/\varepsilon}.$$
In addition, we have the following inequality:
\begin{equation}
\label{cosinus} \exists C>0/\quad
\left|\cos\frac{\widetilde{\Phi}(E,\varepsilon)}{\varepsilon}\right|\geq\frac{C}{\varepsilon}d(E,\bigcup\limits_{l\in\{L_{-},\cdots,L_{+}\}}E^{(l)}(\varepsilon)).
\end{equation}
Actually, there exists a constant $c>0$ such that:
$$|\cos\theta|\geq c d(\theta,\pi\Z+\pi/2).$$ By using \eqref{phasemodifdiff}, we obtain the relation:
$$|\widetilde{\Phi}(E,\varepsilon)-\widetilde{\Phi}(E^{(l)}(\varepsilon),\varepsilon)|\geq m |E-E^{(l)}(\varepsilon)|$$
and finally:
$$\left|\cos\frac{\widetilde{\Phi}(E,\varepsilon)}{\varepsilon}\right|\geq\frac{C}{\varepsilon}d\left(E,\bigcup\limits_{l\in\{L_{-},\cdots,L_{+}\}}E^{(l)}(\varepsilon)\right).$$
For $z_{0}\in\C$ and $r>0$, we define $$D(z_{0},r)=\{z\in\C\ ;\
|z-z_{0}|<r\}.$$ Inequality (\ref{cosinus}) implies that there are
no zeros of $F$ outside exponentially small neighborhoods of the
points $E^{(l)}(\varepsilon)$. Precisely, there exists a positive
constant $D$ such that, if $r\geq D\varepsilon
e^{-\beta/\varepsilon}$, then for any $E\in
\partial D(E^{(l)}(\varepsilon),r)$, we have:
$$|F(\varphi,E,\varepsilon)-a_{0}(E,\varepsilon)|<|a_{0}(E,\varepsilon)|.$$
Rouch{\'e}'s Theorem implies that, for any $l$, $F$ has exactly
one zero $E_{l}(\varphi,\varepsilon)$, in each neighborhood
$D(E^{(l)}(\varepsilon),D\varepsilon e^{-\beta/\varepsilon})$ of
$E^{(l)}(\varepsilon)$. The relation $F=F^{*}$ allows us to
recover that the eigenvalues are real. Indeed, if $F(E)=0$, $\overline{E}$ is also a zero of $F$. By uniqueness, we obtain that $E=\overline{E}$.\\
We set:
$$E_{l}(\varphi,\varepsilon)=E^{(l)}(\varepsilon)+\varepsilon\lambda_{l}(\varphi,\varepsilon).$$
We know that $\lambda_{l}(\varphi,\varepsilon)$ is exponentially
small. Now, we compute its asymptotic behavior.
\subsubsection{Second order approximation}
We define:
$$a_{1}(\varphi,E,\varepsilon)=F(\varphi,E,\varepsilon)-a_{0}(E,\varepsilon).$$
We write
$$e^{-S(E_{l}(\varphi,\varepsilon))/\varepsilon}=e^{-S(E^{(l)}(\varepsilon))/\varepsilon}(1+O(\lambda_{l}(\varphi,\varepsilon))).$$
Similarly, with the help of the modified phase
$\widetilde{\Phi_{d}}$, we obtain the expansion:
$$\cos\left(\frac{\widetilde{\Phi_{d}}(E_{l}(\varphi,\varepsilon))+2\pi\varphi+\rho\varepsilon}{\varepsilon}\right)=\cos\left(\frac{\widetilde{\Phi_{d}}(E^{(l)}(\varepsilon))+2\pi\varphi+\rho\varepsilon}{\varepsilon}\right)+O(e^{-\beta/\varepsilon})$$
The expansion of $a_{1}$ can be written:
$$a_{1}(\varphi,E_{l}(\varphi,\varepsilon),\varepsilon)=a_{1}(\varphi,E^{(l)}(\varepsilon),\varepsilon)(1+r(\varphi,E^{(l)}(\varepsilon),\varepsilon)).$$
Moreover, we use the first order Taylor's expansion of the
function $E\mapsto a_{0}(E,\varepsilon)$:
$$a_{0}(E_{l}(\varphi,\varepsilon),\varepsilon)=(-1)^{l+1}\widetilde{\Phi}'(E^{(l)}(\varepsilon),\varepsilon)\lambda_{l}(\varphi,\varepsilon)(1+r(\varphi,E^{(l)},\varepsilon))=(-1)^{l+1}\Phi'(E^{(l)})\lambda_{l}(\varphi,\varepsilon)(1+o(1)).$$
By combining these computations, we finally obtain:
$$\lambda_{l}(\varphi,\varepsilon)=\frac{(-1)^{l+1}}{\Phi'(E^{(l)}(\varepsilon))}e^{-S(E^{(l)}(\varepsilon))/\varepsilon}\left(\cos\left(\frac{\widetilde{\Phi_{d}}(E^{(l)}(\varepsilon),\varepsilon)+2\pi\varphi}{\varepsilon}+\rho\right)+o(1)\right).$$
\subsection{Application to the trace formula} \label{trform2} In
\cite{Di1}, the author proves the existence of an asymptotic
expansion of $\tr [f(H_{\varphi,\varepsilon})]$, for $f\in
C_{0}^{\infty}$, when $\textrm{Supp }f$ is disjoint from the bands
of $H_{0}$; in addition, he computes explicitly the first and
second
terms of this expansion.\\
Corollary \ref{cordim} allows us to recover these terms.
\subsubsection{}
Let $J$ be an interval satisfying $(H_{J})$. Particularly, $J$ is
such that $J\cap(\sigma_{ac}\cup\sigma_{sc})=\emptyset$. For $f\in
C_{0}^{\infty}$, with $\textrm{Supp }f\subset J$, we compute:
$$\tr[f(H_{\varphi,\varepsilon})]=\sum\limits_{l\in\{L_{-}(\varepsilon),...,L_{+}(\varepsilon)\}}f(E_{l}(\varphi,\varepsilon)).$$
Let $\beta>0$ be such that $S(E)>\beta$ for any $E\in J$;
according to Theorem \ref{eigenloc}, we know that there exists a
constant $C>0$ such that:
$$\forall u\in[0,\varepsilon],\quad\left|\tr[f(H_{\varphi,\varepsilon})]-\tr[f(H_{u,\varepsilon})]\right|<C\sum\limits_{l\in\{L_{-}(\varepsilon),...,L_{+}(\varepsilon)\}}\varepsilon e^{-\beta/\varepsilon}.$$
By integrating with respect to $u$, we obtain that:
$$\textrm{tr
}[f(H_{\varphi,\varepsilon})]=\frac{1}{\varepsilon}\int_{0}^{\varepsilon}\textrm{tr
}[f(H_{u,\varepsilon})]du+O(e^{-\beta/\varepsilon}).$$ According
to Theorem \ref{eigenloc}, we know that there exists a constant
$C$ such that
$$\forall u\in[0,\varepsilon],\quad\left|\tr
[f(H_{u,\varepsilon})]-\sum\limits_{l\in\{L_{-}(\varepsilon),...,L_{+}(\varepsilon)\}}f(E^{(l)}(\varepsilon))\right|<C
e^{-\beta/\varepsilon}$$ By integration, we obtain:
\begin{equation}
\label{formtrace}
\frac{1}{\varepsilon}\int_{0}^{\varepsilon}\textrm{tr
}[f(H_{u,\varepsilon})]du=\sum\limits_{l\in\{L_{-}(\varepsilon),...,L_{+}(\varepsilon)\}}f(E^{(l)}(\varepsilon))+O(e^{-\beta/\varepsilon})
\end{equation}
Now, we estimate:
$$\sum\limits_{l\in\{L_{-}(\varepsilon),...,L_{+}(\varepsilon)\}}f(E^{(l)}(\varepsilon))=\sum\limits_{l\in\{L_{-}(\varepsilon),...,L_{+}(\varepsilon)\}}f\circ\widetilde{\Phi}^{-1}(\varepsilon(l\pi+\pi/2))$$
\subsubsection{} Now, we compute this last term.
\begin{lem}
\label{trace} Let $f$ be a function in $C_{0}^{\infty}$ such that
$\textrm{Supp }f\subset J$. The trace of $H_{\varphi,\varepsilon}$
has the following asymptotic behavior:
$$\int_{0}^{\varepsilon}\tr
[f(H_{u,\varepsilon})]du=\frac{1}{\pi}\int_{J}
f(\E)\widetilde{\Phi}'(\E,\varepsilon)d\E+O(\varepsilon^{\infty})$$
\end{lem}
\begin{dem}
The proof of this Lemma is based on elementary results of real
analysis.
\begin{itemize} \item We apply the Poisson formula
to the function $f\circ\widetilde{\Phi}^{-1}\in C_{0}^{\infty}$:
$$\varepsilon\sum\limits_{l\in\Z}f\circ\widetilde{\Phi}^{-1}(\varepsilon(l\pi+\pi/2))=2\sum\limits_{n\in\Z}(-1)^{n}\widehat{(f\circ\widetilde{\Phi}^{-1})}\left(\frac{2n}{\varepsilon}\right).$$
Besides, the Fourier transform of $f\circ\widetilde{\Phi}^{-1}$
satisfies the estimates:
$$\forall\nu>1,\quad\exists\ C_{\nu}>0,\textrm{ such that }\left|\widehat{(f\circ\widetilde{\Phi}^{-1})}\left(\frac{2n}{\varepsilon}\right)\right|\leq
C_{\nu}\frac{\varepsilon^{\nu}}{n^{\nu}}.$$ Actually, since
$f\circ\widetilde{\Phi}^{-1}$ is $C^{\nu}$,
$|\xi^{\nu}\widehat{f\circ\widetilde{\Phi}^{-1}}(\xi)|$ is bounded.\\
This leads to:
$$\varepsilon\sum\limits_{p\in\Z}f\circ\widetilde{\Phi}^{-1}(\varepsilon(p\pi+\pi/2))=2\widehat{(f\circ\widetilde{\Phi}^{-1})}(0)+O(\varepsilon^{\infty}).$$
\item It remains to prove that:
$$2\widehat{(f\circ\widetilde{\Phi}^{-1})}(0)=\frac{1}{\pi}\int
f\circ\widetilde{\Phi}^{-1}(u)du.$$ With the substitution
$u=\widetilde{\Phi}(\E)$, we obtain that:
$$2\widehat{(f\circ\widetilde{\Phi}^{-1})}(0)=\frac{1}{\pi}\int_{J}
f(\E)\widetilde{\Phi}'(\E,\varepsilon)d\E$$ This completes the
proof of Lemma \ref{trace}.
\end{itemize}
\end{dem}
\subsubsection{Conclusion}
To get an asymptotic expansion of the trace at any order, it
suffices to know an asymptotic expansion of the modified phase at
any order. Our computations are not accurate enough, but we know
that $\widetilde{\Phi}'(\E)=\Phi'(E)+o(\varepsilon)$, hence:
$$\frac{1}{\pi}\int_{J} f(\E)\widetilde{\Phi}'(\E)d\E=\frac{1}{\pi}\int_{J}
f(\E)\Phi'(\E)d\E+o(\varepsilon).$$ To transform the right member
of previous equality, we do the substitution
$(\kappa,u)\mapsto(E(\kappa)+W(u),u)$, which implies:
$$\frac{1}{\pi}\int_{J}
f(\E)\Phi'(\E)d\E=\frac{1}{2\pi}\int_{[-\pi,\pi]}\int_{\varphi_{r}^{-}}^{\varphi_{r}^{+}}f(E(\kappa)+W(u))d\kappa
du$$ We finally obtain:
$$\int_{0}^{\varepsilon}\textrm{tr
}[f(H_{u,\varepsilon})]du=\frac{1}{2\pi}\int_{[-\pi,\pi]}\int_{\varphi_{r}^{-}}^{\varphi_{r}^{+}}f(E(\kappa)+W(u))d\kappa
du+o(\varepsilon)$$ This ends the proof of Corollary \ref{cordim}.
\subsection{Asymptotic behavior of the eigenvalues}
Now, we give a second application of Theorem \ref{eigenloc} for
the computation of the asymptotic behavior of the eigenvalues of
$H_{\varphi,\varepsilon}$. Such a computation is outlined in
\cite{CDS}, in the case $V=0$. We obtain an explicit result at
first order.\\
Under the assumptions of Theorem \ref{eigenloc}, $E_{r}$ is the
only end of $\sigma(H_{0})$ belonging to $(E-W)(\R)$. We define:
\begin{equation}
\label{rnp} d_{p}(E_{n})=\lim\limits_{\stackrel{E\rightarrow
E_{n},}{ E\in[E_{n},E_{p}]}}\frac{k(E)-k(E_{n})}{\sqrt{E-E_{n}}}
\end{equation}
\begin{cor}
Let $H_{\varphi, \varepsilon}$ verify the assumptions of Theorem
\ref{eigenloc}. The eigenvalues $E^{(l)}(\varphi,\varepsilon)$ of
$H_{\varphi,\varepsilon}$ have the following asymptotic behavior:
$$E^{(l)}(\varphi,\varepsilon)=\widetilde{\Phi}^{-1}(\varepsilon(l\pi+\pi/2))+0(\varepsilon^{\infty}).$$
Particularly, $E^{(l)}(\varphi,\varepsilon)$ has the following
Taylor expansion at first order in $\varepsilon$ :
$$E^{(l)}(\varphi,\varepsilon)=E_{r}+W(0)+\sqrt{\frac{W"(0)}{2}}\frac{1}{d_{i}(E_{r})}(2l+1)\varepsilon+o(\varepsilon),$$
where $d_{i}(E_{r})$ is defined by \eqref{rnp}.
\end{cor}
\begin{dem}
The first equality is obvious. It suffices to give an expansion of
$\widetilde{\Phi}^{-1}(\varepsilon(l\pi+\pi/2))$. To do that, we
compute an expansion at first order of:
$$\Phi(E_{r}+W(0)+\alpha)=\int_{\varphi_{r}^{-}(E_{r}+W_{-}+\alpha)}^{\varphi_{r}^{+}(E_{r}+W_{-}+\alpha)}k(E_{r}+W(0)+\alpha-W(u))du.$$
The mapping $W$ is a bijection from $[0,\varphi_{r}^{+}]$ to
$[W(0),E_{r}]$. By the substitution $\alpha v=W(0)+\alpha-W(u)$,
we get that:
$$\int_{0}^{\varphi_{r}^{+}(E_{r}+W(0)+\alpha)}=\alpha\int_{0}^{1}\frac{k(E_{r}+\alpha v)}{W'\circ W^{-1}(W(0)+\alpha(1-v))}dv.$$
But, $\lim\limits_{\alpha\rightarrow 0}\frac{k(E_{r}+\alpha
v)}{W'\circ W^{-1}(W_{-}+\alpha(1-v))}=\frac{d_{i}(E_{r})}{\sqrt{2
W"(0)}}\frac{\sqrt{v}}{\sqrt{1-v}}$.\\
Similarly, on $[\varphi_{r}^{-},0]$, we have:
$$\Phi(E_{r}+W(0)+\alpha)=d_{i}(E_{r})\frac{\pi}{2}\sqrt{\frac{2}{W"(0)}}\alpha[1+o(1)].$$
Consequently, by inverting the expansion of $\widetilde{\Phi}$ in
the neighborhood of $E_{r}+W(0)$, we prove the result.
\end{dem}\\
We point out that, as in \ref{trform2}, a more accurate asymptotic
expansion of $\widetilde{\Phi}$ would give a better result on the
eigenvalues.
\bibliographystyle{plain}
\bibliography{bibliographieart}
\end{document}